\documentclass[a4paper,10pt]{elsarticle}

\usepackage{amsmath}       
\usepackage{amsfonts}
\usepackage{amssymb}
\usepackage{textcomp}
\usepackage{subcaption}
\usepackage[title]{appendix}
\pdfoutput=1

\newcommand{\dint}{\displaystyle\int}
\newcommand{\dsum}{\displaystyle\sum}

\newcommand{\jump}[1]{\ensuremath{[\![#1]\!]} }

\renewcommand\appendix{\par
  \setcounter{section}{0}
  \setcounter{subsection}{0}
  \setcounter{figure}{0}
  \setcounter{table}{0}
  \renewcommand\thesection{Appendix \Alph{section}}
  \renewcommand\thefigure{\Alph{section}\arabic{figure}}
  \renewcommand\thetable{\Alph{section}\arabic{table}}
}

\begin{document}

\title{Phase field modeling of partially saturated deformable porous media}
\author{Giulio Sciarra}
\address{ Dipartimento Ingegneria Chimica Materiali Ambiente,\\
Universit\`a di Roma La Sapienza, Via Eudossiana 18, 00184 Rome, Italy\\
\emph{giulio.sciarra@uniroma1.it}}
\date{}

\begin{abstract}
A poromechanical model of partially saturated deformable porous media is proposed based on a phase field approach at modeling the behavior of the mixture of liquid water and wet air, which saturates the pore space, the phase field being the saturation (ratio). While the standard retention curve is expected still to provide the intrinsic retention properties of the porous skeleton, depending on the porous texture, an enhanced description of surface tension between the wetting (liquid water) and the non-wetting (wet air) fluid, occupying the pore space, is stated considering a regularization of the phase field model based on an additional contribution to the overall free energy depending on the saturation gradient.
The aim is to provide a more refined description of surface tension interactions.

An enhanced constitutive relation for the capillary pressure is established together with a suitable generalization of Darcy's law, in which the gradient of the capillary pressure is 
replaced by the gradient of the so-called generalized chemical potential, which also accounts for the \lq\lq force\rq\rq\, associated to the local free energy of the phase field model. A micro-scale heuristic interpretation of the novel constitutive law of capillary pressure is proposed, in order to compare the envisaged model with that one endowed with the concept of average interfacial area. 

The considered poromechanical model is formulated within the framework of strain gradient theory in order to account for possible effects, at laboratory scale, of the micro-scale hydro-mechanical couplings between highly-localized flows (fingering) and localized deformations of the skeleton (fracturing).
\end{abstract}


\maketitle

\section*{Introduction}
The constitutive characterization of partially saturated porous media became of interest at the beginning of the last century when scientific research started to face fundamental problems in geotechnics and petroleum engineering, concerning modeling the response of partially imbibed soils, during imbibition/drainage cycles (see \cite{Buckingham1907,Richards1931}), or modeling the behavior of sedimentary reservoir rocks, when a multi-phase fluid flows through the porous network. Starting from the analysis of basic static problems, it became clear that the balance between capillary and driving forces, in particular gravitational forces, would have been the central subject of modeling efforts. This pushed the research in the direction of finding out simple relations between the curvature of the wetting/non-wetting fluid interface and the average content of the wetting fluid, over a suitably defined Representative Volume Element (RVE). The main ideas of this identification have been clearly sketched by \cite[Chapter 6]{Coussy_book10} for a partially saturated truncated conical pore, specifying the infinitesimal variation of the interfacial energy in terms of the infinitesimal variation of the volume occupied by the wetting fluid (saturation). This naturally provided the well-known definition of macro-scale capillary pressure, as the derivative of a macro-scale capillary energy. 
In \cite{Leverett1940}, \cite{Brooks1964}, in the summarizing contribution by \cite{Bear_book1972}, in \cite{VanGenuchten1980} etc., different semi-empirical relations, between macro-scale capillary pressure and saturation have been stated, in order to specify the retention properties of the porous skeleton. These retention curves typically exhibit hysteresis during imbibition/drainage cycles, so as to fit with experimental evidence.

At the same time  the pioneering papers by Cahn and Hilliard \cite{CahnHilliard1958,Cahn1959,CahnHilliard1959} established the basic framework within which modeling of multi-phase fluid flow is formulated in terms of space and time evolution of the mass concentration or, in the general case, of a phase field which can vary continuously over thin interfacial layers. Beyond the original formulation due to \cite{CahnHilliard1958} one can refer to \cite{Jacqmin2000} and \cite{Kim12} for general reviews, as well as  to \cite{Lowengrub98}, \cite{LaPuerta06}, \cite{Boyer2010} or \cite{Lamorgese2011}. Surface tension is recovered, in this context, considering the integral, through the thickness of the layer, of the concentration gradient (or the gradient of the phase field). This approach progressively attracted more and more interest, in particular within the framework of fluid mechanics, because of its advantages for numerical calculations which do not necessitate adaptive interface fitting grids, see \textit{e.g.} \cite{Jacqmin2000,Kim12,Gomez2011}. Surprisingly however limited contributions attempted at incorporating these ideas into modeling of unsaturated porous media, one can refer for instance to \cite{Papatzacos2002} and \cite{Papatzacos2004} and more recently to \cite{Cueto2008, Cueto2009, Cueto-Juanes09} and \cite{Gomez2013}.

This paper follows the research path traced by these authors to model partially saturated deformable porous media, considering the pore network infused with a two-phase fluid. Also the constitutive characterization of the solid will be generalized with respect to the classical poromechanical model, in order to describe the coupling between highly localized flows and (possibly localized) strains.

For long time the above mentioned constitutive characterization of the macro-scale capillary pressure in terms of the corresponding hysteretic retention curve has been the only relation used for describing the hydraulic flow through partially saturated porous media, being also the pivot of the hydro-mechanical coupling with the constitutive law of the porous skeleton. Due to the coarse simplification provided by this model, however, it was finally recognized the existence of several problems in which the relation between capillary pressure and saturation is not sufficient to describe at the same time surface tension, between the wetting and the non-wetting fluid, and the retention properties of the porous skeleton, due to its texture. As a matter of fact what is understood by this constitutive law is, as already mentioned, that no variation of the capillary pressure can be observed if no change of the saturation is induced by external forces. This is due to the rough upscaling rule, from the micro to the macro-scale, which assumes the variation of the area of interfaces proportional to the volume occupied by the fluids, see \cite{Coussy_book10}. On the other hand, as originally pointed out by \cite{Morrow1970}, redistribution of the fluids within a porous medium can occur also keeping constant the saturation, and instabilities can even arise from fluid interfaces that are unable to change curvature smoothly with variations in pressure (\lq\lq Haines jumps\rq\rq). In other words, different micro-scale configurations of interfaces between a wetting and a non-wetting phase are possible for a given saturation.

In more refined macro-scale formulations the specific cumulative measure of interfaces, between the wetting and the non-wetting phase, say the specific interfacial area, is also introduced to account for the micro-scale features of the fluids within the RVE, see \cite{HassanizadehGray93} and the related literature, \textit{e.g.} \cite{Reeves1996, Papatzacos2002, Niessner2008}. In order to clarify the meaning of this additional macro-scale state parameter, one can think of the effect of the saturation degree, on the macro-scale capillary pressure, as that which coarsely accounts for the retention characteristics of the porous skeleton, depending on its texture, and consider on the other hand the effect of the cumulative measure of interfaces as a suitable corrective term, which allows for describing different admissible coexistence configurations of the saturating fluids. Within this framework, several experimental campaigns have been carried out, based on laboratory tests, see \textit{e.g.} \cite{Dalla2002, Culligan2004, Costanza2011}, and also pore network  micro-scale numerical simulations have been implemented, see \textit{e.g.} \cite{Joekar2010}, in order to characterize the constitutive relation among capillary pressure, saturation and specific interfacial area. 

Consider now the behavior of the porous skeleton, it is almost standard in continuum poromechanics to introduce a constitutive prescription of stress in terms of strain and saturation, both in the case of elastic and elasto-plastic deformations; one can refer among others to the seminal paper by \cite{Alonso1990} and the related literature, see \textit{e.g.} \cite{Olivella1996, Alonso1999}, to the systematic formulation by \cite{Schrefler_book1998}, see also \cite{Sanavia2002,Sanavia2006}, to the approach based on the generalized Bishop effective stress, see \textit{e.g.} \cite{Houlsby1997, Gens2006, Buscarnera2012,Jommi2000,Tamagnini2004} etc., or to some recent contributions within the range of finite deformations of the porous skeleton, see \cite{Borja2013} and \cite{Borja2014}. Possible modifications of these constitutive models, in order to account for the effect of the specific interfacial area, concern the improvement of the retention curve provided within the approach to multi-phase mechanics proposed by \cite{Hassanizadeh1990,HassanizadehGray93}, in which velocities of phases and interfaces are employed to describe fluid flow and skeleton deformation, see also \cite{Gray2001,Gray2009, Nikooee2013}. Within this framework, however, several phenomena, related to micro-structure remodeling, can not be captured, exactly as several phenomena related to the spatial distribution of fluids within the pores could not be described only in terms of saturation evolution. The typical case is that of localized deformations which have been observed in laboratory scale experimental tests, since the beginning of the 70s, initially using medical scanners for imaging sand bodies, see \textit{e.g.} \cite{Vardoulakis1978} and \cite{Desrues1985}, and later with the aid of RX computerized tomography and Digital Image Correlation,  see \textit{e.g.} \cite{Desrues2004,Lenoir2007,Hall2010} etc. 
Thanks to the improvements in the resolving power of tomographs as well as to the progressively increasing skills of scientists in the treatment of experimental data, dry and (partially) saturated specimens have been imaged under triaxial loading conditions to observe localized deformations and porosity change, see \textit{e.g.} \cite{Ando2012_1,Ando2012_2} and \cite{Desrues2015}.

In order to model the effects at the scale of the specimen of these micro-scale grain displacements, and in particular to obtain the regularization of localized deformation patterns, the gradient theory of elasto-plasticity, as well as the gradient approach to damage and fracture mechanics, have been adopted in the literature. Typical references are \cite{Sulem2004}, \cite{Chambon2001} and \cite{Matsushima2002}, as well as \cite{Marigo1998} and \cite{Bourdin2000,Bourdin2012}. In all these cases however the hydro-mechanical coupling, if any, still remains restricted to the constitutive prescription of the effective or net stress, see \textit{e.g.} \cite{Collin2006} and \cite{Sieffert2009}.

As already mentioned, in this paper a novel general approach is developed, within the framework of continuum poromechanics, which aims at merging phase field modeling of multi-phase fluid flow with unsaturated strain gradient poromechanics. While the standard retention curve is expected still to provide the intrinsic retention properties of the porous skeleton, depending on the porous texture, an enhanced description of surface tension between the wetting and the non-wetting fluid, occupying the pore space, is stated considering a regularized phase field model (in the sense of \cite{CahnHilliard1958}). Similarly to the model which introduces
the specific interfacial area as a corrective term of the standard retention curve, this alternative formulation allows for detecting variations of capillary forces even when no change in saturation is caused by the external loading. Indeed a similar approach has already been addressed in the literature, attempting at modeling gravitational fingering and saturation overshooting, through an undeformable porous skeleton, or viscous fingering through a Hele-Shaw cell, see \cite{Cueto-Juanes09,Cueto-Juanes12,Cueto-Juanes14}. However no account has been taken of the possible coupling with the deformations of the porous skeleton. In the model which is going to be presented in this paper, on the other hand, the behavior of the solid skeleton is described by means of a strain gradient model, to predict localized strains and fracture, when considering irreversible processes. To the best of author's knowledge, no general model accounting for both the above mentioned items, say localization in fluid flow and strain, can be retrieved from the literature, except a preliminary study by the author \cite{Sciarra2013}, stemming from previous results on modeling of porous media saturated by quasi-incompressible fluids, see \cite{dellisola2003,Sciarra07,Sciarra08} and \cite{CIS2009,CIS2010,CIS2011,CIS2012,CIS2013}. Within this enhanced framework, the hydro-mechanical coupling is therefore responsible not only for the effects of average variations of the hydraulic regime within the RVE on the skeleton deformation, and vice-versa, 
see among others \cite{Olivella1996, Alonso2010, Tamagnini2004, Casini2012,Rotisciani2014,Rotisciani2015}, but it can also account for the effects of capillary fingering on damaging and fracturing of the skeleton and vice-versa for the effects of strain localization on permeability variations and heterogeneous/anisotropic fluid flow. 

The paper is organized as follows: in section 1 the basics of kinematics of (partially) saturated porous media are summarized. In section 2 the expression of the virtual working of external forces, relative to a second gradient multi-phase continuum, is manipulated to obtain the virtual working of internal forces, which is consistent with the balance of the overall momentum. In section 3 the first and the second principle of thermodynamics are introduced in order to get the generalized form of the Clausius-Duhem inequality, prescribing the solid, fluid and thermal dissipations. In section 4, the constitutive laws for the effective stress and hyper-stress are established, considering the contribution of strain and saturation gradients to the overall free energy of the porous continuum; the generalized Darcy law is also deduced. Section 5 is devoted to discuss the properties of the so-called pore-fluid, constitutively characterized in terms of a suitable double well potential, specifying the two phases of the fluid within the pores. In section 6 the generalized prescription of the macro-scale capillary pressure is placed, accounting for its dependence on the saturation gradient. In section 7 the governing equations, say the overall balance law and the generalized Darcy law, are deduced by means of a variational approach which also provides the proper boundary conditions. Finally in section 8 some conclusions are stated.

\section{Kinematics}\label{S:Kin}
A macroscopic description of kinematics is adopted, treating the partially saturated porous medium as the superposition of a skeleton and a binary mixture of liquid water and wet air. The current placement of the skeleton particles is provided by the deformation of a reference configuration $\mathcal D_0$. The partial saturation of the pore space is accounted for thinking of the fluid mixture as a non-uniform fluid in the sense of  \cite{CahnHilliard1958}, \textit{i.e.} a fluid possibly having a spatial variation in one of its intensive scalar properties. 

Let $\chi_{\alpha}: \mathcal D_0\times \mathcal I\rightarrow \mathtt{I}\!\mathtt{E}$ be the $\alpha$-th component of the deformation (placement) of the skeleton particles with respect to a fixed orthonormal frame in the Euclidean space $\mathtt{I}\!\mathtt{E}$ of positions; $\mathcal D_0$ is the reference configuration of the solid constituent and $\mathcal I$ a time interval. The image of $\mathcal D_0$ under the deformation map is the current configuration of the porous medium, say $\mathcal D \subset \mathtt{I}\!\mathtt{E}$. Moreover let $u_{\alpha}$ indicate the displacement of the solid particles, $F_{\alpha i}=\chi_{\alpha,i}$ the deformation gradient, along the $i$-th direction of an orthonormal frame in the reference configuration of the solid, and $E_{ij}:=\left(F_{\alpha i} F_{\alpha j}-\delta_{ij}\right)/2$ the associated strain tensor, $\delta_{ij}$ being the Kronecker delta. From now on Greek indices label the components of any $n$-th order tensor with respect to a frame in the Euclidean space, whilst Latin indices label the components of any $n$-th order tensor with respect to a frame into the reference configuration of the skeleton $\mathcal D_0$. Finally let $v_s^{\alpha}$ be the $\alpha$-th component of the velocity of the skeleton particles. 
As usual within the framework of poromechanics, the notions of Eulerian porosity $n$ and  Lagrangian porosity $\phi$ are introduced, as the current pore volume density per unit volume of the porous continuum, and the current pore volume density per unit reference volume, respectively. Clearly $\phi=J\, n$; $J$ being the determinant of $F$. 

Concerning the non-uniform fluid the intensive scalar property, used from now on to characterize the biphasic characteristic of the fluid, is not the mass concentration but the volume density of the liquid phase with respect to the volume of the pores, say the degree of saturation. The liquid phase of the mixture (water) is assumed incompressible and the gaseous phase (wet air) is assumed to be passive (that is, have infinite mobility) so that its density can be neglected with respect to that of the liquid, $\rho_L$; the apparent density of the fluid mixture, say the mass density of the non-uniform fluid per unit volume of the porous medium, is therefore definitely prescribed in terms of the volumetric liquid content $\theta$. If $S_r$ indicates the saturation degree (saturation ratio), measuring the current volume occupied by the liquid per unit volume of the pores, the liquid content is $\theta=n\, S_r$ and the fluid apparent density is $(n\,\rho_f)=\rho_L (n\, S_r)=\rho_L\, \theta$. The saturation degree ranges within the interval $[0,1]$, where $S_r=0$ and $S_r=1$ indicate the gaseous and the liquid phase, respectively. The kinematics of the non-uniform fluid is specified as in \cite{Sciarra08}; the reference placement of the non-uniform fluid particles is therefore prescribed by means of a regular map $g: \mathcal D_0\times \mathcal I\rightarrow \mathcal D_0^f$ defined on the reference configuration of the skeleton. It identifies the fluid material particle within a reference domain $\mathcal D_0^f$, which occupies, at time $t$, the same current place as the solid particle, chosen in $\mathcal D_0$. Accordingly the fluid velocity coincides with the time derivative of $\chi^f(\cdot,t):=\chi(\cdot,t)\circ g(\cdot,t)^{-1}$, which means for every reference particle  $X\in\mathcal D_0$ and every current place $x=\chi(X,t)$:
\begin{equation}\label{fluid velocity}
v^f_{\alpha}(x,t)=v^s_{\alpha}(X,t) - F_{\alpha i}(X,t)\, \left(G(X,t)^{-1}\right)_{i k} \left.\dfrac{\partial g_k}{\partial t}\right|_{(X,t)},
\end{equation}
where $G_{k i}:=g_{k,i}$; the Einstein summation for repeated indices is understood in equation \eqref{fluid velocity}. Similarly to the deformation gradient of the solid skeleton also the gradient of $\chi^f(\cdot,t)$ can be defined in terms of the gradient of $\chi$ and $g$ as $F^f=FG^{-1}$.

Both the saturation degree and the liquid content are naturally defined on the current configuration of the porous medium, however the corresponding pull-back in the reference configuration of the skeleton can easily be defined by means of the inverse of the deformation map $\chi(\cdot,t)$. 
The mass conservation of the fluid mixture can be stated, following \cite{Coussy_book04}, in the form 
\begin{equation}\label{fluid mass Coussy}
 \dfrac{d\, m_f}{dt} + M_{k,k}=0, \quad \text{with}\quad m_f=\phi \rho_f \quad \text{and}\quad M_k:= \phi \rho_f \left(F^{-1} \right)_{k \alpha} \left(v^f_{\alpha}-v^s_{\alpha} \right),
\end{equation}
where $m_f$ is the Lagrangian fluid mass content and $M_k$ is the Lagrangian filtration vector; the time derivative is computed keeping fixed the reference placement of the solid particle. According with previous remarks, the incompressibility of the liquid, say $\rho_L=\textrm{const}$, implies equation \eqref{fluid mass Coussy} to reduce to:
\begin{equation}\label{fluid mass incomp}
\dfrac{d}{dt} \left(J \theta \right)-\left[J \theta \left(G^{-1} \right)_{i k}\dfrac{d g_k}{d t} \right]_{,i}=0\quad \Leftrightarrow \quad J\, \theta=\phi\, S_r=\textrm{det}\, G.
\end{equation}
In equations \eqref{fluid mass Coussy}-\eqref{fluid mass incomp}, for the sake of simplicity, no explicit dependence of the considered fields on the current position or the corresponding placement in the skeleton configuration has been specified. It is worth to underline that even if the liquid is incompressible the non-uniform fluid is not, consequently no restriction on the fluid velocity is placed.

\section{External \& strain working: from a mixture model towards a Biot-like model}
Here an approach much similar to that developed by \cite{Sciarra07} is adopted, in order to deduce the strain working relative to a porous medium within the framework of gradient continuum mechanics. Starting from the standard formulation of the mechanics of superimposed continua, and using the almost classical formulation of gradient theories introduced by \cite{Toupin1962}, \cite{Mindlin1964} and \cite{germain73}, the external working of the solid-fluid mixture is defined as a continuous linear functional of the velocity fields $v^s$ and $v^f$:
\begin{equation}\label{Wext}
W^{\mathrm{ext}}\left(v^s,v^f\right)=
\dsum_c \left\lbrace \dint_{\mathcal D}b^c_{\alpha}\, v^c_{\alpha}+\dint_{\partial\mathcal D}\left(t^c_{\alpha}\, v^c_{\alpha}+\tau^c_{\alpha}\, v^c_{\alpha,\beta}\,m_{\beta}\right)+\dint_{{\mathcal E}}f^c_{\alpha}\, v^c_{\alpha}\right\rbrace,\quad c=\{s,f\}
\end{equation}
where $\partial\mathcal D$ is the boundary of the current configuration of the porous continuum, assumed differentiable almost everywhere, and ${\mathcal E}$ is the union of the edges of $\partial\mathcal D$, on which the normal $m$ to the boundary suffers a jump. Equation \eqref{Wext} implies that not only bulk forces $b^c_{\alpha}$ and surface tractions $t^c_{\alpha}$, but also double forces $\tau^c_{\alpha}$ and tractions per unit line $f^c_{\alpha}$ on both constituents $c=\{s,f\}$, are expected to be be balanced by proper internal tractions. This balance can be achieved considering the extended Cauchy theorem, see \cite{germain73,fdigssv}, which for the $c$-constituent reads
\begin{equation}\label{Cauchy}
\begin{array}{rlll}					
\left(\Sigma^c_{\alpha \beta}-\Pi^c_{\alpha \beta \gamma ,\gamma }\right)m_{\beta}-\left(\mathcal Q_{B \beta} \Pi^c_{\alpha \beta \gamma } m_{\gamma } \right)_{,B} & = t^c_{\alpha}, & & \textrm{on}\,\, \partial\mathcal D,\medskip\\ 
\Pi^c_{\alpha \beta \gamma} m_{\gamma } m_{\beta} & = \tau^c_\alpha &  & \textrm{on}\,\, \partial\mathcal D,\medskip\\ 
\jump{ \mathcal Q_{B \beta} \Pi^c_{\alpha \beta \gamma } m_{\gamma } \mu_B  } & =f^c_{\alpha}  & & \textrm{on}\,\, \mathcal E.
\end{array}
\end{equation}
$\Sigma^c_{\alpha \beta}$ and $\Pi^c_{\alpha \beta \gamma}$ are tensorial quantities defined over the current configuration, which represent the stress and the so-called hyper-stress, acting on the solid and the fluid. Let $\hat{x}_B$, $B=1,2$ indicate the local parametrization of the boundary $\partial \mathcal{D}$, the tensor field $\mathcal Q_{B \beta}:= \partial \hat{x}_B/\partial x_{\beta}$ is the projection tensor onto the tangent space of $\partial\mathcal D$, while the partial derivatives with respect to $\hat{x}_B$-coordinates indicate the surface-gradient; $\mu_B$ are the components in the surface coordinate system of the Darboux tangent-normal vector to each edge of $\partial\mathcal D$. Finally $\jump{ \cdot}=(\cdot)^+ - (\cdot)^-$ indicates the jump through the edge. Equations \eqref{Cauchy} imply that surface tractions depends on the curvature of $\partial \mathcal D$, say the surface gradient of the normal unit vector $m$, and, in the limit when this curvature tends to infinity, say over the edges $\mathcal E$ of $\partial \mathcal D$, that a line density of forces must arise.  Thinking of the pure solid, double forces generalize the concept of skew-symmetric couples, which is typical of Cosserat or couple-stress theories, bearing into account not only surface density of moment but also surface density of symmetric couples. The former, working on a combination of differential shearing, bending and torsion, the latter, on differential elongation. On the other hand, thinking of non-viscous but non-uniform fluids, characterized by internal capillarity, say Cahn-Hilliard fluids, double forces are associated only to symmetric couples working on differential liquid content or, within our framework, saturation ratio, which allows to describe surface tension effects, see \textit{e.g.} \cite{Seppecher1989,Seppecher1996}. In this case, which is the one considered in this paper, the hyper-stress tensor relative to the fluid reduces to $\Pi^f_{\alpha \beta \gamma}=\delta_{\alpha \beta}\, \pi^f_{\gamma}$.

In order to deduce a Biot-like poromechanical model the overall balance of momentum must hold true and the corresponding strain working must be characterized in terms of strains and stresses defined over the reference configuration of the skeleton.  Using equations \eqref{Cauchy}$_2$ and \eqref{Cauchy}$_3$ together with the orthogonal decomposition of the velocity gradients, given by $v^c_{\alpha,\beta}=\left(\mathcal Q_{B \beta} v^c_{\alpha,B} + v^c_{\alpha,\eta} m_{\eta} m_{\beta} \right)$, the external working \eqref{Wext} reduces to the following form:
\begin{equation}\label{Wext_red}
\begin{array}{rl}
W^{\mathrm{ext}}\left(v^s,v^f\right)&=\dsum_c\left\lbrace\dint_{\mathcal D} b^c_{\alpha} v^c_{\alpha} + \dint_{\partial\mathcal D} \left[\left(\Sigma^c_{\alpha \beta}-\Pi^c_{\alpha \beta \gamma ,\gamma }\right)m_{\beta}v^c_{\alpha} + \Pi^c_{\alpha \beta \gamma } m_{\gamma} v^c_{\alpha,\beta}\right]\right\rbrace\medskip \\
& =\dsum_c\left\lbrace \dint_{\mathcal D} \left[ b^c_{\alpha} v^c_{\alpha} +\left( \Sigma^c_{\alpha \beta}-\Pi^c_{\alpha \beta \gamma ,\gamma }\right)_{,\beta} v^c_{\alpha}+ \Sigma^c_{\alpha \beta} v^c_{\alpha,\beta} +\Pi^c_{\alpha \beta \gamma} v^c_{\alpha, \beta \gamma} \right]\right\rbrace.
\end{array}
\end{equation}
Let $\Sigma_{\alpha \beta}:=\Sigma^s_{\alpha \beta}+\Sigma^f_{\alpha \beta}$ and $\Pi_{\alpha \beta \gamma}:=\Pi^s_{\alpha \beta \gamma}+\Pi^f_{\alpha \beta \gamma}$ be the $(\alpha,\beta)$-th and the $(\alpha,\beta,\gamma)$-th components of the overall stress and hyper-stress tensor, respectively, let moreover $b_{\alpha}=b^s_{\alpha}+b^f_{\alpha}$ be the $\alpha$-th component of the overall bulk force per unit volume; in order for the overall porous medium to be balanced the following equation must hold true
\begin{equation}\label{overall_eq}
 \left(\Sigma_{\alpha \beta}-\Pi_{\alpha \beta \gamma,\gamma}\right)_{,\beta} +b_{\alpha}=0,
\end{equation}
for every place $x\in\mathcal{D}$. A consistent definition of the strain working of the porous skeleton can deduced from equation \eqref{Wext_red}:
\begin{equation}\label{W_def}
W^{\mathrm{ext}}=\dint_{\mathcal D}\left[b^f_{\alpha} w_{\alpha} +\left(\Sigma^f_{\alpha \beta} w_{\alpha} \right)_{,\beta} - \Pi^f_{\alpha \beta \gamma,\gamma \beta}\, w_{\alpha} + \Pi^f_{\alpha \beta \gamma}\, w_{\alpha, \beta \gamma} +\Sigma_{\alpha \beta} v_{\alpha,\beta} + \Pi_{\alpha \beta \gamma} v_{\alpha, \beta \gamma} \right] =: W^{\mathrm{in}},
\end{equation}
where, from now on, the following positions are assumed: $v_{\alpha}:=v^s_{\alpha}$ and $w_{\alpha}:=v^f_{\alpha}-v^s_{\alpha}$.

Following \cite{fdigssv} the generalized Piola-Kirchhoff overall stress tensors $S_{ij}$ and $P_{ijk}$ are defined as the pull-back in the reference configuration of the porous skeleton of the current stress and hyper-stress tensors:
\begin{equation}\label{Cauchy-Piola}
 \begin{array}{ll}
\Sigma_{\alpha \beta}&=J^{-1}\left[S_{i j} F_{\alpha i} F_{\beta j}+ P_{i j k} \,\left(F_{\alpha j} F_{\beta i, k}+F_{\alpha i, k} F_{\beta j}\right)\right],\medskip\\
\Pi_{\alpha \beta \gamma} &=J^{-1} \, P_{i j k} \, F_{\alpha j} F_{\beta i} F_{\gamma k}.
\end{array}
\end{equation}
Accordingly the strain working is re-written as
\begin{equation}\label{W_def_cap}
W^{\mathrm{in}}=\dint_{\mathcal D_0} \left\lbrace J b^f_{\alpha} w_{\alpha} + J \left[\left(\Sigma^f_{\alpha \beta} - \Pi^f_{\alpha \beta \gamma,\gamma} \right) w_{\alpha} \right]_{,\beta} + J  \left(\Pi^f_{\alpha \beta \gamma} w_{\alpha,\beta} \right)_{,\gamma}+ S_{i j} \dot E_{i j} + P_{i j k} \dot E_{i j, k}\right\rbrace,
\end{equation}
where $\dot {(\,)}$ indicates the time derivative, keeping the placement of the solid particles fixed in the reference configuration. As already mentioned the fluid is a non-uniform Cahn-Hilliard fluid, consequently the fluid hyper-stress and its pull-back in the reference configuration of the skeleton reduce to vectors: $\pi^f_{\gamma}=\left(\Pi^f_{\alpha \alpha \gamma}\right)/3$ and $\gamma_k=\left(P^f_{ijk} F_{\alpha j} F_{\alpha i}\right)/3$; which implies $\pi^f_{\beta}=J^{-1} F_{\beta k}\,\gamma_k$, with an obvious definition of the third order tensor $P^f_{ijk}$. On the other hand the fluid stress is split into a spherical and a deviatoric part, say $\Sigma^f_{\alpha \beta}=-p^f\, \delta_{\alpha \beta} + \Sigma'^f_{\alpha \beta}$, where this last does not a-priori vanish.

Consider the first three terms of equation \eqref{W_def_cap}, the following chains of equalities hold true:
\begin{equation}\label{W_def_ident}
\begin{array}{rl}
J b^f_{\alpha}\, w_{\alpha} &= J b^f_{\alpha}\, F_{\alpha k} \left(F^{-1} \right)_{k \beta} w_{\beta} = \left(J  F_{\alpha k}\,b^f_{\alpha}\right) \dfrac{M_k}{m_f},\medskip\\
J \left[\left(\Sigma^f_{\alpha \beta} - \Pi^f_{\alpha \beta \gamma,\gamma} \right) w_{\alpha} \right]_{,\beta} &= -\left\lbrace J \left(F^{-1} \right)_{k \beta} \left[\left(p^f+\pi^f_{\gamma,\gamma} \right)\delta_{\alpha \beta}-\Sigma'^f_{\alpha \beta}\right] F_{\alpha q} \dfrac{M_q}{m_f}\right\rbrace_{,k}\medskip\\
&= -\left[\left(J p^f + \gamma_{l,l} \right)\dfrac{M_k}{m_f} - J \left(F^{-1} \right)_{k \beta} \Sigma'^f_{\alpha \beta}  F_{\alpha q}\dfrac{M_q}{m_f}\right]_{,k}\medskip\\
J \left(\Pi^f_{\alpha \beta \gamma}\, w_{\alpha,\beta} \right)_{,\gamma} &= \left[J \left(F^{-1} \right)_{k\gamma} \pi^f_{\gamma}\, w_{\alpha,\alpha} \right]_{,k}=\left[\dfrac{\gamma_k}{J} \left(J\dfrac{M_l}{m_f} \right)_{,l}\right]_{,k},
\end{array}
\end{equation}
once considered that for every vector field $u$ one has $J u_{\alpha,\alpha} = \left[J \left(F^{-1} \right)_{i \alpha} u_{\alpha}\right]_{,i}$. Identities \eqref{W_def_ident} imply the strain working \eqref{W_def_cap} to be rewritten in terms of the pull-back of stresses in the reference configuration $\mathcal D_0$ as:
\begin{equation}\label{W_def_Lagr}
\begin{array}{rl}
W^{\mathrm{in}}=\dint_{\mathcal D_0} &\!\!\! \left\lbrace b^{0f}_k \dfrac{M_k}{m_f} -\left(J p^f \dfrac{M_k}{m_f} \right)_{,k} +\left(J \left(F^{-1} \right)_{k \beta} \Sigma'^{f}_{\alpha \beta} F_{\alpha q}\, \dfrac{M_q}{m_f}\right)_{,k} +S_{i j} \dot E_{i j} \right. \medskip\\
 &\left.-\left[\gamma_{l,l}\dfrac{M_k}{m_f}-\dfrac{\gamma_k}{J} \left(J \dfrac{M_l}{m_f} \right)_{,l}\right]_{,k} + P_{i j k} \dot E_{i j, k}\right\rbrace.
\end{array}
\end{equation}
where $b^{0f}_k:= J F_{\alpha k} b^f_{\alpha}$ is the $k$-th component of the pull-back of the bulk forces acting on the fluid, in the reference configuration of the skeleton.

\section{Thermodynamics}\label{S:Thermodynamics}
Thermodynamics of porous media has been summarized by \cite{Coussy_book04,Coussy_book10} for two or more monophasic superimposed interacting continua, say the solid skeleton and the fluids saturating the porous space.  In that case, the specific internal energies of the fluids, which fill the pores, are separately defined, whether they are in the liquid or in the gaseous phase; whilst the energy due to interfacial interactions between the fluids and among the solid and the fluids are incorporated into the macroscopic energy of the skeleton. This contribution to the energy accounts for the micro-scale adhesion properties among the constituents, coherently with the Young-Dupr\'e equation, which at the micro-scale states the equilibrium of the triple line keeping in contact the solid wall (of the pore), with the wetting and the non-wetting phase, see \textit{e.g.} \cite{deGennes1985}. Considering a geometrically simple configuration of the porous space and a rough upscaling rule, as that already mentioned in the Introduction, which assumes the variation of the area of interfaces proportional to the volume occupied by the phases, a macroscopic interfacial energy, per unit volume, can be explicitly separated from the energy of the skeleton and regarded as a function of the saturation ratio $S_r$ only, see \cite{Coussyetal2004}. Afterwards the micro-scale surface tension can be related to the derivative of this energy, with respect to $S_r$, which is interpreted a-posteriori as the so-called macro-scale capillary pressure, see \textit{e.g.} \cite{Bear_book1972,VanGenuchten1980} and \cite{Coussy_book04}. The macro-scale capillary pressure, defined in this way, accounts at the same time for two distinct features, say the effect of surface tension between the liquid and the gas, stored within any possible reservoir, and the effect of the retention characteristics of the porous material.

Several criticisms have been moved to this definition of the macroscopic capillary pressure; in particular Gray and Hassanizadeh \cite{Gray1989,Hassanizadeh1990, HassanizadehGray93} demonstrated that the spatial distribution of interfaces within a multi-phase system is fundamental in order to characterize the intrinsic state of the system. 

Here a novel approach is adopted in order to incorporate, into the macroscopic constitutive prescription, the role of the interface between the two fluids stored in the pores, describing the fluid mixture as a non-uniform diphasic fluid in the spirit of \cite{CahnHilliard1958}. A phase-field model with possible diffuse interface is considered, which is analogous to that one of (more than) two immiscible components incompressible flows, see \cite{Lowengrub98}, \cite{LaPuerta06}, or \cite{Kim12} for a general review. As already noticed in \S~\ref{S:Kin} and following \cite{Cueto-Juanes12,Cueto-Juanes14} the phase field here is the degree of saturation $S_r$.

\begin{figure}
\centering
\begin{subfigure}[b]{.45\textwidth}
\includegraphics[width=\textwidth]{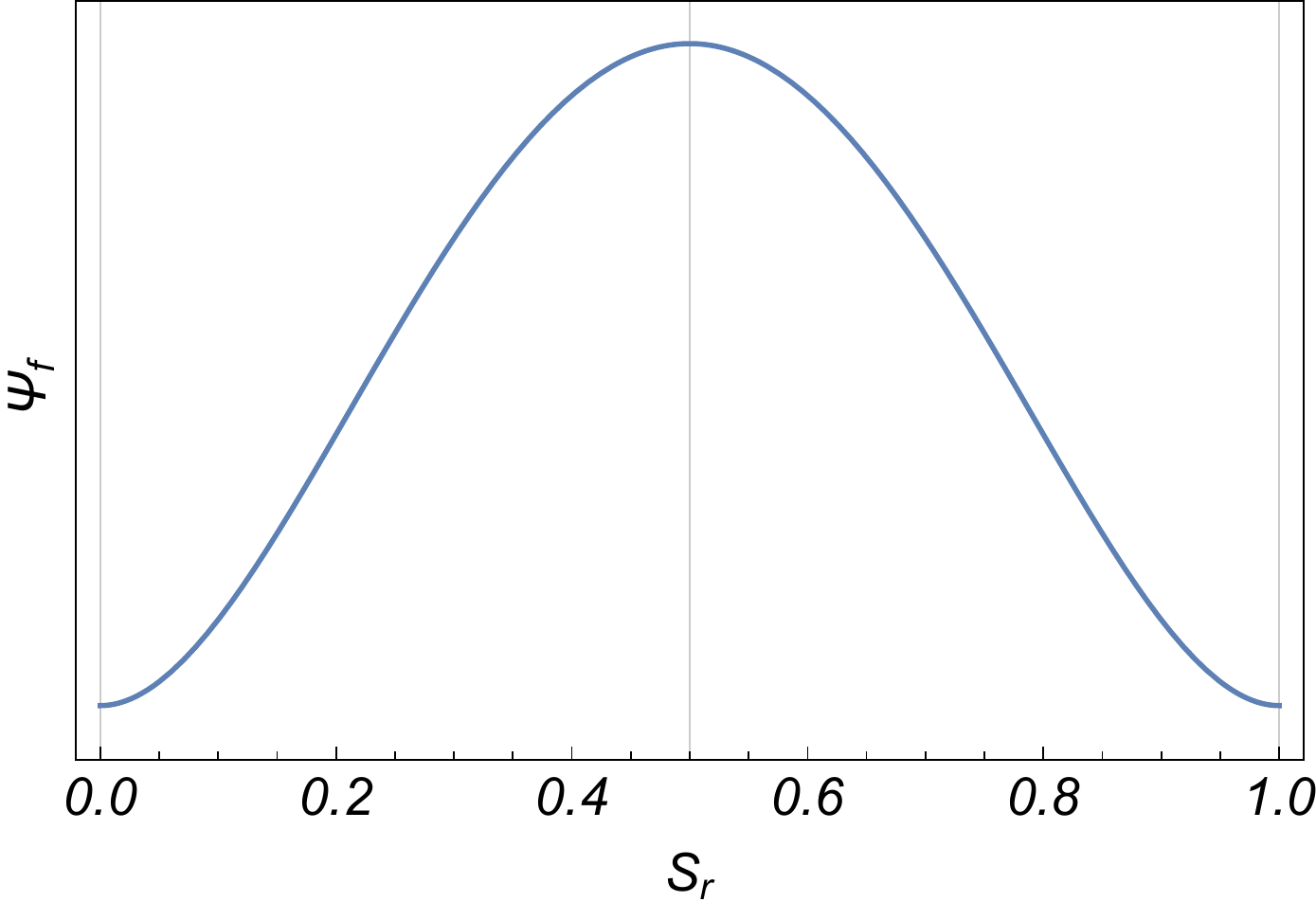}
\caption{Van der Waals' like potential}
\label{fig:fluidenergyplot}
\end{subfigure}
\quad
\begin{subfigure}[b]{.45\textwidth}
\includegraphics[width=\textwidth]{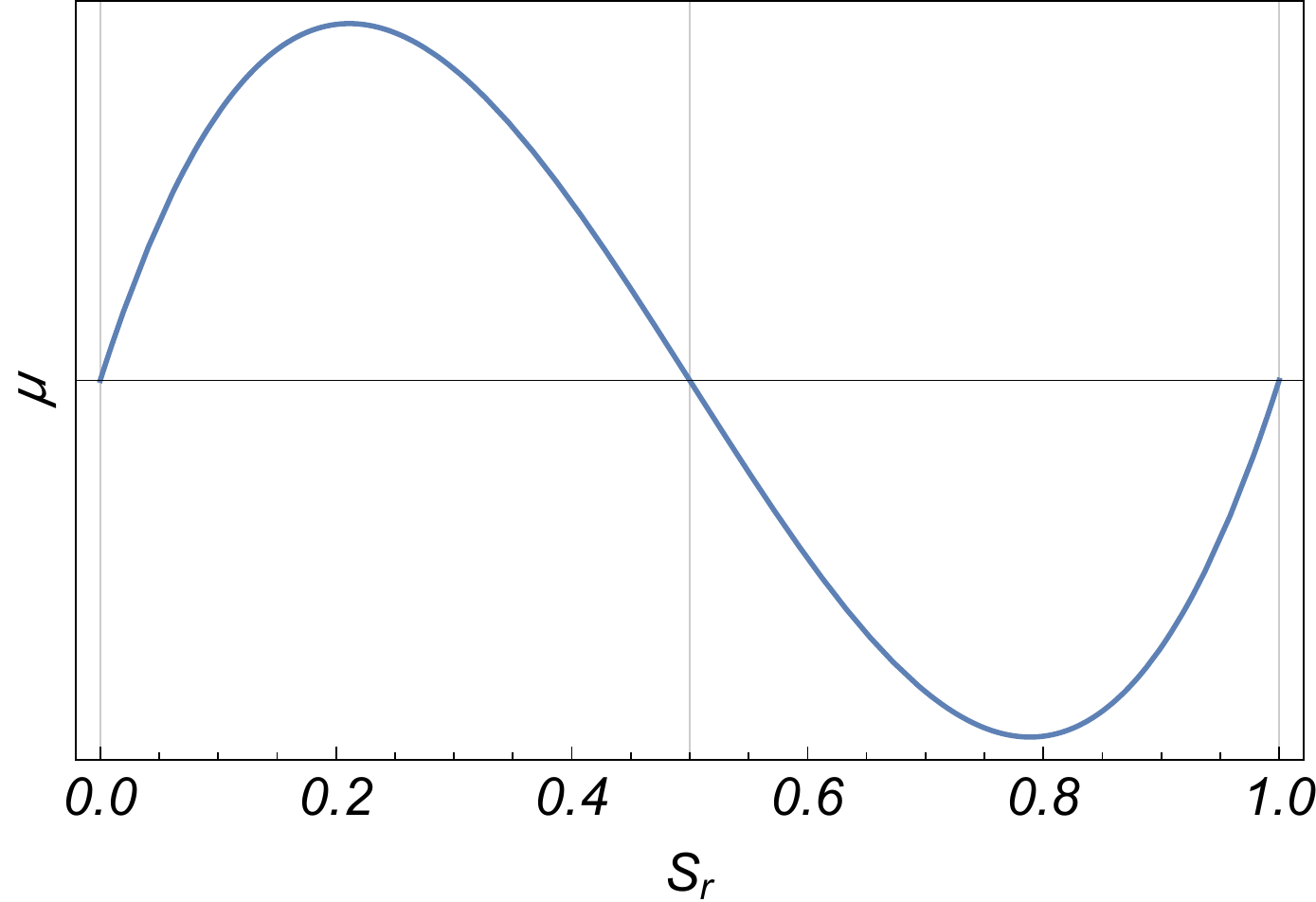}
\caption{Chemical potential}
\label{fig:fluidchemicalplot}
\end{subfigure}
\caption{The liquid and the gaseous phase coexist at equilibrium as they are isopotential minima of the fluid free energy}
\label{fig:fluid}
\end{figure}

The internal energy of the fluid is given by
\begin{equation}\label{fluid_energy}
\mathcal{E}_f=n\rho_f\, e_f\left(1/\rho_f,s_f \right) + \kappa_f\left(f_{\rho_f} \right), \quad f_{\rho_f}:=\delta_{\alpha \beta}\left(n\rho_f\right)_{,\alpha} \left(n\rho_f\right)_{,\beta},
\end{equation}
where the term $\rho_f e_f$ is a double-well potential depending on the specific density $\rho_f$ and parametrized by the specific entropy $s_f$. The non-local energy $\kappa_f\left( f_{\rho_f} \right)$ penalizes the formation of interfaces and provides a regularization of the non-convex energy contribution. 
As usual the state equation of the fluid defines a relation between conjugate variables; in particular the so-called fluid thermodynamic pressure $\mathcal P$, and the fluid chemical potential $\mu$ can be defined in terms of the  the density $\rho_f$:
\begin{equation}
\label{pressure_chempot}
\mathcal P:=-\left.\dfrac{\partial e_f}{\partial (1/\rho_f)}\right|_{s_f=\textrm{const}},\quad \mu:=\left.\dfrac{\partial \left(\rho_f e_f\right)}{\partial \rho_f}\right|_{T=\textrm{const}}.
\end{equation}
$T$ in equation \eqref{pressure_chempot} is the absolute temperature, which is conjugate to the specific entropy by $T=\partial e_f/\partial s_f$. The double-well volumetric energy resembles that of Van der Waals' model and allows for describing coexistence of the immiscible phases, even if no mass exchange is considered. No information on the domains occupied by the phases in the current configuration is supplied by the model if this contribution to the fluid energy is the only non-vanishing one. According to Maxwell's rule, an affine term, with respect to $\rho_f$, can be added to the Van der Waals-like energy to make arise two isopotential phases, characterized by the same (vanishing) value of the chemical potential $\mu$, see Figure \ref{fig:fluidenergyplot} and \cite{Coussy_book10}.
The non-local contribution allows for governing the coarsening of the domains occupied by the fluid phases and pattern formation. Indeed it provides just a correction of the bulk energy, when considering the so-called flat interface limit, see \cite{CahnHilliard1958}, whilst it constitutes the main part of the energy of the non-uniform fluid, when describing phenomena in which a clear scale separation can not be assumed, as in topological transitions in multi-phase fluids \cite{Lowengrub98} or gravity driven fingering through porous media \cite{Cueto-Juanes09}. 

Because of the constitutive law \eqref{fluid_energy}, the following prescriptions on the fluid stress and hyper-stress hold true
\begin{equation}\label{fluid_const}
\begin{array}{ll}
\left\lbrace
\begin{array}{l}
p^f = -\left[n\, \dfrac{\partial e_f}{\partial (1/\rho_f)}+\kappa_f -2 \left(1+\dfrac{1}{\mathrm{tr} I} \right)\dfrac{\partial \kappa_f}{\partial f_{\rho_f}} f_{\rho_f}\right],\medskip\\
\Sigma'^{f}_{\alpha \beta} = -2\, \dfrac{\partial \kappa_f}{\partial f_{\rho_f}}\left[\left(n\rho_f\right)_{,\alpha} \left(n\rho_f\right)_{,\beta}-\dfrac{1}{\mathrm{tr} I}\left(n\rho_f\right)_{,\gamma} \left(n\rho_f\right)_{,\gamma} \delta_{\alpha \beta}\right],
\end{array}
\right.
\qquad
&
\pi^f_{\alpha} = -2\, \dfrac{\partial \kappa_f}{\partial f_{\rho_f}} n \rho_f \left(n\rho_{f}\right)_{,\alpha},
\end{array}
\end{equation}
$I$ being the identity tensor and $\mathrm{tr} I=n$ the dimension of the Euclidean space $\mathtt{I}\!\mathtt{E}$. The thermodynamic pressure $\mathcal P$ and the chemical potential $\mu$ given in  \eqref{pressure_chempot} are just a part of these constitutive prescriptions. $\Sigma'^f$ represents the deviatoric stress acting on the fluid, which is non-vanishing because of the gradient contribution to energy. 

As already noticed incompressibility of the liquid phase means that the variations of the density $\rho_f$ are univocally determined by the variations of the degree of saturation, so that equation \eqref{fluid_const} can be rephrased in terms of the saturation ratio $S_r$, see \textit{e.g.} \cite{Cueto-Juanes12}. In particular the double well shape of the energy potential can be prescribed assuming the fluid energy per unit volume of the porous medium to be a kind of Duffing potential
\begin{equation}
\Psi_f:=\rho_f e_f=\mathrm{C}\, \dfrac{\gamma_{nw}}{R} S_r^2 \left(1-S_r \right)^2,
\label{Psif}
\end{equation}
whilst the non-local term is typically assumed quadratic in the gradient of $n S_r$:
\begin{equation}
\kappa_f\left(f_{\rho_f} \right)= \dfrac{\mathrm{C}_{\kappa}}{2}\, (n S_r)_{,\alpha} (n S_r)_{,\beta}\, \delta_{\alpha\beta}.
\label{kappa_f}
\end{equation}
Here $\gamma_{nw}$ is the surface tension between the non-wetting and the wetting phase, whilst $R$ is the characteristic size of the channel through which the fluid can pass, see \textit{e.g.} \cite{LaPuerta06}. In Figure~\ref{fig:fluid} the energy per unit volume and the corresponding chemical potential are plotted. The energy of the non-uniform fluid could also be prescribed in terms of a different phase field as for instance the mass concentration of the liquid without its double well feature being modified. With an abuse of notation from now on $\mu$ will indicate the derivative of $\Psi_f$ with respect to $S_r$ rather than to $\rho_f$, as indicated in equation \eqref{pressure_chempot}. 

As saturation can not overwhelm the limit $S_r=1$, a suitable constraint should be stated when formulating the minimization of the fluid energy potential $\Psi_f$ with respect to $S_r$, say $S_r\leq 1$ which corresponds to $S_r-1=-\alpha^2$, with $\alpha \geq 0$ a slack variable used to transform the inequality constraint into equality. This yields the introduction of a Lagrangian multiplier, in the functional to be made stationary, representing the reactive chemical potential, $\mu^{\mathrm{r}}$, which allows to account for the transition from partially-saturated to fully-saturated conditions. In partially-saturated states $S_r<1$ implies $\alpha\neq 0$ and consequently $\mu^{\mathrm{r}}=0$; on the other hand when saturation is attained $\alpha$ vanishes and the reactive chemical potential plays the role of the liquid pressure in standard saturated poromechanics. Looking at equation \eqref{kappa_f} it is worth to notice that in the case of saturation the non-local contribution to energy still does not vanish, but reduces to a function of the porosity gradient.

The effect of confinement of the air-water mixture into the porous space, say the retention characteristics of the porous material, will be accounted for by an additional energy $\mathcal E_{sf}$. Within the considered framework, no a-priori constitutive characterization is assumed on this functional, but suitable restrictions on it are deduced from the first and the second principle of thermodynamics, summarized in the generalized form of the Clausius-Duhem (dissipation) inequality.

In the following paragraphs the explicit form of the dissipation relative to the partially saturated porous medium is stated, taking in due account the expression of the strain working given by equation \eqref{W_def_Lagr}. The corresponding restrictions on the constitutive law of the porous skeleton and the local diffusion fluid mass flux $M_k$ are deduced.

%

\subsection{The first principle of thermodynamics}\label{first_principle}
Modifying the form of the energy equation, stated by \cite[equation (3.14)]{Coussy_book04} for a standard porous continuum, so as to separate the interaction energy between the solid and the non-uniform fluid implies the first principle of thermodynamics to read as
\begin{equation}\label{1st law}
\dfrac{d_s}{dt}\dint_{\mathcal D} \left(\mathcal E_s+\dfrac{1}{2} \mathcal E_{s f}\right)+\dfrac{d_f}{dt}\dint_{\mathcal D} \left(\mathcal E_f+\dfrac{1}{2} \mathcal E_{s f}\right)= W^{\mathrm{in}} + \mathring{\mathbb Q}.
\end{equation}
where $d_c/dt$ indicates the time derivative following the motion of the particle $c=\{s,f \}$. In a similar way to the fluid, the energy of the solid is defined in terms of the corresponding intrinsic energy as $\mathcal E_s=(1-n)\rho_s e_s$. Moreover the rate of change of the coupling energy $\mathcal{E}_{sf}$ is associated partly to the motion of the solid grains, partly to the motion of the molecules of the fluid, say the liquid and the gas. $\mathring{\mathbb Q}$ accounts for the rate of heat externally supplied and can be prescribed in terms of a surface rate of heat due to conduction,
\begin{equation}
 \mathring{\mathbb Q} = - \dint_{\partial \mathcal D} q_{\alpha} m_{\alpha} = - \dint_{\partial\mathcal D_0} \mathfrak{q_i}\, m^0_i=-\dint_{\mathcal D_0} \mathfrak{q}_{i,i};\quad \mathfrak q_i:= J \left(F^{-1} \right)_{i \alpha} q_{\alpha},
\end{equation}
finally $W^{\textrm{in}}$ is the strain working, given by equation \eqref{W_def_Lagr}. The pull-back of the energy balance \eqref{1st law} into the reference configuration of the skeleton yields
\begin{equation}\label{1st law ref}
\dint_{\mathcal D_0}\left\lbrace \dfrac{d\, \mathbb E}{d t}+\left[\dfrac{1}{\rho_f n}\left(\mathcal E_f+\dfrac{1}{2} \mathcal E_{sf} \right) M_k +\mathfrak{q}_k \right]_{,k}\right\rbrace = W^{\mathrm{in}}
\end{equation}
where $\mathbb{E}:= J \left(\mathcal E_s + \mathcal E_f +\mathcal E_{s f} \right)$ is the overall Lagrangian density of internal energy; $d/dt$ indicates the time derivative following the motion of the solid particle fixed in the reference configuration of the skeleton. Equations \eqref{W_def_Lagr} and \eqref{1st law ref} imply the following Lagrangian local energy equation to hold true:
\begin{equation}\label{local_energy}
\begin{array}{rl}
\dfrac{d\, \mathbb E}{dt} = & S_{i j} \dot E_{i j} + P_{i j k} \dot E_{i j, k} -\left(J p^f \dfrac{M_k}{m_f} \right)_{,k} -\left[\gamma_{l,l} \dfrac{M_k}{m_f}-\dfrac{\gamma_k}{J}\left(J\dfrac{M_l}{m_f} \right)_{,l}\right]_{,k}+b^{0f}_k \dfrac{M_k}{m_f}\medskip\\
 & + \left[J \left(F^{-1} \right)_{k \beta} \Sigma'^{f}_{\alpha \beta} F_{\alpha q}\, \dfrac{M_q}{m_f}\right]_{,k}
  -\left[\dfrac{1}{\rho_f n} \left(\mathcal E_f+\dfrac{1}{2} \mathcal E_{sf} \right) M_k +\mathfrak{q}_k \right]_{,k}.
\end{array}
\end{equation}

\subsection{The second principle of thermodynamics \& the Causius-Duhem inequality}
In the same way as the first principle, the second principle of thermodynamics (entropy balance) for the porous continuum is stated in the form:
\begin{equation}\label{2nd law}
\dfrac{d_s}{dt}\dint_{\mathcal D} \rho_s (1-n)s_s+\dfrac{d_f}{dt}\dint_{\mathcal D} \rho_f n s_f \geq \dint_{\partial \mathcal D} \dfrac{q_{\alpha} m_{\alpha}}{T},
\end{equation}
where $s_c$ stands for the specific entropy of the $c$-th constituent and $q_{\alpha}$ again indicates the surface rate of heat due to conduction. The pull-back of the entropy balance in the reference configuration of the skeleton reads therefore:
\begin{equation}\label{2nd law ref}
\dint_{\mathcal D_0} \left[\dfrac{d\, \mathbb S}{dt} +\left(s_f M_k + \dfrac{\mathfrak{q}_k}{T} \right)_{,k}\right]\geq 0,
\end{equation}
$\mathbb S:=J \left(\rho_s (1-n) s_s +\rho_f n s_f \right)$ being the overall Lagrangian entropy. Using the Legendre transform $\Psi=\mathbb E -T\, \mathbb S$; the local form of equation \eqref{2nd law ref} can be written as:
\begin{equation}\label{CD}
\dfrac{d\, \mathbb E}{dt}-\mathbb S\dfrac{d T}{dt} - \dfrac{d \Psi}{dt}\geq -T\left(s_f M_k + \dfrac{\mathfrak{q}_k}{T}\right)_{,k};
\end{equation}
which using equation \eqref{local_energy} delivers the Clausius-Duhem inequality for a porous material within the framework of gradient poromechanics, see \cite{Sciarra07} for similar results:
\begin{equation}\label{CD_complete}
\begin{array}{l}
S_{i j} \dot E_{i j} + P_{i j k} \dot E_{i j, k} -\left(J p^f \dfrac{M_k}{m_f} \right)_{,k} -\left[\gamma_{l,l} \dfrac{M_k}{m_f}-\dfrac{\gamma_k}{J}\left(J\dfrac{M_l}{m_f} \right)_{,l}\right]_{,k}+\left[J \left(F^{-1} \right)_{k \beta} \Sigma'^{f}_{\alpha \beta} F_{\alpha q}\, \dfrac{M_q}{m_f}\right]_{,k}\medskip\\
 +\,b^{0f}_k \dfrac{M_k}{m_f}-\left[\dfrac{1}{\rho_f n} \left(\mathcal E_f+\dfrac{1}{2} \mathcal E_{sf} \right) M_k +\mathfrak{q}_k \right]_{,k}+ T\left(s_f M_k + \dfrac{\mathfrak{q}_k}{T} \right)_{,k} - \mathbb S\,\dfrac{dT}{dt}-\dfrac{d\Psi}{dt}\geq 0.
\end{array}
\end{equation}

\subsection{Dissipation}
A characterization of the dissipation relative to the solid and the non-uniform fluid is obtained, starting from the Clausius-Duhem inequality \eqref{CD_complete}, separating, within the overall free energy $\Psi$ and the overall entropy $\mathbb S$, the contributions of the bulk fluid, say $m_f \psi_f$ and $m_f s_f$, from the residual terms relative to the energy and the entropy of the bulk skeleton and the interfaces, say $\Psi_s$ and $\mathbb S_s$:
\begin{equation}\label{Helm_skel}
\Psi_s=\Psi-m_f\, \psi_f\left(\rho_f,T \right), \quad \mathbb S_s=\mathbb S -m_f s_f.
\end{equation}
Remind that $\psi_f=e_f- T\, s_f$ is the Legendre transformation of $e_f$, such that $\partial \psi_f/\partial T=-s_f$. This approach stems from \cite{Biot72}, who defined the wetted solid as a system composed of the solid skeleton and a thin layer of fluid attached to the internal walls of the pore. The aim is proving that $\Psi_s$ is a state function which prescribes the constitutive behavior of the skeleton in thermo-poroelasticity, say in the case when just reversible processes are involved by the deformation of the porous continuum. In this case the residual part of the dissipation will depend just on the Lagrangian filtration vector $M_k$ and the gradient of the absolute temperature.

The fluid mass conservation \eqref{fluid mass Coussy} together with the constitutive law relative to the fluid stress and hyper-stress \eqref{fluid_const}, allow to rephrase the Clausius-Duhem inequality \eqref{CD_complete} in the following form:
\begin{equation}\label{CD_upgraded}
\begin{array}{l}
\Phi:=S_{ij} \dot E_{ij}+P_{ijk} \dot E_{ij,k}+\left(\mathcal P +\dfrac{1}{2 n}\, \mathcal E_{sf}+\dfrac{\gamma_k}{\phi}\, \dfrac{J_{,k}}{J} \right)\dfrac{d\phi}{dt}+\dfrac{\phi}{S_r} \left(\dfrac{1}{2 n}\, \mathcal E_{sf}+\dfrac{\gamma_k}{\phi}\, \dfrac{J_{,k}}{J} \right)\dfrac{d S_r}{dt}-\left(\dfrac{\gamma_k}{\phi S_r}\right) \dfrac{d(\phi S_r)_{,k}}{dt}\medskip\\
-\mathbb S_s \dfrac{dT}{dt}-\dfrac{d\Psi_s}{dt} -\dfrac{1}{\rho_L}\left\lbrace\dfrac{1}{S_r}\mathcal P_{,k} +\left[\left(\dfrac{\gamma_l}{\phi S_r}\right)_{,l} +\dfrac{1}{S_r}\left(\dfrac{1}{2n} \mathcal E_{sf} +\dfrac{\gamma_{l}}{\phi} \dfrac{J_{,l}}{J}\right)\right]_{,k}-\dfrac{b_k^{0f}}{\phi S_r}\right\rbrace M_k-\dfrac{\mathfrak q_k}{T} T_{,k}\geq 0.
\end{array}
\end{equation}
The coefficients of $d\phi/dt$, $d S_r/dt$, $d\left(\phi S_r \right)_{,k}/dt$ and $M_k$ reported in equation \eqref{CD_upgraded} are explicitly deduced from equation \eqref{CD_complete} in Appendix \ref{section_App_A}. Following \cite{Coussy_book04} the dissipation \eqref{CD_upgraded} can be split into three terms, one related to the solid skeleton, one to the non-uniform fluid and the last one referred to thermal effects. These three contributions are separately assumed non-negative:
\begin{align}
\Phi_s = & S_{ij} \dot E_{ij}+P_{ijk} \dot E_{ij,k}+\mathcal P \dfrac{d\phi}{dt}+\dfrac{1}{S_r} \left(\dfrac{\mathcal E_{sf}}{2 n}+\dfrac{\gamma_k}{\phi}\, \dfrac{J_{,k}}{J} \right)\dfrac{d (\phi S_r)}{dt}-\dfrac{\gamma_k}{\phi S_r} \dfrac{d(\phi S_r)_{,k}}{dt}-\mathbb S_s \dfrac{dT}{dt}-\dfrac{d\Psi_s}{dt}\geq 0,\label{Phi_s}\\
\Phi_f = & -\dfrac{1}{\rho_L}\left\lbrace\dfrac{1}{S_r}\mathcal P_{,k} +\left[\left(\dfrac{\gamma_l}{\phi S_r}\right)_{,l} +\dfrac{1}{S_r}\left(\dfrac{1}{2n} \mathcal E_{sf} +\dfrac{\gamma_{l}}{\phi} \dfrac{J_{,l}}{J}\right)\right]_{,k}-\dfrac{b_k^{0f}}{\phi S_r}\right\rbrace M_k\geq 0,\label{Phi_f}\\
\Phi_{th} = & -\dfrac{\mathfrak q_k}{T} T_{,k}\geq 0\label{Phi_th}.
\end{align}
As previously remarked, equation \eqref{Phi_s} states that within thermo-poroelasticity the Helmholtz free energy of the skeleton is a state function: 
\begin{equation}
\Psi_s=\widehat\Psi_s(E_{ij},E_{ij,k},\phi,S_r,(\phi S_r)_{,k},T),
\label{statefunct}
\end{equation}
which depends not only on the strain of the skeleton, the Lagrangian porosity and the degree of saturation, as in standard unsaturated poromechanics, but also on the gradient of strain, $E_{ij,k}$, and the gradient of Lagrangian water content, $(\phi S_r)_{,k}$. It is worth to notice that equation \eqref{statefunct} could be also formulated considering a free energy function depending on the gradient of strain, $E_{ij,k}$, and the gradient of water content, $(n S_r)_{,k}$, in addition to the dependence on $(E_{ij},\phi,S_r)$, so modifying the definition of conjugate variables. 

As the fluid dissipation \eqref{Phi_f} does not vanish except at equilibrium, the filtration force must be related to the filtration vector itself, in a way to fulfill the dissipation inequality during evolution. Finally equation \eqref{Phi_th} recalls the well-known result that heat flows through materials along the direction of the negative gradient of temperature (from higher to lower). 

Now, assuming isothermal conditions, the poro-elastic constitutive equations of the solid skeleton and the generalized Darcy law relative to the fluid are deduced and compared with those ones typically adopted in modeling the behavior of partially saturated porous media.

\section{Poroelastic constitutive relations}
According with equation \eqref{Phi_s} and its corollary remarks, the poroelastic constitutive relations of the solid skeleton must fulfill the following restriction:
\begin{equation}
\label{Solid_diss_constr}
\begin{array}{l}
\left(S_{ij}-\dfrac{\partial \Psi_s}{\partial E_{ij}}\right)\dfrac{d E_{ij}}{dt}+\left(P_{ijk}-\dfrac{\partial \Psi_s}{\partial E_{ij,k}}\right)\dfrac{d E_{ij,k}}{dt}+\left[\mathcal P +\left(\dfrac{1}{2 n}\mathcal E_{sf}+\dfrac{\gamma_k}{\phi}\, \dfrac{J_{,k}}{J} \right)-\dfrac{\partial \Psi_s}{\partial \phi}\right]\dfrac{d\phi}{dt}\,+\medskip\\
\left[\dfrac{\phi}{S_r} \left(\dfrac{1}{2 n}\mathcal E_{sf}+\dfrac{\gamma_k}{\phi}\, \dfrac{J_{,k}}{J} \right)-\dfrac{\partial \Psi_s}{\partial S_r}\right] \dfrac{d S_r}{dt}+\left(-\dfrac{\gamma_k}{\phi S_r} -\dfrac{\partial \Psi_s}{\partial (\phi S_r)_{,k}} \right)\dfrac{d(\phi S_r)_{,k}}{dt}
= 0,
\end{array}
\end{equation}
if no frozen contribution to the free energy has been taken into account, see \textit{e.g.} \cite{Collins2005} and \cite{CPV2010}. Within this framework one gets the constitutive relations for the overall stresses $S_{ij}$ and $P_{ijk}$,
\begin{equation}
\label{Overall_const}
S_{ij}= \dfrac{\partial \Psi_s}{\partial E_{ij}},\qquad  P_{ijk}= \dfrac{\partial \Psi_s}{\partial E_{ij,k}},
\end{equation}
the constitutive prescription for the coupling energy term $\mathcal E_{sf}$,
\begin{equation}
\label{E_sf_const}
\dfrac{\phi}{S_r} \left(\dfrac{1}{2n} \mathcal E_{sf}+\dfrac{\gamma_k}{\phi}\dfrac{J_{,k}}{J} \right)=\dfrac{\partial \Psi_s}{\partial S_r},
\end{equation}
and the constitutive constraints to be satisfied by the skeleton free energy once the thermodynamic pressure $\mathcal P$ and the fluid hyper-stress $\gamma_k$ have been assigned according with equation \eqref{Psif}, say
\begin{equation}
\label{energy_s_const}
\mathcal P+\left(\dfrac{1}{2n} \mathcal E_{sf}+\dfrac{\gamma_k}{\phi}\dfrac{J_{,k}}{J} \right)=\dfrac{\partial \Psi_s}{\partial \phi}, \qquad \dfrac{\gamma_k}{\phi S_r}=-\dfrac{\partial \Psi_s}{\partial (\phi S_r)_{,k}}.
\end{equation}
When the gradient contribution to the fluid energy vanishes, equation \eqref{E_sf_const} reduces to the prescription of the macro-scale capillary pressure, postulated in classical unsaturated poromechanics, see \cite{Coussy_book04}, say $\phi\mathcal P_c=-\partial\Psi_s/\partial S_r$, if one assumes
\begin{equation}
\label{Pc}
\mathcal P_c := - \dfrac{1}{S_r}\left(\dfrac{1}{2 n}\mathcal E_{sf}\right).
\end{equation}
As a consequence equations \eqref{E_sf_const} and \eqref{energy_s_const}$_1$ can be rephrased as the generalized constitutive characterizations of the macro-scale capillary pressure $\mathcal P_c$ and the fluid thermodynamic pressure $\mathcal P$, which account for the spatial distribution of interfaces, within the porous network, considering the additional contribution of the fluid hyper-stress:
\begin{equation}
\label{retention_&_porosity}
\phi \mathcal P_c = -\dfrac{\partial \Psi_s}{\partial S_r}+\dfrac{\gamma_k}{S_r}\, \dfrac{J_{,k}}{J},\qquad \mathcal P -S_r \mathcal P_c+\dfrac{\gamma_k}{\phi}\, \dfrac{J_{,k}}{J} = \dfrac{\partial \Psi_s}{\partial \phi}.
\end{equation}
Equation \eqref{Overall_const} is standard, within the framework of gradient theories of continuum mechanics; on the other hand equations \eqref{retention_&_porosity} deserve a deeper discussion in order to understand in which sense they can be considered an enhanced constitutive prescription of the macro-scale capillary pressure and the thermodynamic pressure of the fluid. Developing a micro-scale analysis this model will be compared, in \S~\ref{micro}, with the micro-structured one which accounts for the dependence of the capillary pressure on the so-called interfacial area between the non-wetting and the wetting phase, see \textit{e.g.} \cite{Hassanizadeh1990, HassanizadehGray93, Joekar2010, Reeves1996}. 

\subsection{Generalized effective stresses}
Some additional remarks can be deduced from the thermodynamical restrictions stated by equation \eqref{Phi_s} if the solid grains, which constitute the matrix of the porous medium, undergo negligible volume change. In this case the Lagrangian porosity is definitely prescribed in terms of volume changes of the porous elementary volume, say: $J=1+\phi-\phi_0$, $\phi_0$ being the reference value of porosity, see \textit{e.g.} \cite[Chapter 3]{Coussy_book04}. The solid dissipation can therefore be rewritten and the generalization of Bishop's effective stress naturally arises, within the framework of the considered gradient model. 
Assuming small strains, but allowing the porous skeleton to suffer strain gradients of order $O(1)$, with respect to the considered small perturbation parameter, the simplified expression of $\Phi_s$ reads as follows:
\begin{equation}
\label{Bishop_diss}
\Phi_s=S'_{ij} \dot E_{ij} + P'_{ijk} \dot E_{ij,k} -\phi\mathcal P_c\, \dot S_r-\dfrac{\gamma_k}{\phi S_r}\, \dfrac{d(n S_r)_{,k}}{dt}-\dot\Psi_s.
\end{equation}
Here $S'_{ij}$ and $P'_{ijk}$ are the generalized Bishop stress and hyper-stress defined by
\begin{equation}
\label{Bishop_stress_small_def}
S'_{ij}:=S_{ij}+\left(\mathcal P -S_r \mathcal P_c \right)\delta_{ij}+\gamma_k\left(2 E_{ij,k}-\delta_{ij} \dfrac{(n S_r)_{,k}}{\phi S_r} \right),\qquad P'_{ijk} :=P_{ijk} -\gamma_k \delta_{ij}.
\end{equation}
%
As a consequence the corresponding free energy of the solid skeleton can be regarded as a state function of strain, strain gradient, saturation degree and gradient of liquid content, only. The general form of equations \eqref{Bishop_stress_small_def}, when no restriction to small strains is assumed, is reported in Appendix \ref{section_App_B}.

If one aims at investigating non-reversible processes affecting the behavior of the solid skeleton, different dissipative mechanisms should be analyzed, which could concern interactions among the grains of the solid skeleton, or between the grains and the fluid. In the first case inelastic constitutive relations for the solid stress and (eventually) hyper-stress should be introduced, in the second one hysteresis during wetting-drying cycles should modify the prescriptions of the retention properties, see \textit{e.g.} \cite{Hassanizadeh2002}. It is not the goal of this paper to address modeling inelastic processes, however this can be done specifying the way in which dissipation depends on a suitable set of internal variables.

\subsection{Generalized Darcy law}\label{S:gen_Darcy}
As usual in poromechanics the dissipation inequality \eqref{Phi_f} relative to the fluid can be satisfied requiring $\Phi_f$ to be a quadratic function of $M_k$, which means: $\Phi_f=\left(A_{k l} M_l M_k \right)/\rho^2_L.$ This implies the coefficient of $M_k$ in equation \eqref{Phi_f} to be constitutively constrained to
\begin{equation}
\label{Darcy_gen}
-\dfrac{1}{S_r}\mathcal P_{,k} -\left[\left(\dfrac{\gamma_l}{\phi S_r}\right)_{,l} +\dfrac{1}{S_r}\left(\dfrac{1}{2n} \mathcal E_{sf} +\dfrac{\gamma_{l}}{\phi} \dfrac{J_{,l}}{J}\right)\right]_{,k}+\dfrac{b_k^{0f}}{\phi S_r} = \dfrac{1}{\rho_L} A_{kl} M_l,
\end{equation}
which reads as a generalization of classical Darcy's law. $A_{kl}$ is the inverse of permeability, its dimension is therefore $\left[A \right]=L^{-3}\, M\, T^{-1}$. According to equations \eqref{pressure_chempot} and \eqref{retention_&_porosity} equation \eqref{Darcy_gen} can be reformulated as follows
\begin{equation}
\label{Darcy_gen_energy}
-\left[\dfrac{\partial \left(S_r\, \rho_L e_f \right)}{\partial S_r} - \left(\dfrac{\partial \Psi_{s}}{\partial (\phi S_r)_{,l}} \right)_{,l} +\dfrac{\partial U}{\partial S_r}\right]_{,k} +\dfrac{b_k^{0f}}{\phi S_r} = \dfrac{1}{\rho_L} A_{kl} M_l,
\end{equation}
where $S_r \rho_L e_f$ is the specific free energy of the fluid per unit volume of the pores and therefore coincides with the double-well energy potential $\Psi_f$, possibly describing coexistence of the liquid and the gaseous phases, see equation \eqref{Psif}. For the sake of simplicity, the capillary energy $U$ has been introduced, following \cite{Coussyetal2004}, so that $\partial\Psi_s/\partial S_r=\phi\, \partial U/\partial S_r$. 

According with classical arguments proposed within the framework of diffuse interface models in fluid mechanics, see \textit{e.g.} \cite{Jacqmin2000, Jamet2001, Boyer2010}, the first two terms in equation \eqref{Darcy_gen_energy} correspond to the so-called generalized chemical potential of the non-uniform fluid defined as
\begin{equation}
\text{\textmu}=\dfrac{\partial \Psi_f}{\partial S_r}-\left(\dfrac{ \partial \Psi_{s}}{\partial \left(\phi S_r \right)_{,l}}\right)_{,l},
\label{generalized_chempot}
\end{equation}
the whole quantity $\text{\textmu}+\partial U/\partial S_r$ can therefore be interpreted as the generalized chemical potential of the pore-fluid, say of the non-uniform fluid within the pore network. This last accounts, on one hand, for surface tension effects which are typical of a diphasic fluid on the other one for the retention properties of the skeleton, due to its texture, by means of $\partial U/\partial S_r$.

Apparently the role of the capillary energy $U$ is that of modifying the double-well potential $\Psi_f$ which prescribes the free energy of the fluid, in order to account for the presence of a confining surface densely distributed within the porous continuum. The new free energy $\left(\Psi_f+U\right)$, which can be called effective energy of the pore-fluid, has not the same minima as $\Psi_f$, as they are shifted inward the interval $(0,1)$ from below or from above whether the solid skeleton is gas or liquid wet, see \textit{e.g.} \cite{Papatzacos2002,Papatzacos2004}. In \S~\ref{S:porefluid} the effects of combining these two energy contributions are discussed, considering the retention properties of different soils. A similar behavior has been discussed by \cite{Cueto-Juanes14}, modeling two-phase flow in a Hele-Shaw cell, introducing a suitable 
symmetry-breaking function.

\section{Constitutive characterization of the pore-fluid}\label{S:porefluid}
Providing a constitutive characterization of the pore-fluid, say of the fluid within the pore network, is generally achieved, in unsaturated poromechanics, assuming a proper prescription of the retention curve, which means prescribing the capillary pressure $\mathcal P_c$, or the capillary energy $U$, as a function of $S_r$. As already noticed in \S~\ref{S:Thermodynamics} however, the air-water mixture, which saturates the pore space, is regarded here as a non-uniform fluid, the corresponding saturation ratio being used to characterize the state of the fluid at any current placement. No distinction is therefore made explicit between the pressure of the liquid and the pressure of the gaseous phase, as well as between the corresponding chemical potentials. 
Moreover no algebraic relation between the saturation degree and the capillary pressure or the chemical potential of the fluid can a-priori be stated, without solving, at least at equilibrium, the following generalized Richards equation:
\begin{equation}
\label{Richards}
\dfrac{d (\phi S_r)}{d t}-\left\lbrace K_{\textrm{sat}}\,k(S_r)\left[ \left(\dfrac{\partial \Psi_f}{\partial S_r} - \left(\dfrac{\partial \Psi_{s}}{\partial (\phi S_r)_{,l}} \right)_{,l} +\dfrac{\partial U}{\partial S_r} \right)_{,k} - \dfrac{b^{0f}_k}{\phi S_r}\right] \right\rbrace_{,k}=0.
\end{equation}
Here the permeability of the porous medium has been assumed isotropic: $(A^{-1})_{mk}=K_{\textrm{sat}}\, k(S_r) \delta_{mk}$, $K_{\textrm{sat}}$ being the so-called saturated permeability and $k(S_r)$ the relative permeability of the wetting phase, see \textit{e.g.} \cite{Coussy_book04,Coussy_book10}. As usual in unsaturated poromechanics equation \eqref{Richards} which provides the distribution in space and time of the saturation degree is deduced merging the mass conservation law \eqref{fluid mass Coussy} and the Darcy law, which in this case takes the form given by equation \eqref{Darcy_gen_energy}. 

At stationary conditions, say when $\partial (\phi S_r)/\partial t=0$, equation \eqref{Richards} reduces to a fourth order partial differential equation in the space variable, which is definitely similar to the one prescribing the mass density distribution of a Cahn-Hilliard fluid at equilibrium. However a fundamental additional term is here accounted for, say the derivative of $U$ with respect to $S_r$, which allows for describing the confining effect on the non-uniform fluid, due to the presence of the porous skeleton. Equation \eqref{Richards} looks also similar to that one stated in \cite[equation (2)]{Cueto2008} and \cite[equation (13)]{Cueto-Juanes09}, who adopted a phase field model for describing gravity fingers and saturation overshoot, during water infiltration through a non-deformable soil. In this case the difference resides in the account for the chemical potential $\mu=\partial \Psi_f/\partial S_r$ in \eqref{Richards}, allowing for possible coexistence between the phases of the non-uniform fluid. Moreover equation \eqref{Richards} is here definitely coupled with the balance equation of the overall continuum. 

Following the same procedure which induced the definition of the generalized chemical potential \eqref{generalized_chempot} when simulating multi-phase flows, see \textrm{e.g.} \cite{Boyer2010}, equation \eqref{Richards} can be rephrased in terms of the functional derivative of the free energy of the pore-fluid 
\begin{equation}
\mathcal F[S_r]=\dint_{\mathcal D_0} \left[\left(\Psi_f(S_r)+U(S_r)\right)+\kappa_f\,(nS_r)_{,k} (nS_r)_{,l}\, \delta_{kl}\right]
\label{pore-fluid_energy}
\end{equation}
with respect to $S_r$, as follows:
\begin{equation}
\left\lbrace
\begin{array}{l}
\dfrac{d (\phi S_r)}{d t} = \left[ K_{\textrm{sat}}\,k(S_r)\left(\text{\textmu}^{\textrm{eff}}_{,k} - \dfrac{b^{0f}_k}{\phi S_r} \right)\right]_{,k},\medskip\\
\text{\textmu}^{\textrm{eff}}=\dfrac{\delta\mathcal F}{\delta S_r}= \dfrac{\partial \Psi_f}{\partial S_r} - \left(\dfrac{\partial \Psi_{s}}{\partial (\phi S_r)_{,l}} \right)_{,l} +\dfrac{\partial U}{\partial S_r},
\end{array}
\right.
\label{Richards_rephrased}
\end{equation}
where the generalized effective chemical potential of the pore-fluid $\text{\textmu}^{\textrm{eff}}$ has been introduced. 
The energy of the skeleton depends on the gradient of the liquid content, so that the gradient contribution to the constitutive law of the non-uniform fluid is not modified, with respect to equation \eqref{fluid_const}. For the sake of completeness the effective chemical potential of the pore-fluid, accounting only for the derivative of the effective energy $\Psi_f+U$ is defined as $\mu^{\textrm{eff}}:=\partial (\Psi_f+U)/\partial S_r=\mu+\partial U/\partial S_r$. As already noticed $\text{\textmu}^{\textrm{eff}}$ is different from \textmu, because of the term involving the derivative of the capillary energy $U$, but it also differs from the flow potential $\Phi$ introduced by \cite[equation (12)]{Cueto-Juanes09}, because of the term involving the derivative of the fluid energy $\Psi_f$.

\begin{figure}[h]
\centering
\begin{subfigure}[b]{.23\textwidth}
\includegraphics[width=\textwidth]{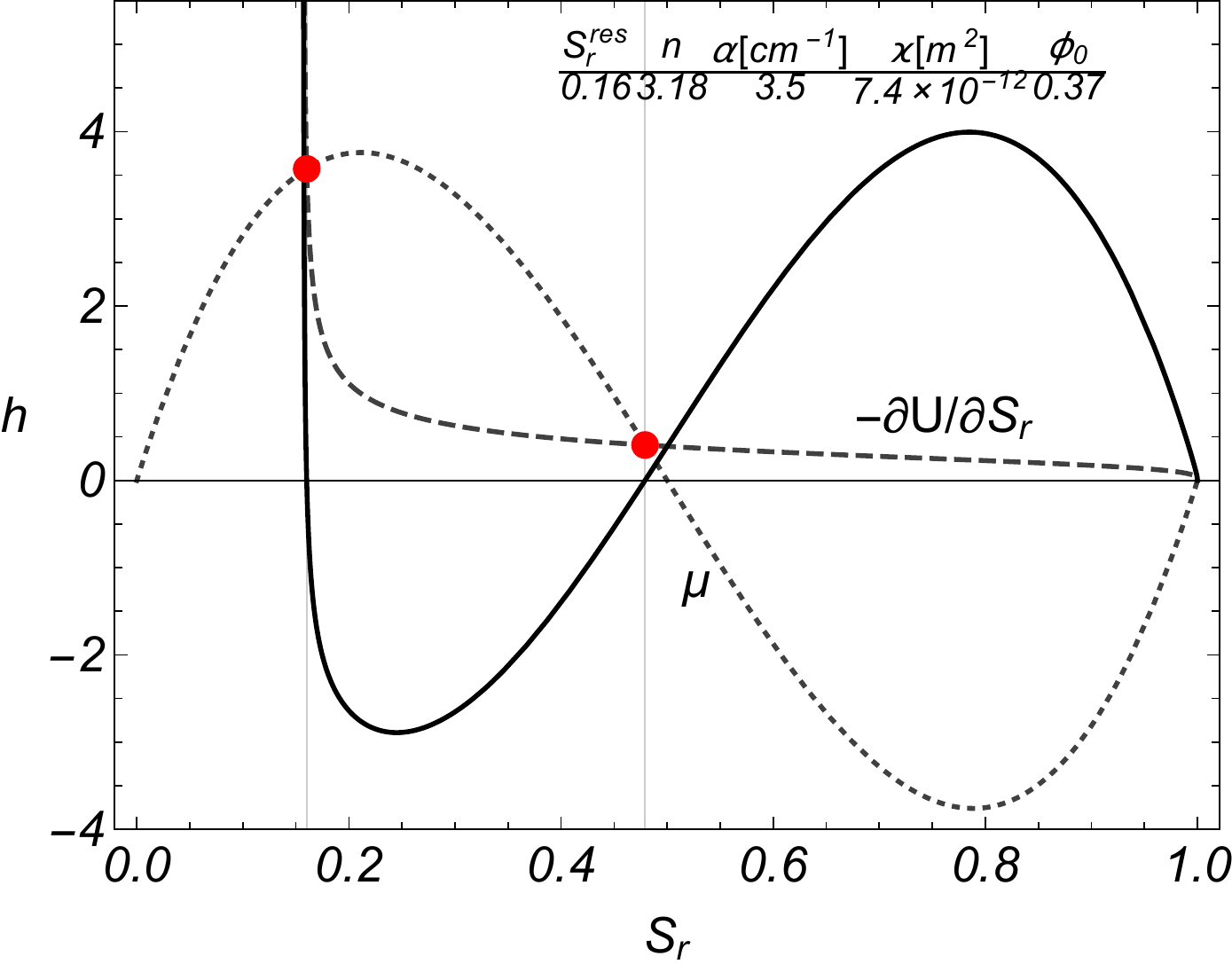}
\caption{Sand}
\label{fig:sandpot}
\end{subfigure}
\,
\begin{subfigure}[b]{.23\textwidth}
\includegraphics[width=\textwidth]{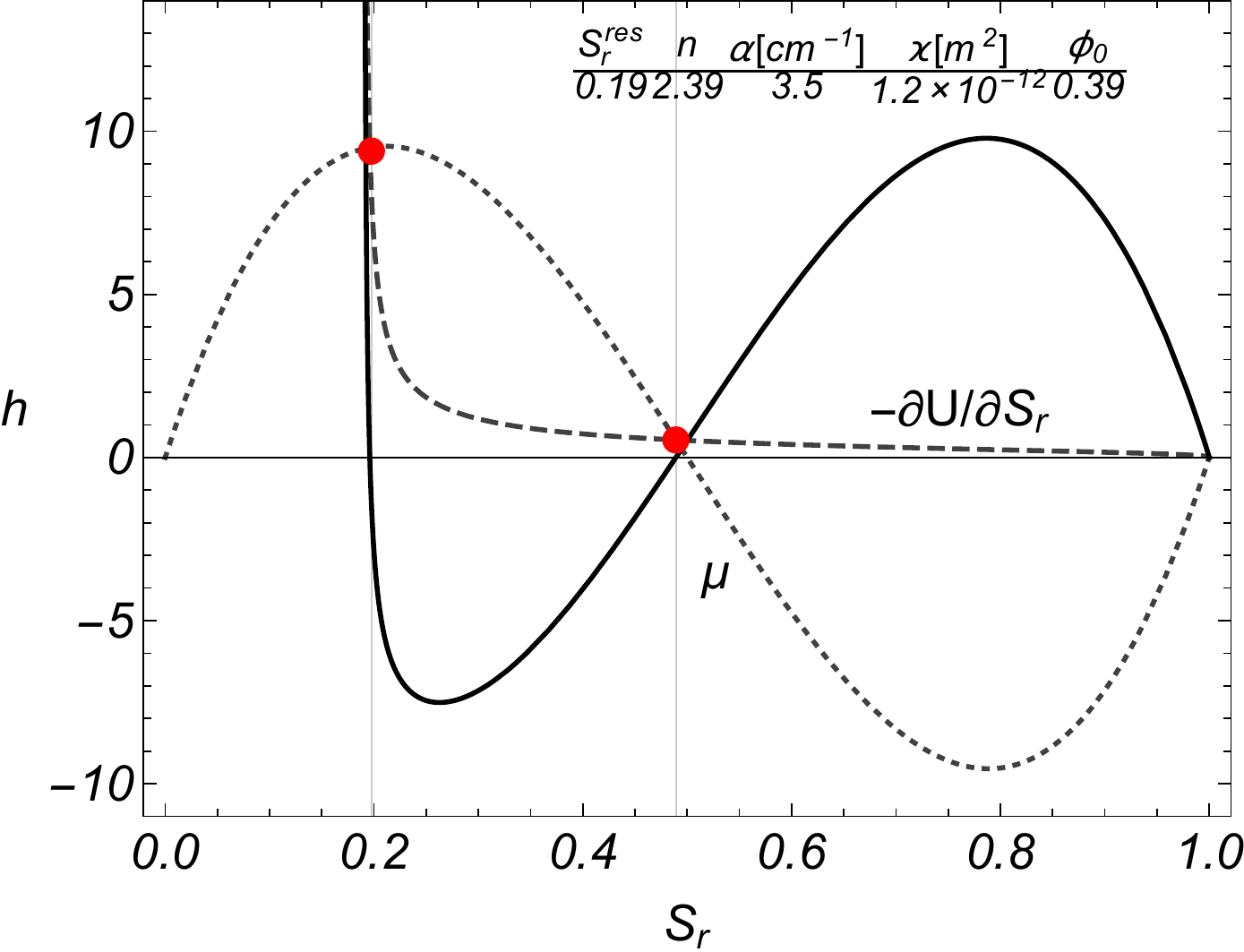}
\caption{Loamy sand}
\label{fig:loamysandpot}
\end{subfigure}
\,
\begin{subfigure}[b]{.23\textwidth}
\includegraphics[width=\textwidth]{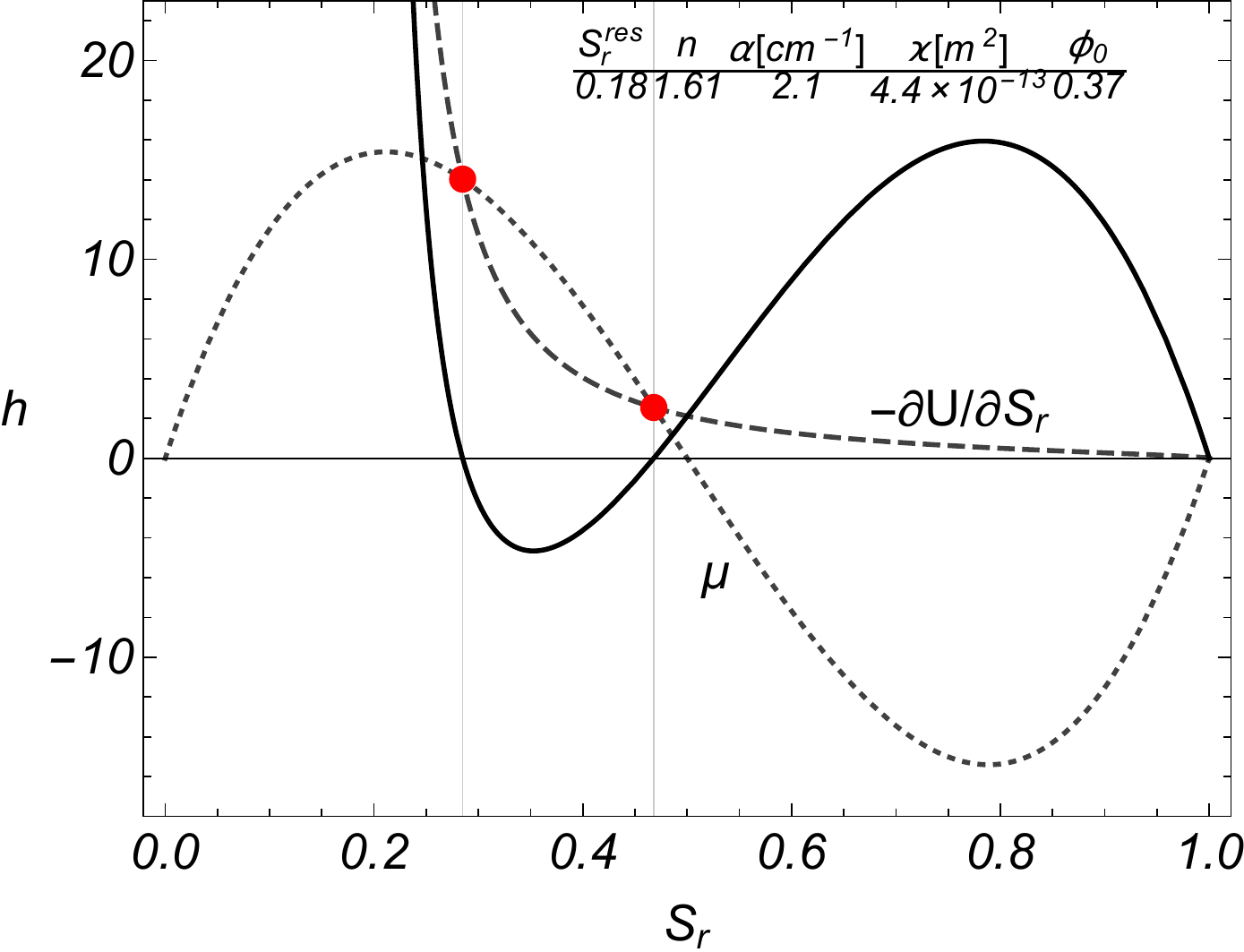}
\caption{Sandy loam}
\label{fig:sandyloampot}
\end{subfigure}
\,
\begin{subfigure}[b]{.23\textwidth}
\includegraphics[width=\textwidth]{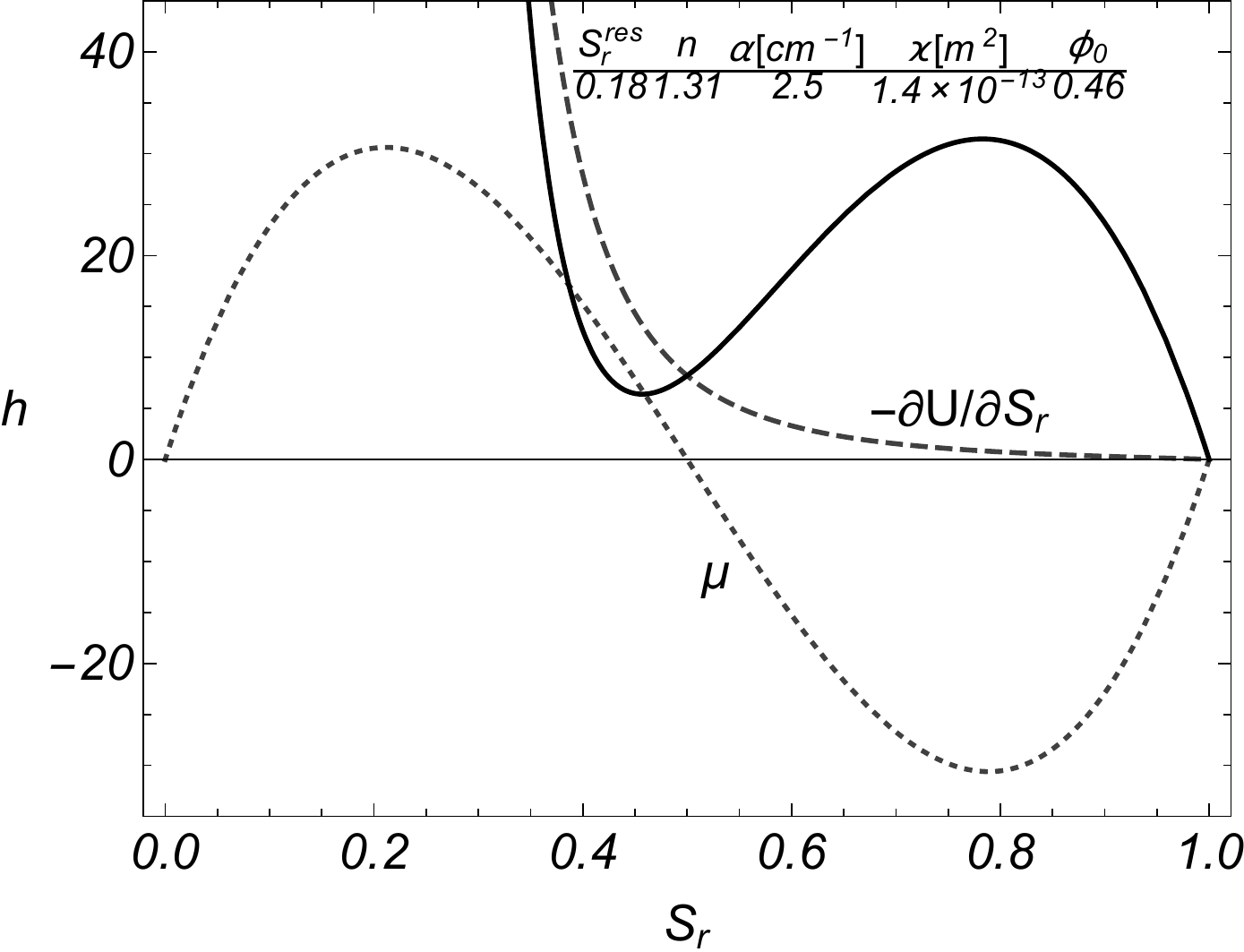}
\caption{Loam}
\label{fig:loampot}
\end{subfigure}
\\
\begin{subfigure}[b]{.23\textwidth}
\includegraphics[width=\textwidth]{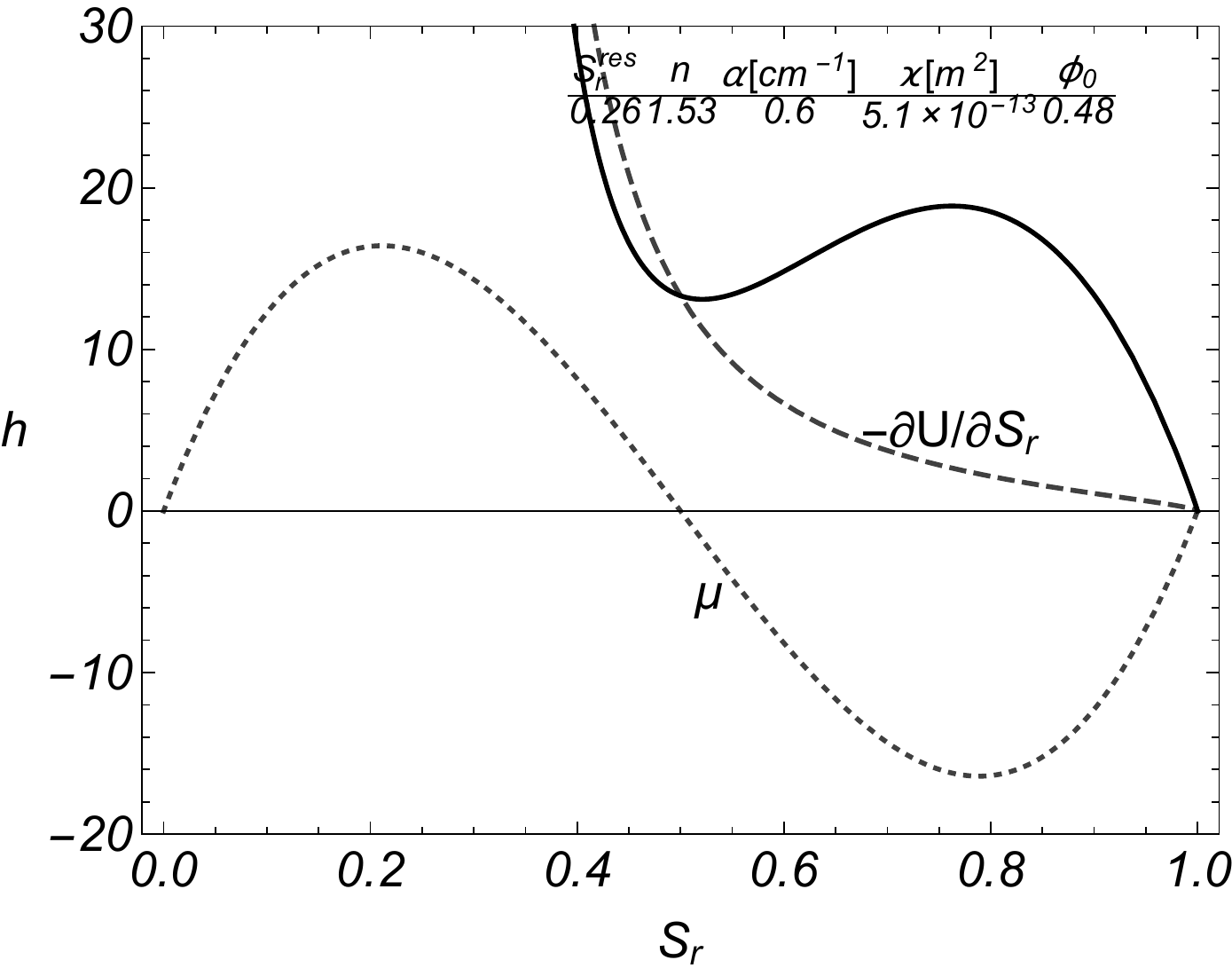}
\caption{Silt}
\label{fig:siltpot}
\end{subfigure}
\,
\begin{subfigure}[b]{.23\textwidth}
\includegraphics[width=\textwidth]{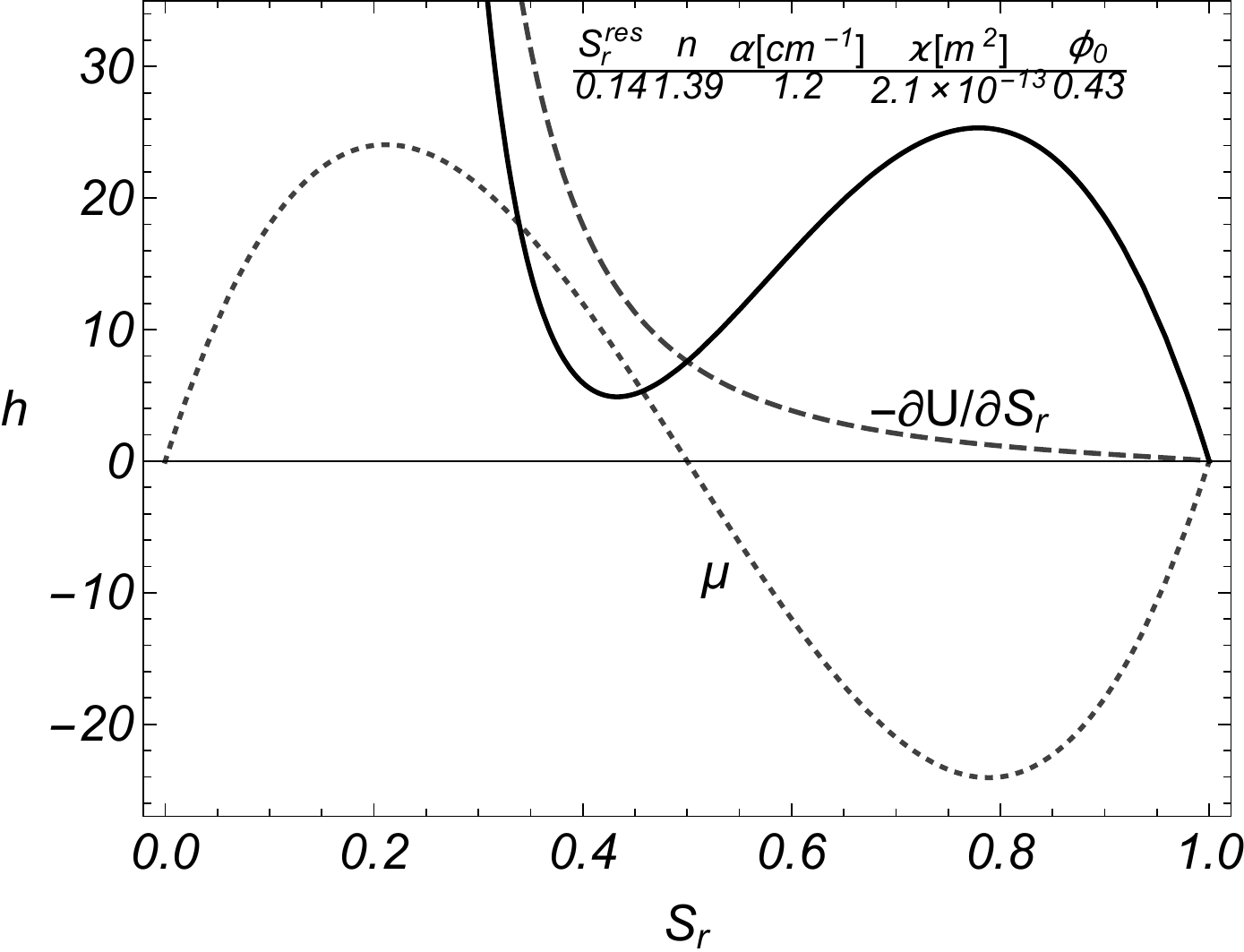}
\caption{Silt loam}
\label{fig:siltloampot}
\end{subfigure}
\,
\begin{subfigure}[b]{.23\textwidth}
\includegraphics[width=\textwidth]{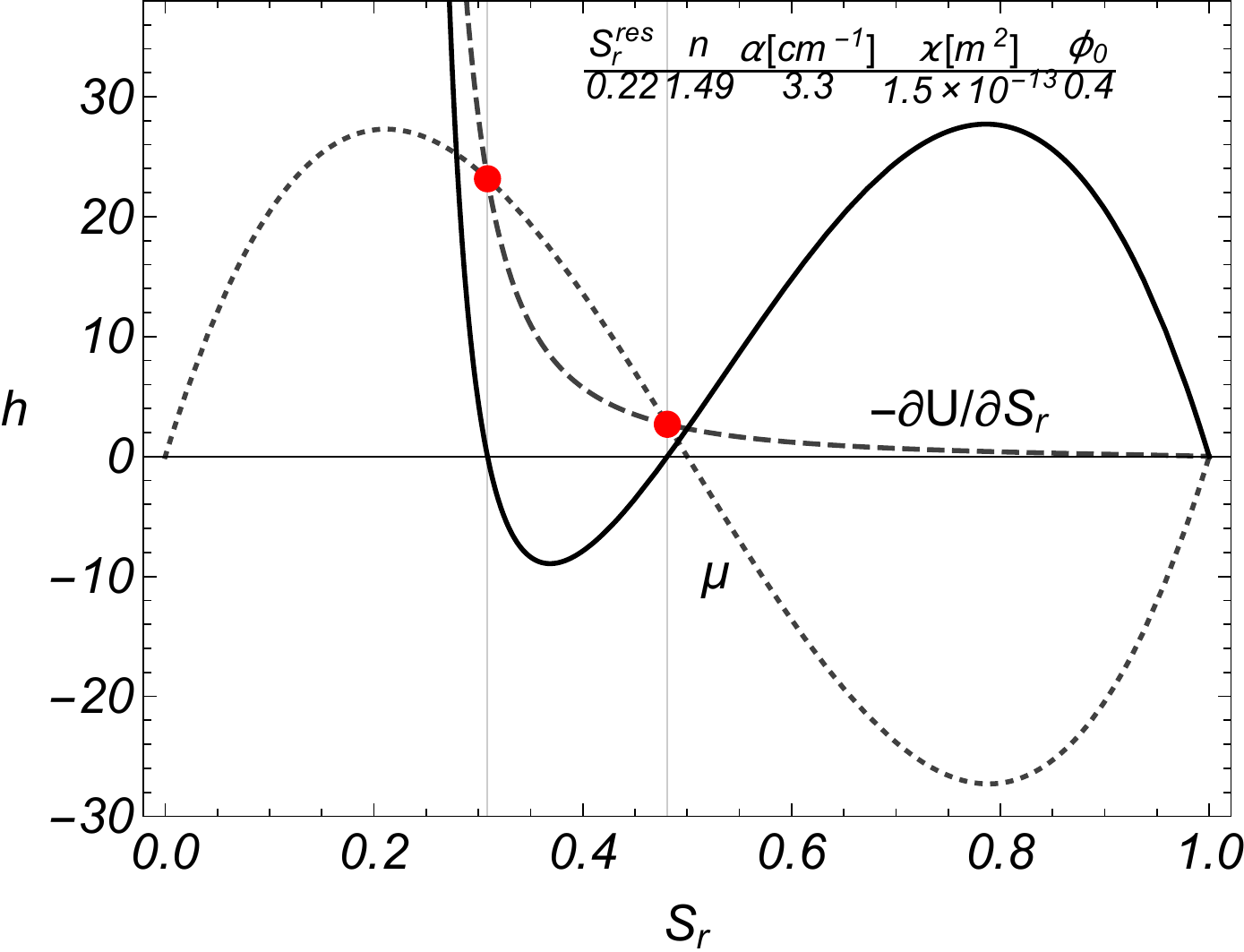}
\caption{Sandy clay Loam}
\label{fig:sandyclayloampot}
\end{subfigure}
\,
\begin{subfigure}[b]{.23\textwidth}
\includegraphics[width=\textwidth]{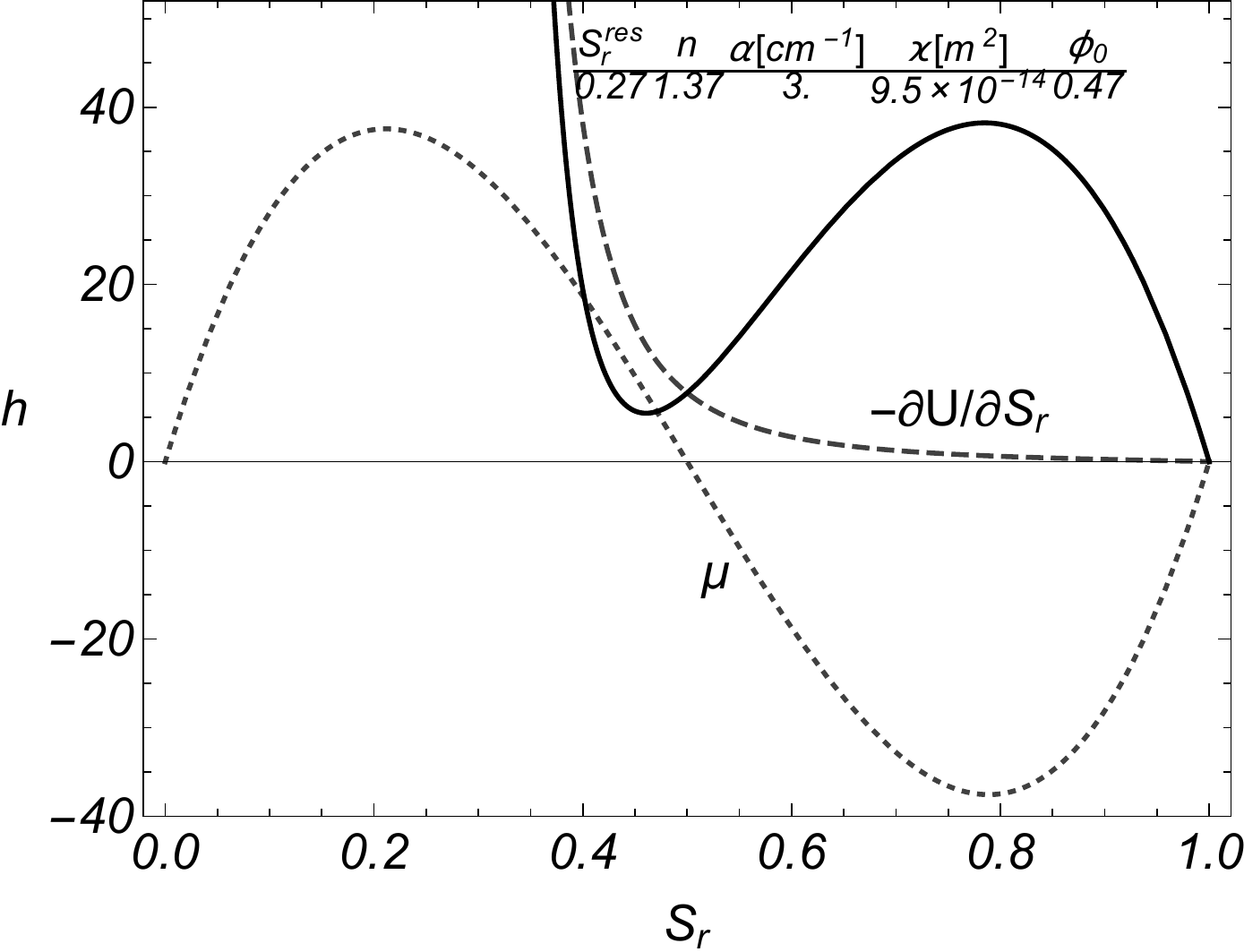}
\caption{Clay loam}
\label{fig:clayloampot}
\end{subfigure}
\smallskip
\\
\begin{subfigure}[b]{.23\textwidth}
\includegraphics[width=\textwidth]{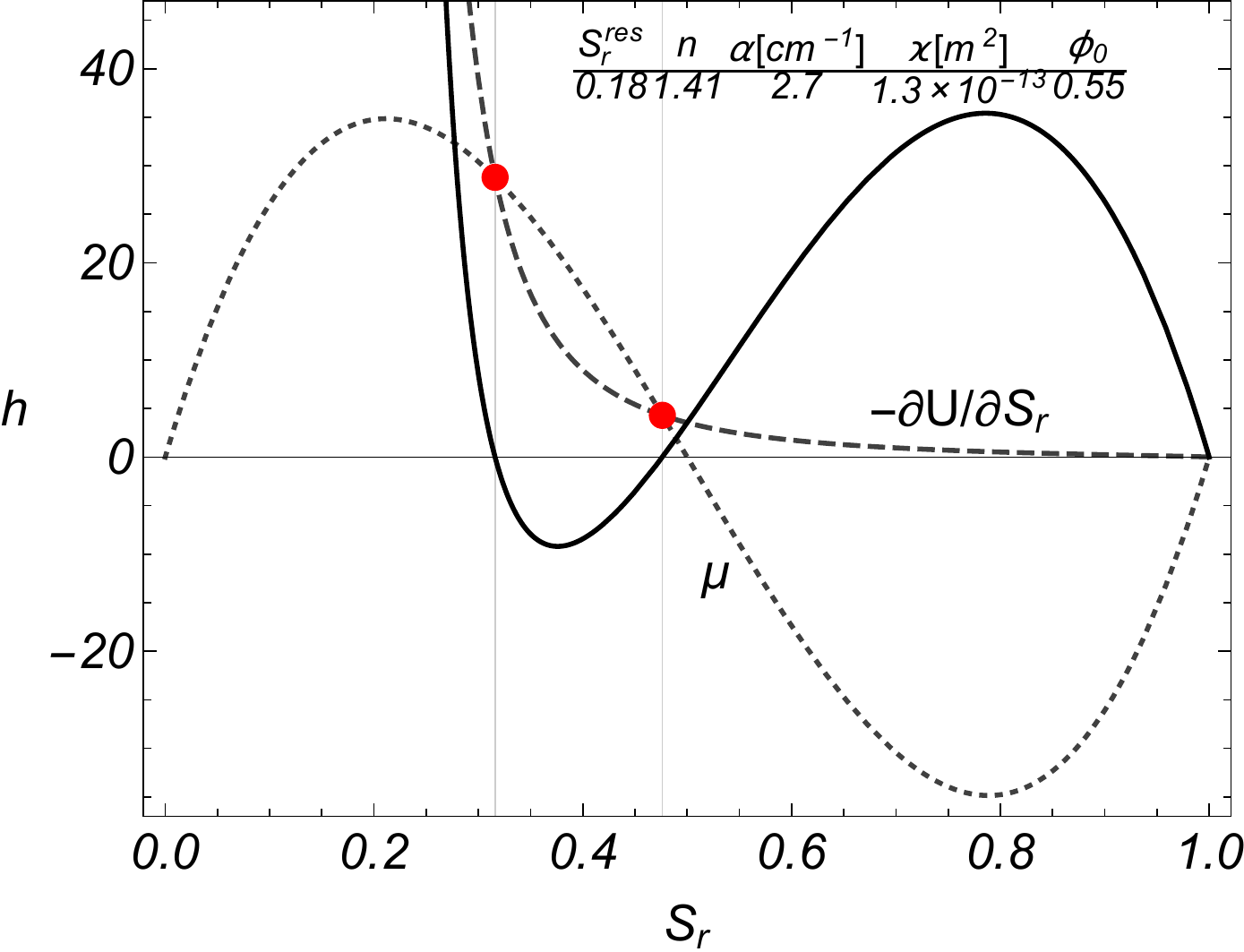}
\caption{Silty clay loam}
\label{fig:siltyclayloampot}
\end{subfigure}
\,
\begin{subfigure}[b]{.23\textwidth}
\includegraphics[width=\textwidth]{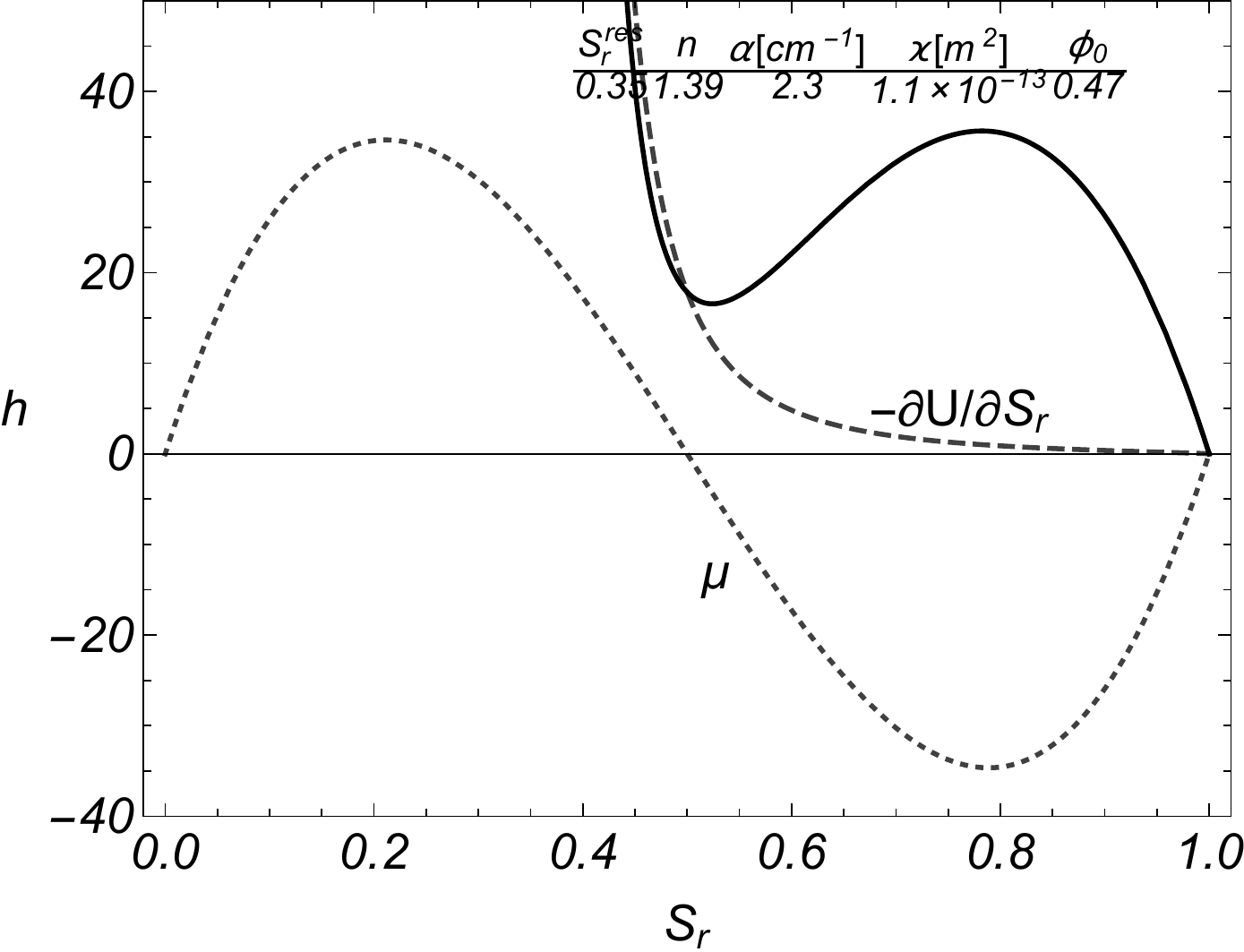}
\caption{Silty clay}
\label{fig:siltyclaypot}
\end{subfigure}
\,
\begin{subfigure}[b]{.23\textwidth}
\includegraphics[width=\textwidth]{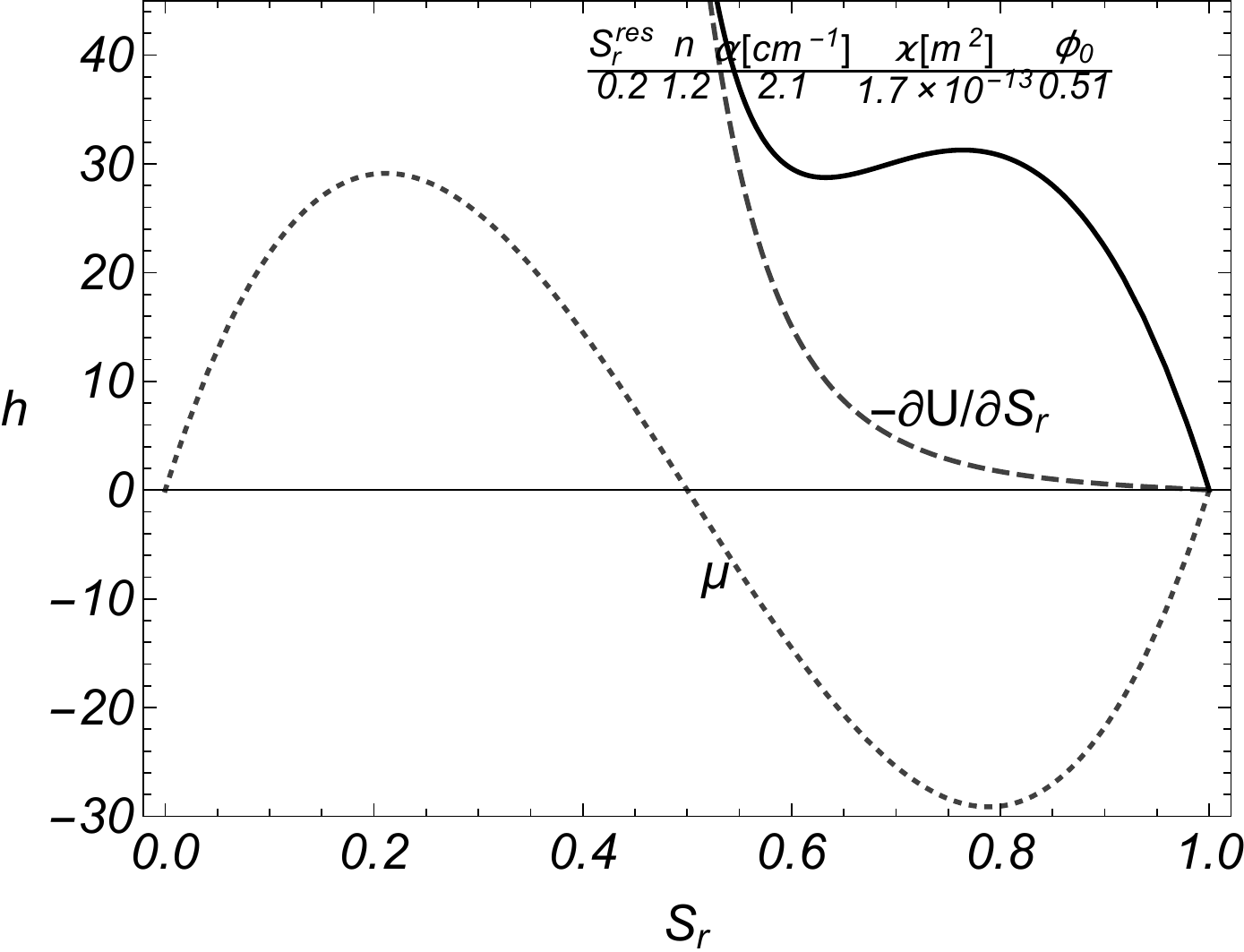}
\caption{Clay}
\label{fig:claypot}
\end{subfigure}
\caption{Heads relative to the chemical potential $\mu$ of the pure fluid (dotted gray line), the derivative of the capillary energy (dashed gray line) and the negative effective chemical potential $\mu^{\textrm{eff}}$ (solid black line). The values of the parameters which characterize the retention curve \eqref{retention_curve} are listed in each panel, together with the intrinsic permeability $\varkappa$ and the referential porosity $\phi_0$, relative to the portion of sand, clay and silt which constitute the soil. The red spots in panels (a), (b), (c), (g) and (i) highlight the real zeros of $\mu+\partial U/\partial S_r$, different from $S_r=1$, which corresponds to the highest relative minimum and the saddle point of the effective free energy $\Psi_f+U$.}
\label{fig:soilpot}
\end{figure}

The role of $\Psi_f$ and $U$ in the characterization of the spatial distribution of $S_r$, $\mu$, $\mathcal P_c$ and $\text{\textmu}^{\textrm{eff}}$ deserves a deeper analysis in order to underline the main differences between the proposed modified Richards equation \eqref{Richards_rephrased} and that one stated by \cite{Cueto-Juanes09} for describing gravity fingering in soils. In Figure~\ref{fig:soilpot} the chemical potential of the pure fluid prescribed, according with equation \eqref{Psif}, as
\begin{equation}
\mu=\dfrac{\partial \Psi_f}{\partial S_r} = (\rho_L g)\, h_f,\qquad h_f= 2 C \dfrac{\gamma_{nw}}{(\rho_L g)\, R}\, S_r \left(1 - 3 S_r + 2 S_r^2 \right),
\end{equation}
and the derivative of the capillary energy, given by a van Genuchten-like curve
\begin{equation}
\dfrac{\partial U}{\partial S_r}=-(\rho_L g)\, h_U,\qquad h_U=\dfrac{1}{\alpha}\left[\left(\dfrac{S_r-S_r^{\textrm{res}}}{1-S_r^{\textrm{res}}}\right)^{-\frac{1}{m}} -1\right]^{\frac{1}{n}},
\label{retention_curve}
\end{equation}
are depicted (gray dotted and gray dashed lines, respectively), considering different soil textures and therefore different choices for the relative partition of sand, clay and silt in the soil. The profile of the negative effective chemical potential of the pore-fluid $\mu^{\textrm{eff}}$ with respect to $S_r$ is also drawn in each panel of Figure~\ref{fig:soilpot} (solid lines). Both the chemical potential and the derivative of th capillary energy are expressed in head units, via the two fields $h_f$ and $h_U$, adopting the classical \cite{Leverett1940} scaling, which prescribes the characteristic length $R$ in terms of the intrinsic permeability $\varkappa$ (of the soil): $R=\sqrt{\varkappa/\phi_0}$. The different values of the parameters $S^{\mathrm{res}}_r$, $\alpha$ and $n$ ($m=1-1/n$), which characterize the retention curve \eqref{retention_curve} for the considered soil textures, are listed in the panels.
\begin{figure}[h]
\centering
\begin{subfigure}[b]{.23\textwidth}
\includegraphics[width=\textwidth]{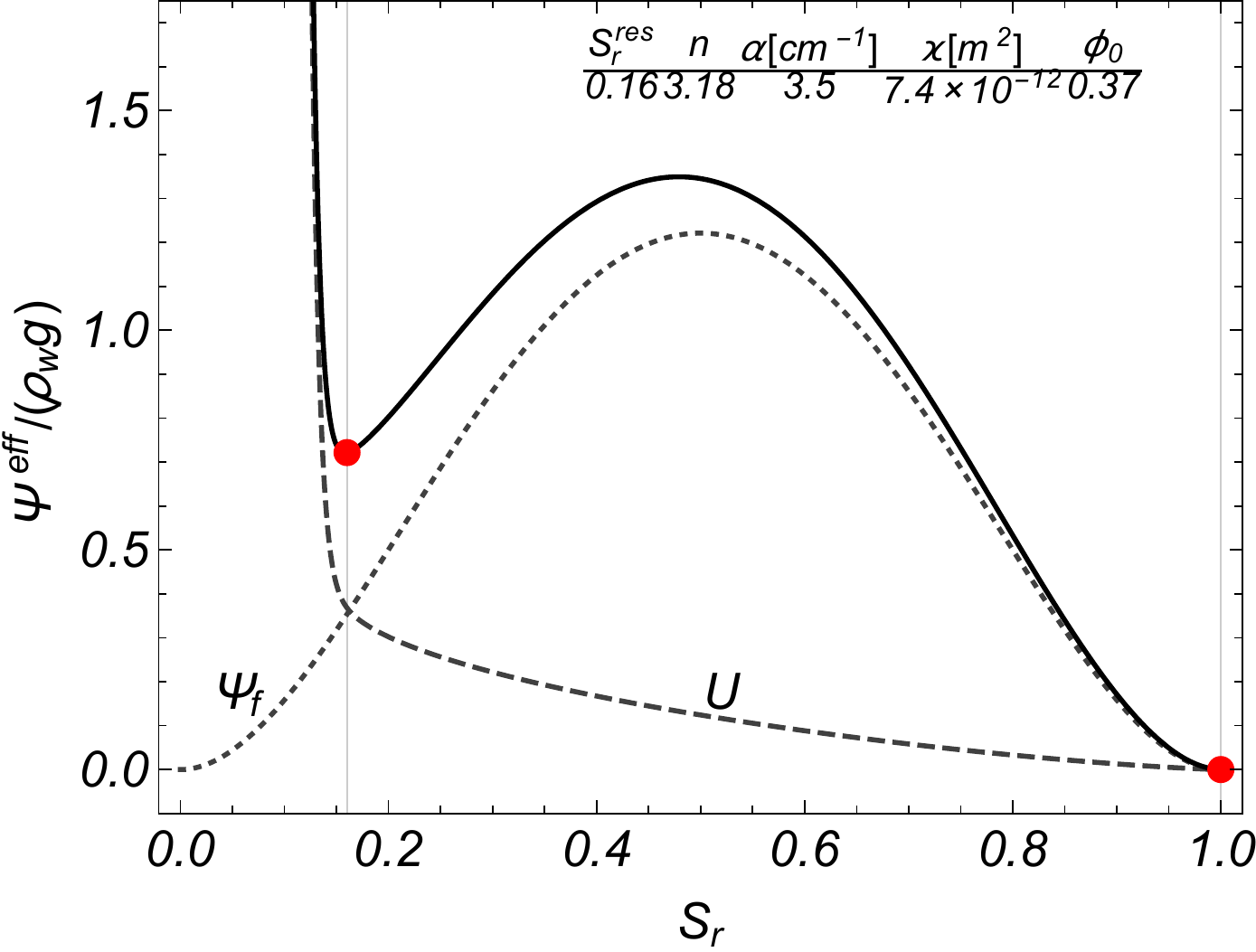}
\caption{Sand}
\label{fig:sandenergy}
\end{subfigure}
\,
\begin{subfigure}[b]{.23\textwidth}
\includegraphics[width=\textwidth]{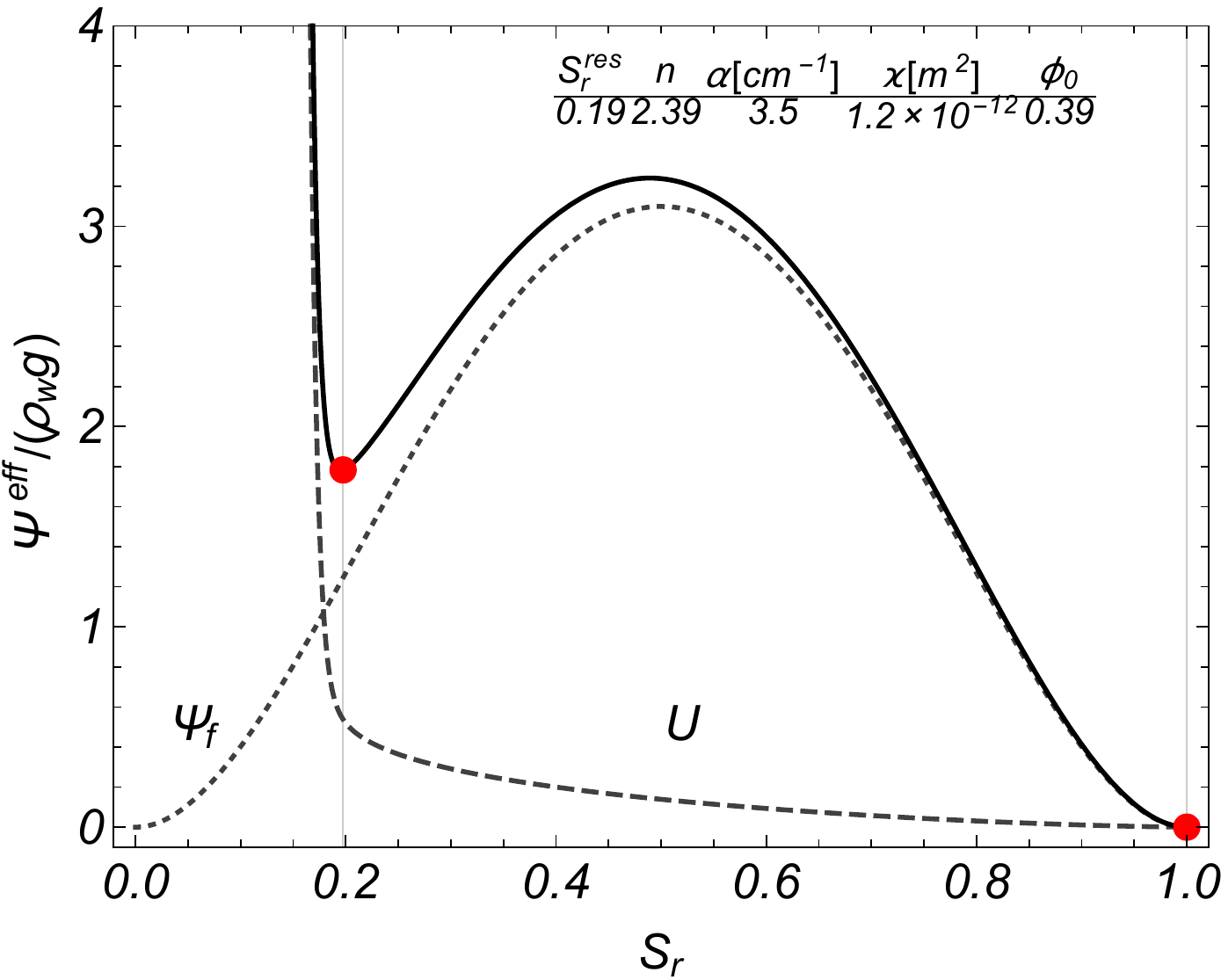}
\caption{Loamy sand}
\label{fig:loamysandenergy}
\end{subfigure}
\,
\begin{subfigure}[b]{.23\textwidth}
\includegraphics[width=\textwidth]{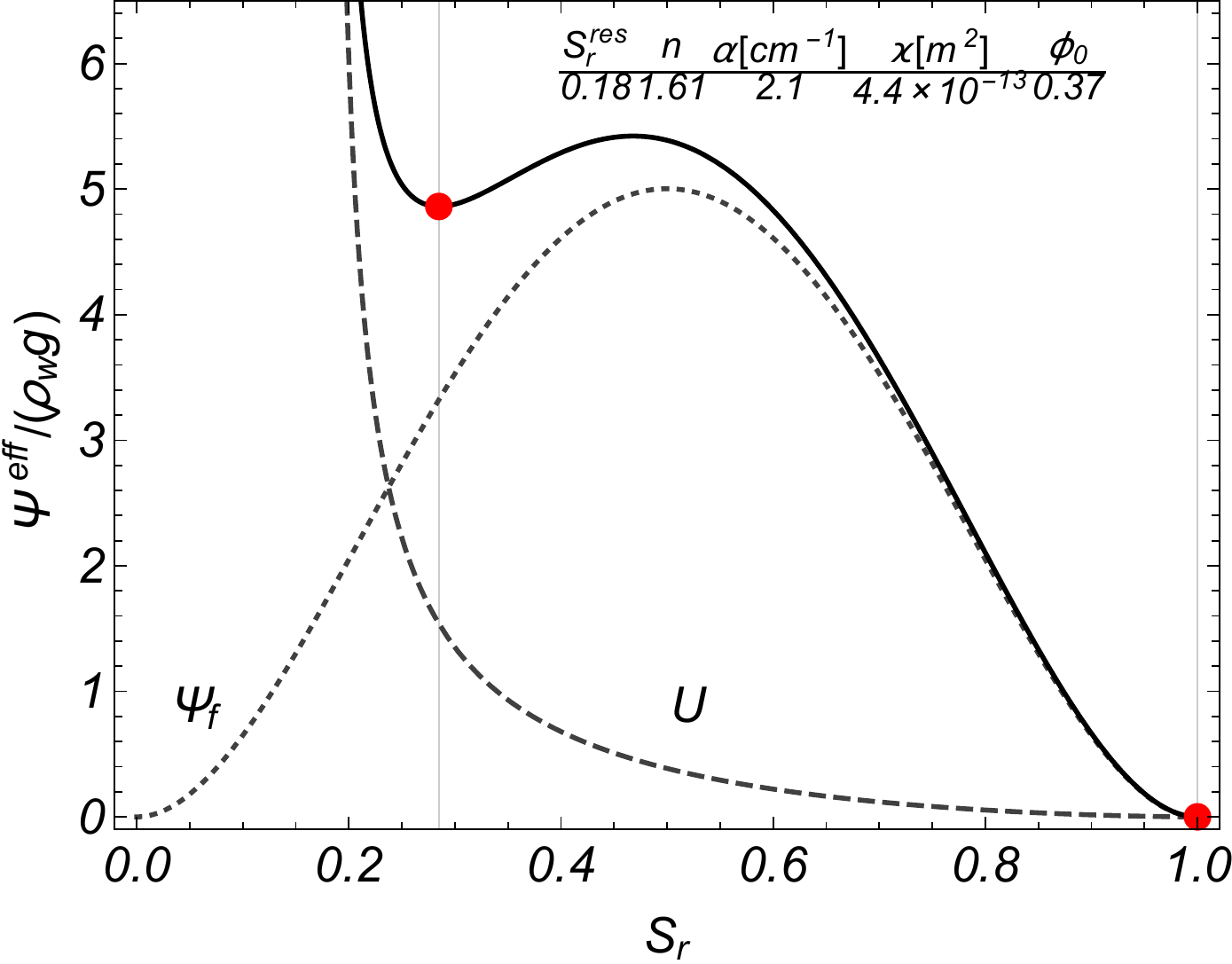}
\caption{Sandy loam}
\label{fig:sandyloamenergy}
\end{subfigure}
\,
\begin{subfigure}[b]{.23\textwidth}
\includegraphics[width=\textwidth]{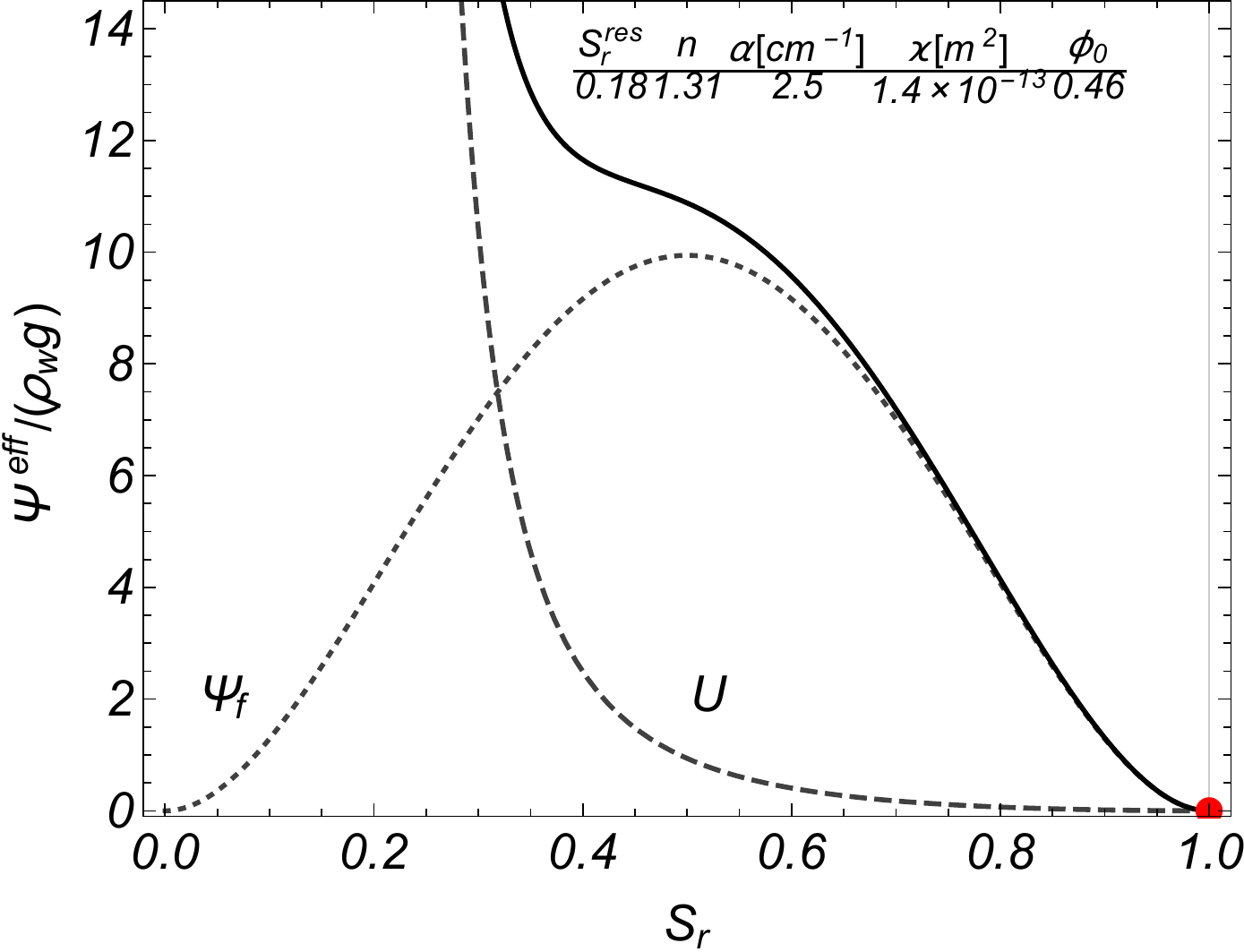}
\caption{Loam}
\label{fig:loamenergy}
\end{subfigure}
\\
\begin{subfigure}[b]{.23\textwidth}
\includegraphics[width=\textwidth]{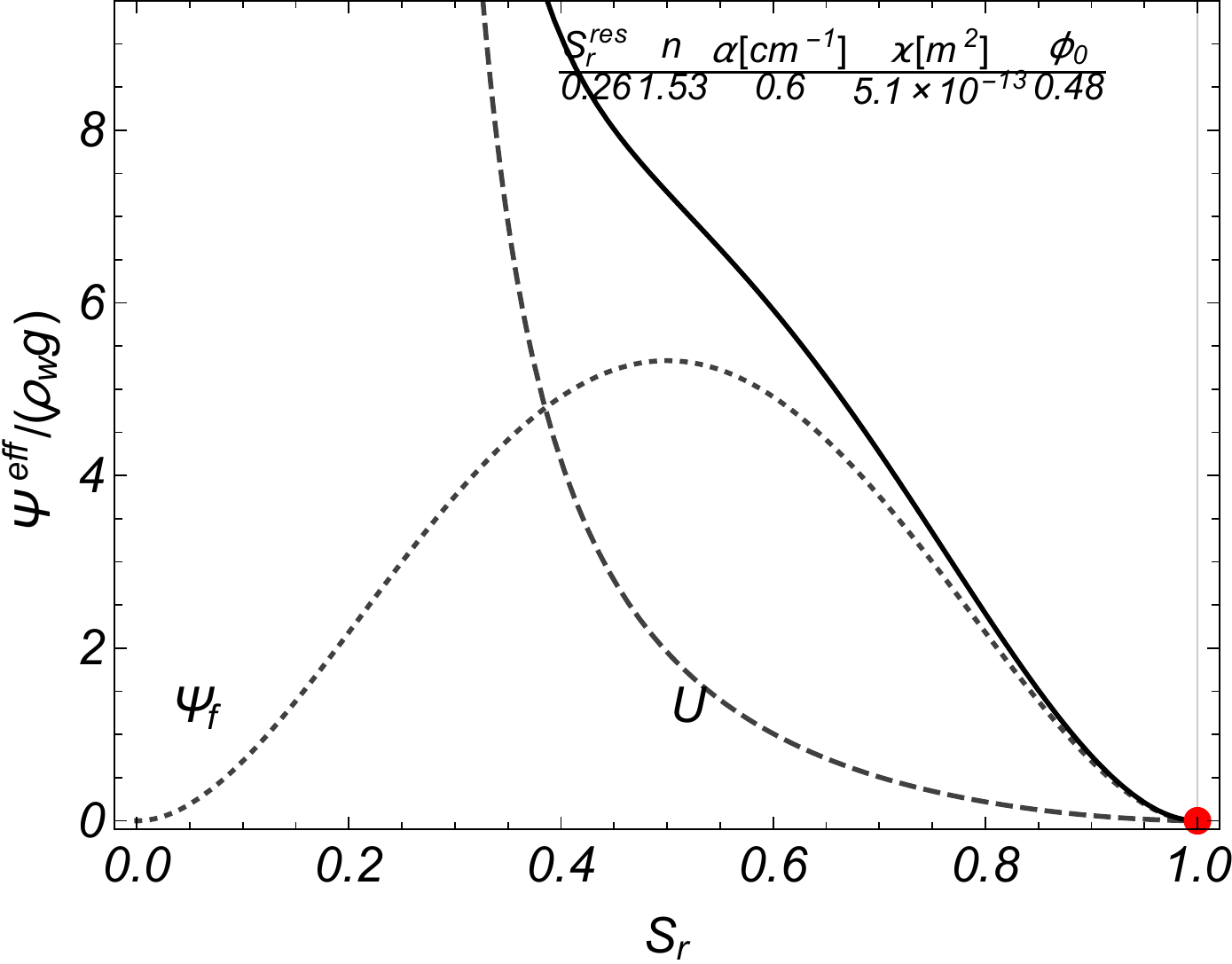}
\caption{Silt}
\label{fig:siltenergy}
\end{subfigure}
\,
\begin{subfigure}[b]{.23\textwidth}
\includegraphics[width=\textwidth]{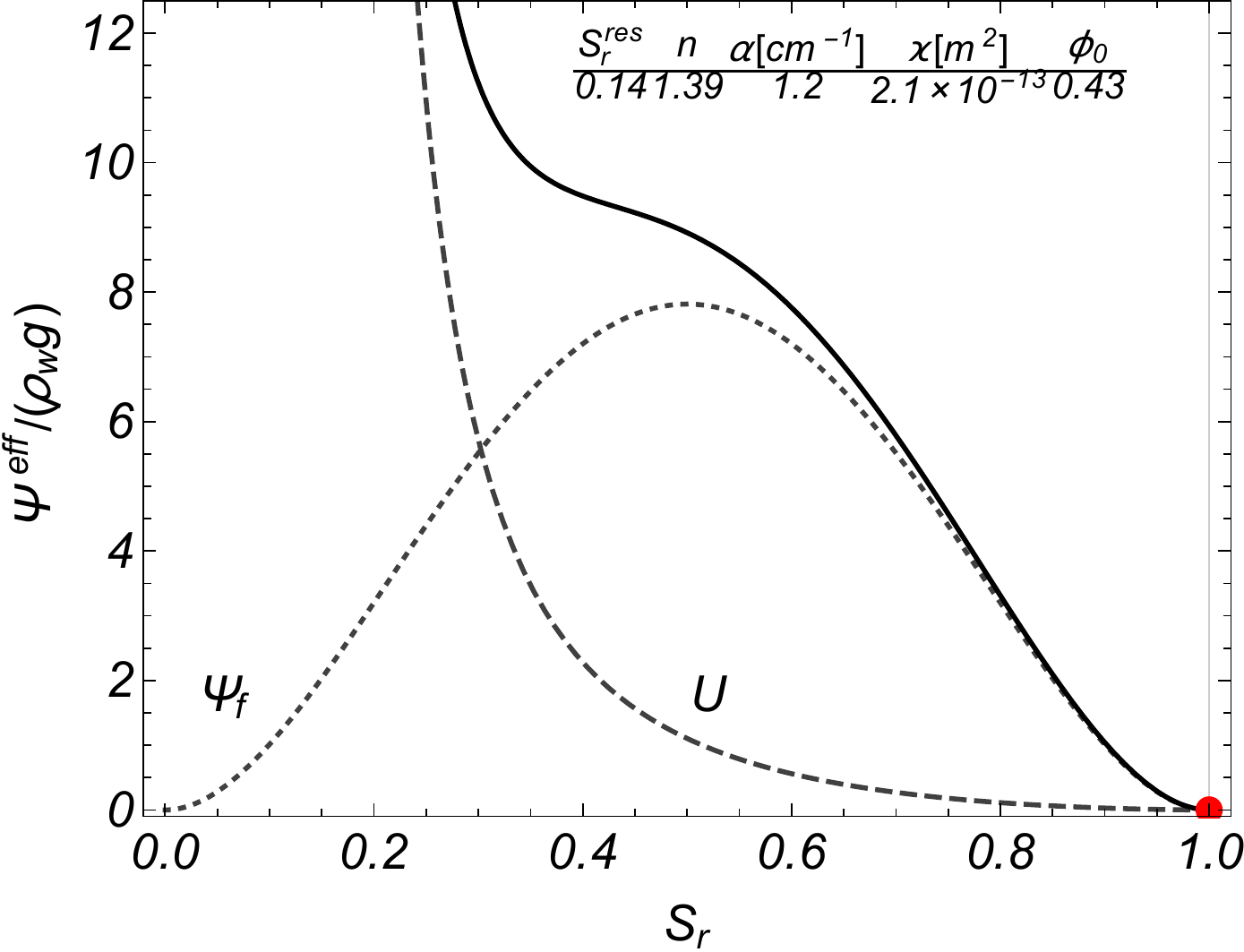}
\caption{Silt loam}
\label{fig:siltloamenergy}
\end{subfigure}
\,
\begin{subfigure}[b]{.23\textwidth}
\includegraphics[width=\textwidth]{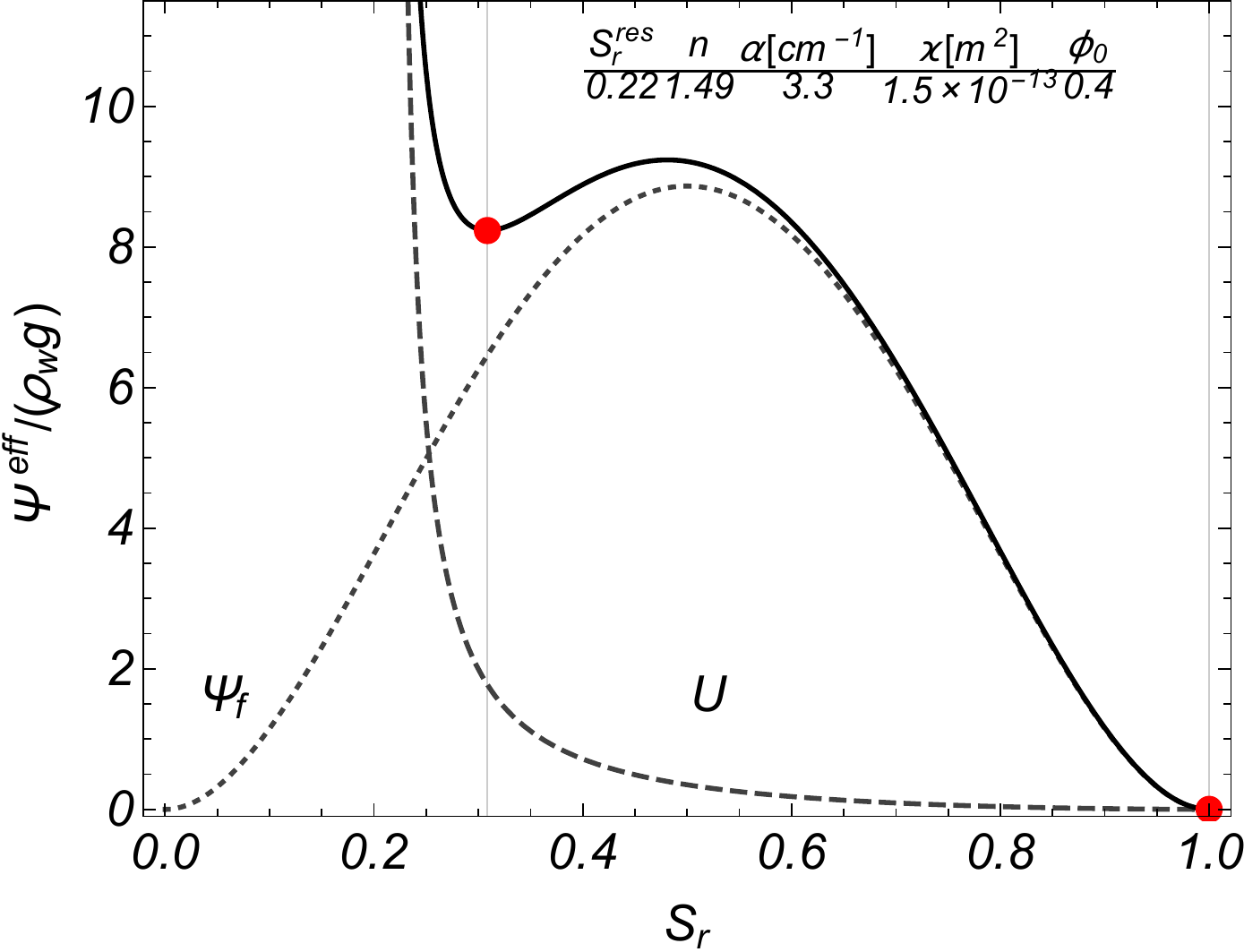}
\caption{Sandy clay Loam}
\label{fig:sandyclayloamenergy}
\end{subfigure}
\,
\begin{subfigure}[b]{.23\textwidth}
\includegraphics[width=\textwidth]{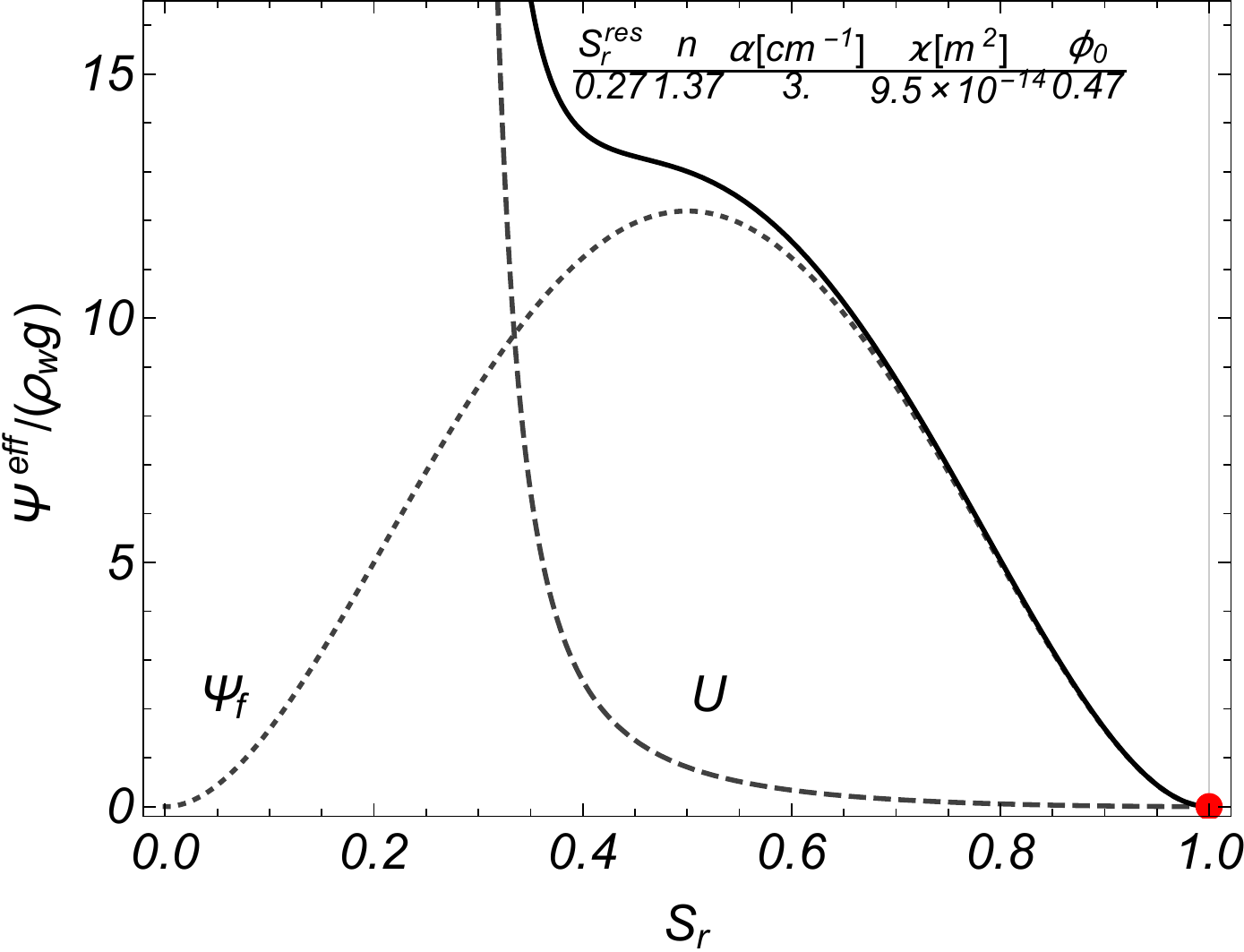}
\caption{Clay loam}
\label{fig:clayloamenergy}
\end{subfigure}
\smallskip
\\
\begin{subfigure}[b]{.23\textwidth}
\includegraphics[width=\textwidth]{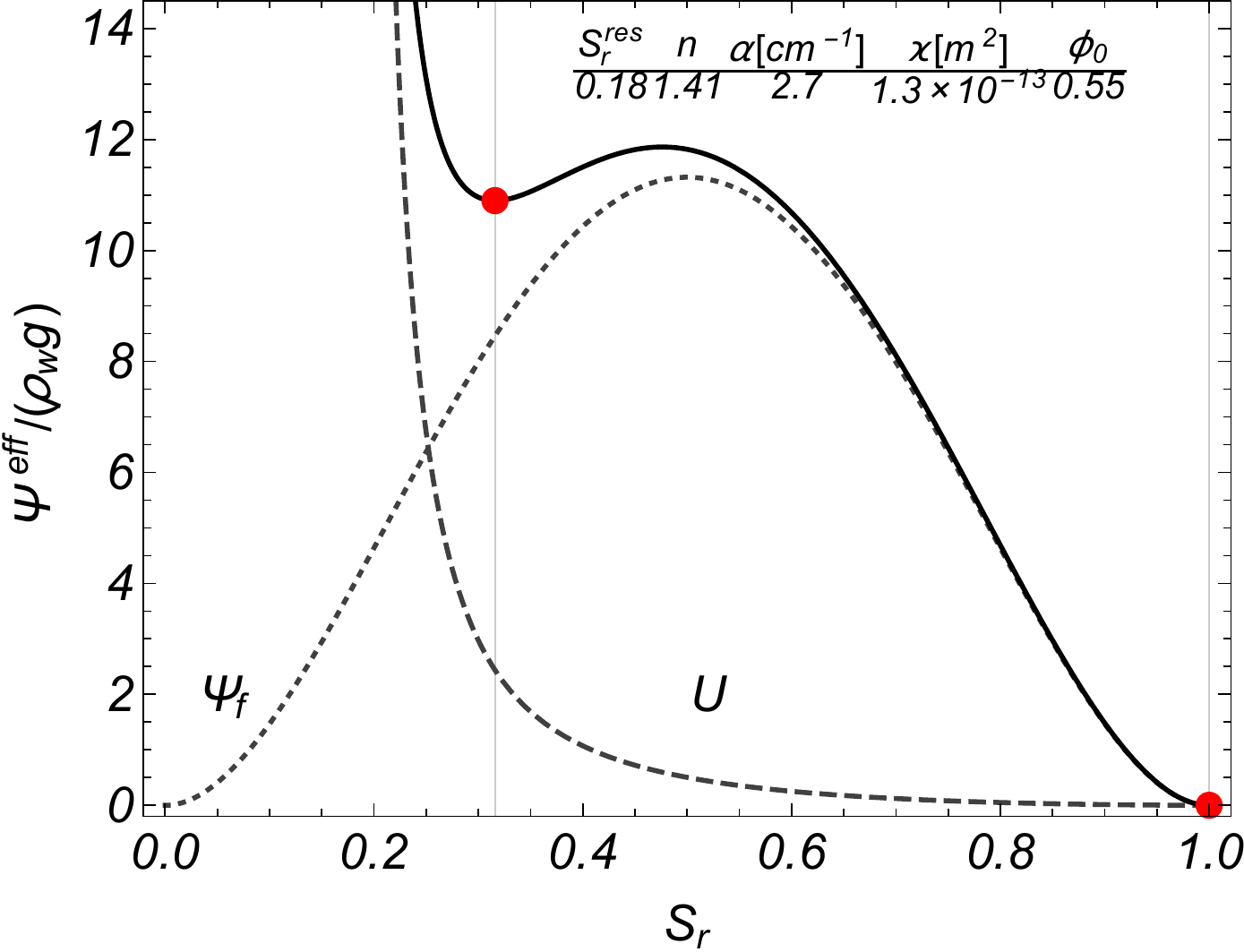}
\caption{Silty clay loam}
\label{fig:siltyclayloamenergy}
\end{subfigure}
\,
\begin{subfigure}[b]{.23\textwidth}
\includegraphics[width=\textwidth]{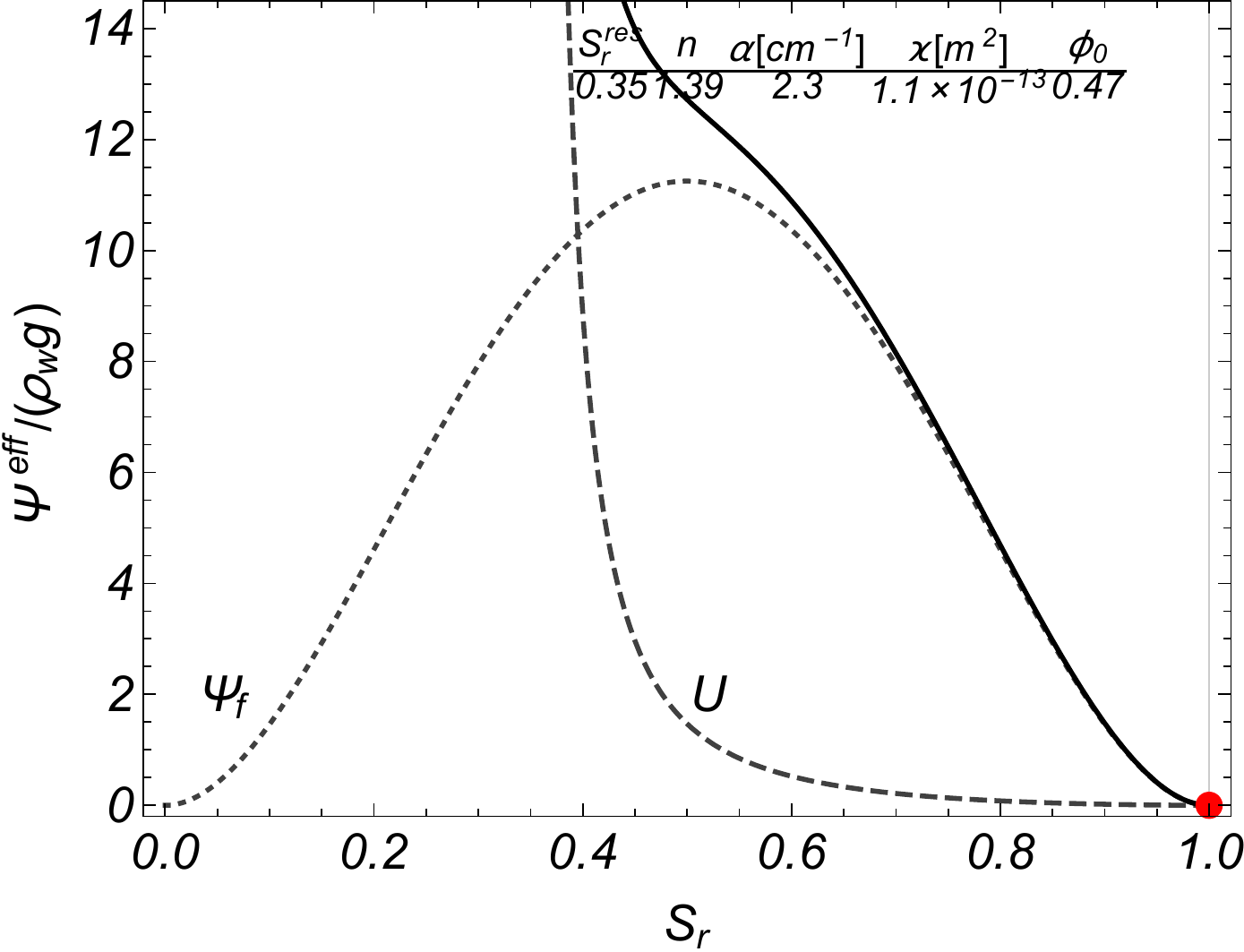}
\caption{Silty clay}
\label{fig:siltyclayenergy}
\end{subfigure}
\,
\begin{subfigure}[b]{.23\textwidth}
\includegraphics[width=\textwidth]{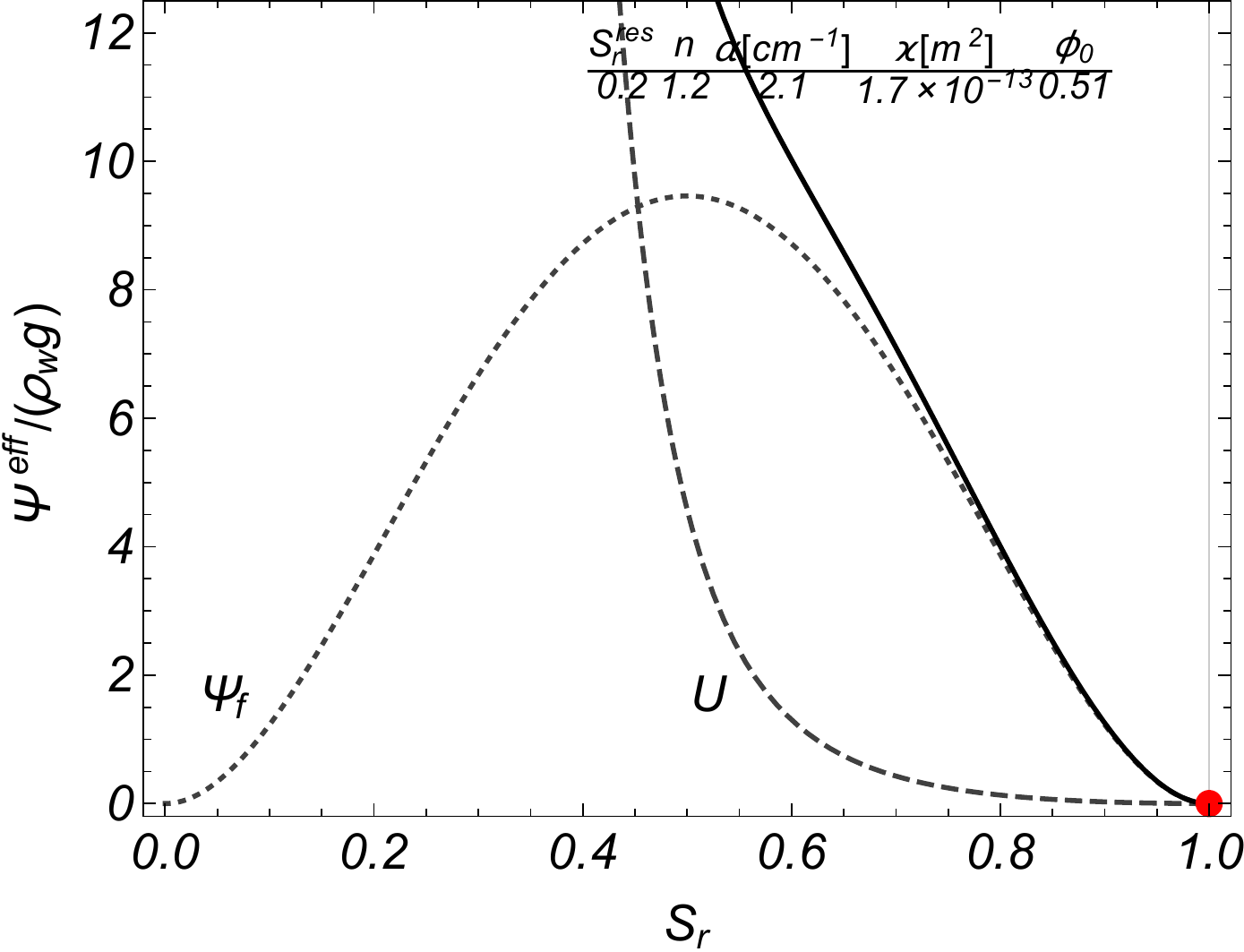}
\caption{Clay}
\label{fig:clayenergy}
\end{subfigure}
\caption{Energy of the pure fluid $\Psi_f$ (dotted gray line), capillary energy $U$ (dashed gray line) and effective pore-fluid energy (solid black line). The values of the parameters which characterize the capillary energy are those of the retention curve \eqref{retention_curve} and are listed in each panel, together with the intrinsic permeability $\varkappa$ and the referential porosity $\phi_0$, relative to the portion of sand, clay and silt which constitute the soil. The red spots indicate the minima of the pore-fluid energy $\Psi_f+U$ which maintains a double-well shape in panels (a), (b), (c), (g) and (i).}
\label{fig:soilenergy}
\end{figure}
It is worth to underline that the negative effective chemical potential exhibits, for all the considered textures, a non-monotonic behavior and in panels $(a)$, $(b)$, $(c)$, $(g)$ and $(i)$ two additional zeros (the spots) with respect to that at $S_r=1$, which can always be found. This feature corresponds to the fact that the effective pore-fluid energy $\Psi_f+U$ maintains the double-well shape typical of $\Psi_f$, see panels $(a)$, $(b)$, $(c)$, $(g)$, $(i)$ of Figure~\ref{fig:soilenergy}. 
The two minima of $\Psi_f+U$ are however no more isopotential and the one associated to the smallest value of $S_r$ has been shifted inwards the interval $(0,1)$. As already noticed the correction provided by $U$ to $\Psi_f$ allows therefore to account for the wetting properties of the grains, which do not permit the whole fluid to escape from the pore space even when a vanishing chemical potential is assumed. In all the other cases the non-monotonic behavior of $\mu^{\textrm{eff}}$ does not correspond to additional stationary states of the pore-fluid as the energy $\Psi_f+U$ exhibits only one minimum.

This non-monotonic behavior of the effective chemical potential resembles that one postulated by \cite{DiCarlo2008} in order to explain the formation of gravity fingers in soils. However it is worth to underline, following \cite{Cueto2010}, that the appropriate form of the potential $-\partial U/\partial S_r$ is captured solving an inverse problem which stems from experimental evidence. As a consequence, the effective generalized chemical potential $\text{\textmu}^{\textrm{eff}}$ can indeed be hydrostatic at equilibrium, and therefore consistent with gravitational loading, even if the effective chemical potential $\mu^{\textrm{eff}}$ is not.

\section{Micro-scale interpretation of the enhanced constitutive prescription of capillary pressure}\label{micro}
In order to compare the enhanced model of capillary pressure \eqref{retention_&_porosity} with the classical one, see \textit{e.g.} \cite{Coussy_book04,Coussy_book10}, the dependence of the free energy $\Psi_s$ on the saturation degree $S_r$ is concentrated, as already done in \S~\ref{S:gen_Darcy} and \S~\ref{S:porefluid}, in the capillary energy $U$ multiplied by the Lagrangian porosity, say $\phi U(S_r)$. Equation \eqref{retention_&_porosity}$_1$ therefore explicitly improves the basic constitutive prescription of the macro-scale capillary pressure by a correction depending on the gradient of the Jacobian determinant $J$ and the gradient of the water content $n S_r$, once the quadratic form \eqref{kappa_f} for the non-local contribution to the energy of the fluid has been assumed:
\begin{equation}
\label{Pc_corrected}
\mathcal P_c = -\dfrac{\partial U}{\partial S_r}-\dfrac{\mathrm{C}_{\kappa}}{J^2} \left(F^{-1} \right)_{m\alpha}\left(F^{-1} \right)_{k\alpha}\left[(\phi S_r)_{,m}-(\phi S_r) \frac{J_{,m}}{J}\right]\,  J_{,k}.
\end{equation}
When the hyper-stress acting on the non-uniform fluid vanishes, this correction vanishes as well; in other words the larger the gradient of the liquid content is, the wider the discrepancy between the standard and the enhanced constitutive prescriptions of $\mathcal P_c$ will be. The enhanced constitutive prescription \eqref{Pc_corrected} is expected therefore to yield significant modifications of the capillary pressure, in the narrow subdomains of the current shape of the porous medium where significant gradients of the liquid content can be detected. This justifies the claim of an interpretation of this contribution in terms of the so-called specific interfacial area.

According to the experimental data, reported among others by \cite{Dalla2002, Culligan2004} and \cite{Costanza2011}, and the pore-network numerical simulation of quasi-static drainage and imbibition of a porous medium, see \textit{e.g.} \cite{Joekar2010}, local variations of the macroscopic capillary pressure are typically accompanied not only by changes in the saturation degree but also by changes in the interfacial area $a_{nw}$, which accounts for the local cumulative measure of the interfaces between the non-wetting and the wetting phase, per unit volume of the RVE. Apparently changes in the interfacial area may or may not result into an incoming or an outcoming flow through the RVE, however, in the spirit of providing an interpretation of the considered macro-scale model of the capillary force, in terms of possible effects of the interfacial area, only the variations of $a_{nw}$, which are associated to a flow through the boundary of the RVE are considered, whilst no account is taken of those variations which yield a pure remodeling of the internal structure of the RVE.  In other words a relation between the interfacial area and the gradient of the saturation degree is expected to be identified, only if the liquid particles are displaced towards the boundary of the RVE. 

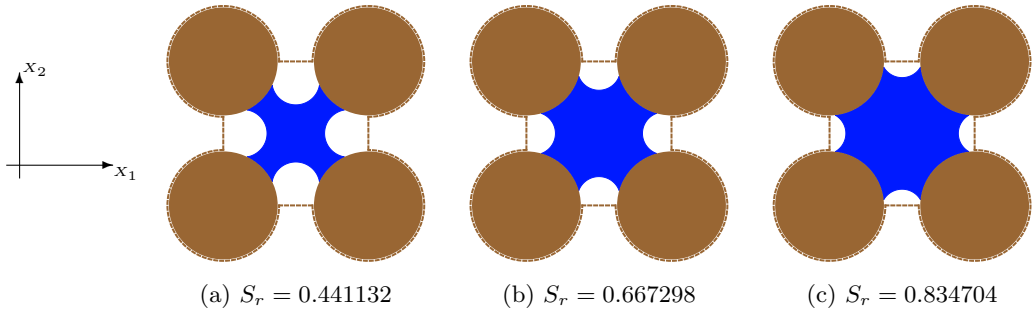
\begin{figure}[h]
\begin{picture}(50,50)(0,-50)
\put(0,0){\vector(0,1){40}}
\put(-5,5){\vector(1,0){40}}
\put(35,1){${\scriptscriptstyle X_1}$}
\put(1,40){${\scriptscriptstyle X_2}$}
\end{picture}
\begin{subfigure}[b]{.22\textwidth}
\includegraphics[width=\textwidth]{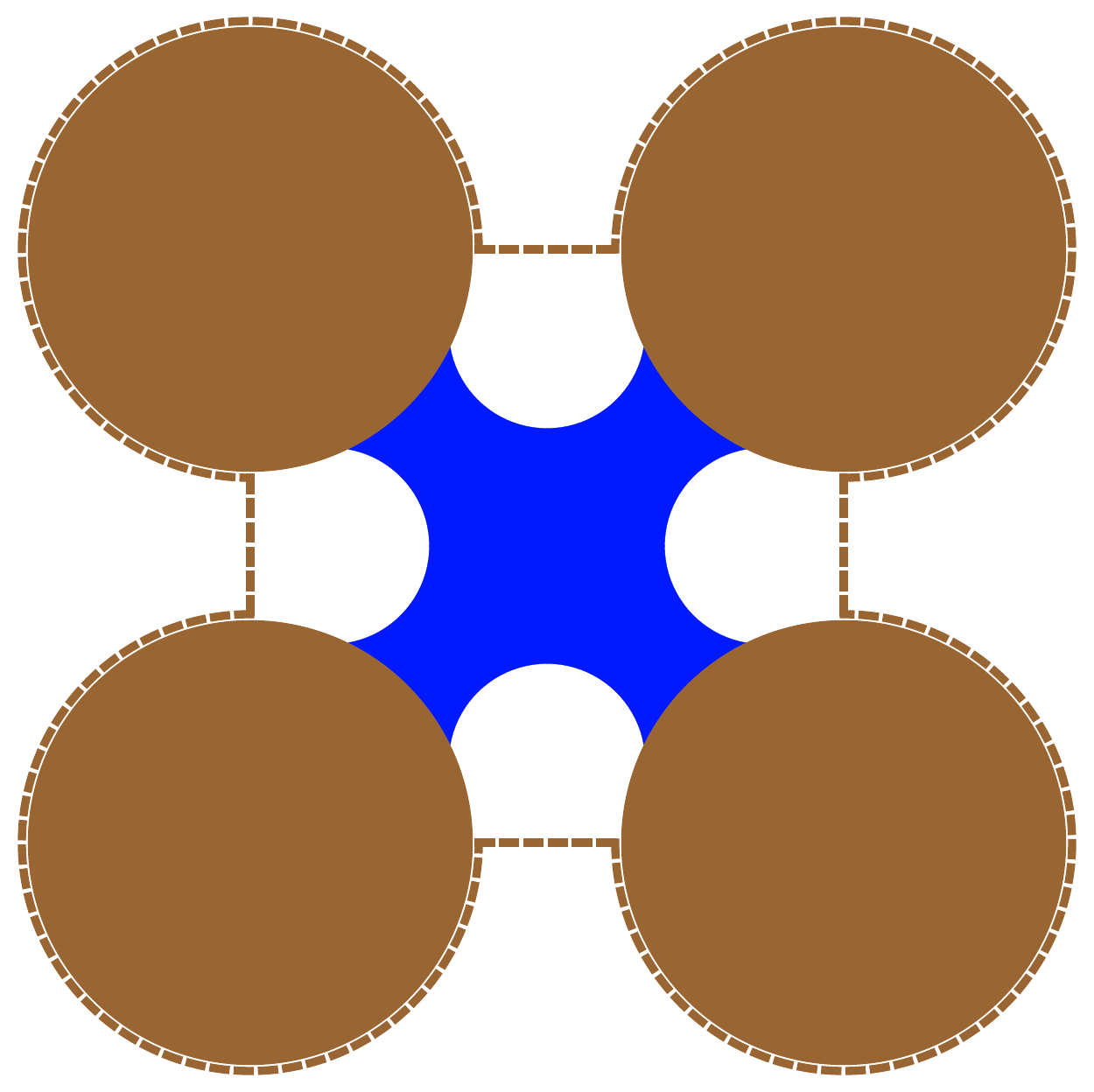}
\caption{$S_r=0.441132$}
\label{fig:microSr1}
\end{subfigure}
\quad
\begin{subfigure}[b]{.22\textwidth}
\includegraphics[width=\textwidth]{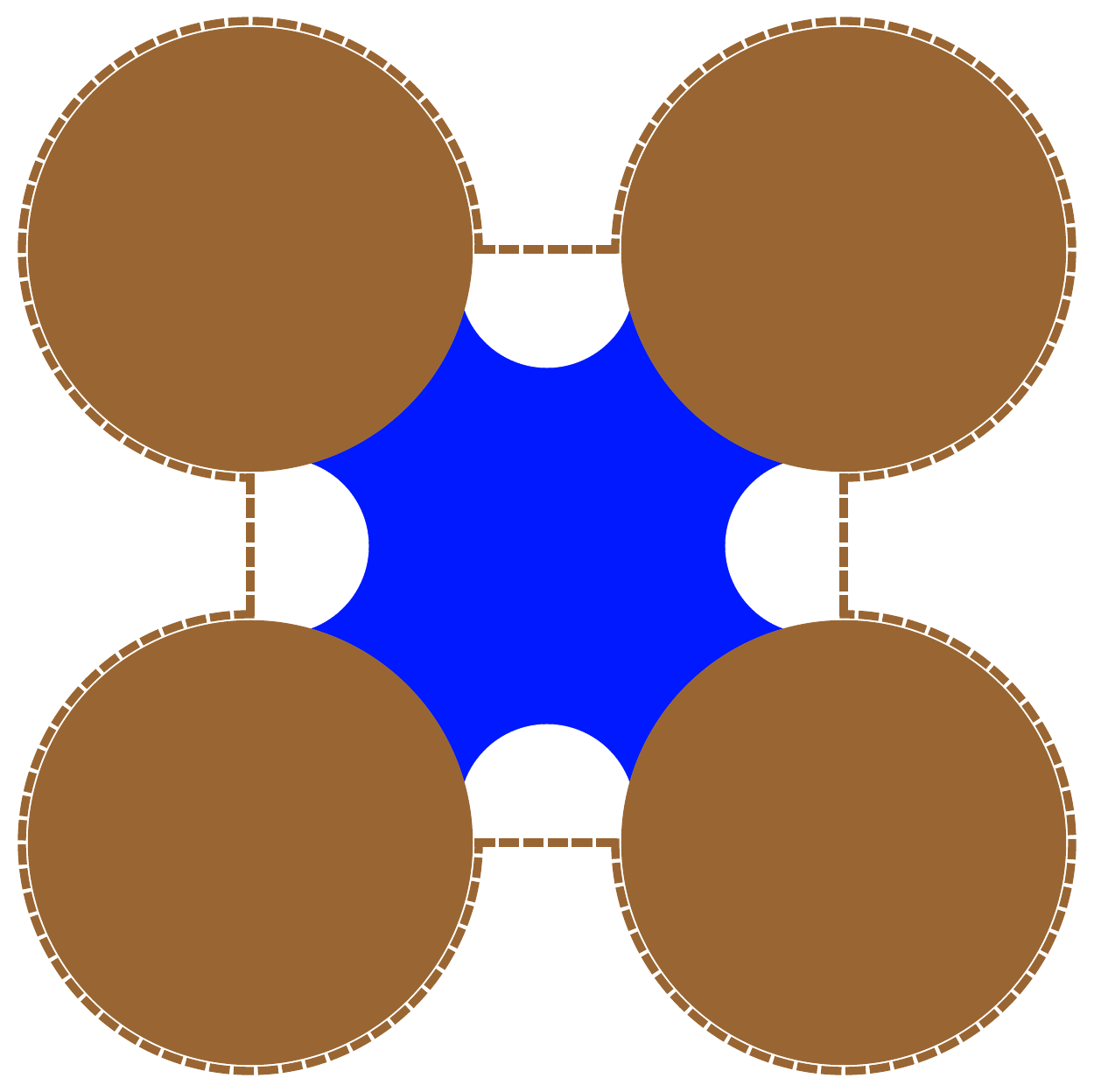}
\caption{$S_r=0.667298$}
\label{fig:microSr2}
\end{subfigure}
\quad
\begin{subfigure}[b]{.22\textwidth}
\includegraphics[width=\textwidth]{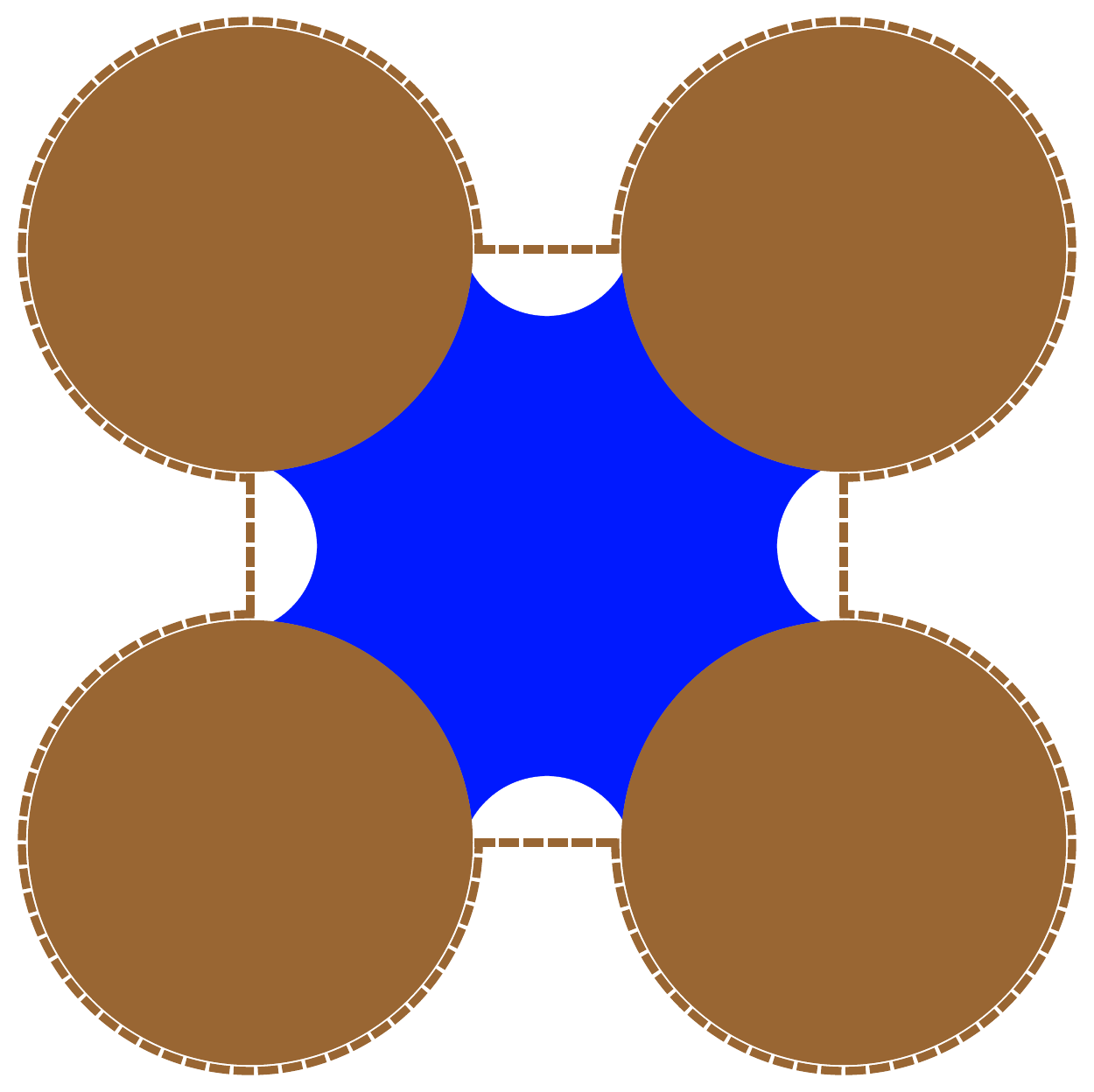}
\caption{$S_r=0.834704$}
\label{fig:microSr3}
\end{subfigure}
\caption{Referential RVEs parametrized by the saturation degree; the dashed line indicates the boundary of the RVE. The distance between the beads, made dimensionless with respect to the characteristic size of the RVE, is $L=0.25$, whilst their radius is $R=(1-L)/2$. $(X_1,X_2)$ indicates coordinate system in the reference configuration.}
\label{fig:microSr}
\end{figure}

To verify the validity of this hypothesis a heuristic micro-scale analysis is developed deforming a prototype initial configuration $\mathcal B^0$ of a RVE by a quasi-static loading path driven by the Lagrangian gradient of the macro-scale Jacobian determinant $J$; the liquid is squeezed out of the porous chamber along the same direction of the gradient of $J$. The saturation degree $S_r$ is assumed to be not affected by the considered micro-scale deformation, which means that the ratio between the volume of liquid and the volume of the pores within the RVE does not vary during the considered deformation process. This assumption allows to capture the correction to the constitutive prescription of $\mathcal P_c$ provided by the gradient term of the fluid energy only. In Figure~\ref{fig:microSr} possible initial configurations of the RVE, parametrized by the degree of saturation are depicted; the dashed lines delimitate the boundary of the reference shapes.

Each RVE corresponds to the deformed current configuration $\mathcal B$ of an initial domain $\mathcal B^0$ constituted by four identical beads among which a suitable amount of liquid water (the wetting phase) is trapped by capillary forces. The amount of liquid (in volume) is prescribed by the value of the degree of saturation; on the other hand its spatial distribution is a function of the local wetting properties of the beads. Here, for the sake of simplicity, the equilibrium value of Young's contact angle is kept constant during deformation and circular interfaces between the non-wetting and the wetting fluid are assumed at each loading step. Tuning the parameterizing value of $S_r$ is expected to provide a-posteriori the relation between the specific interfacial area and the saturation degree, and consequently to compare the results of the model with the benchmark experimental data, relative to drainage-imbibition cycles. 
\begin{figure}[h]
\centering
\includegraphics[width=.6\textwidth]{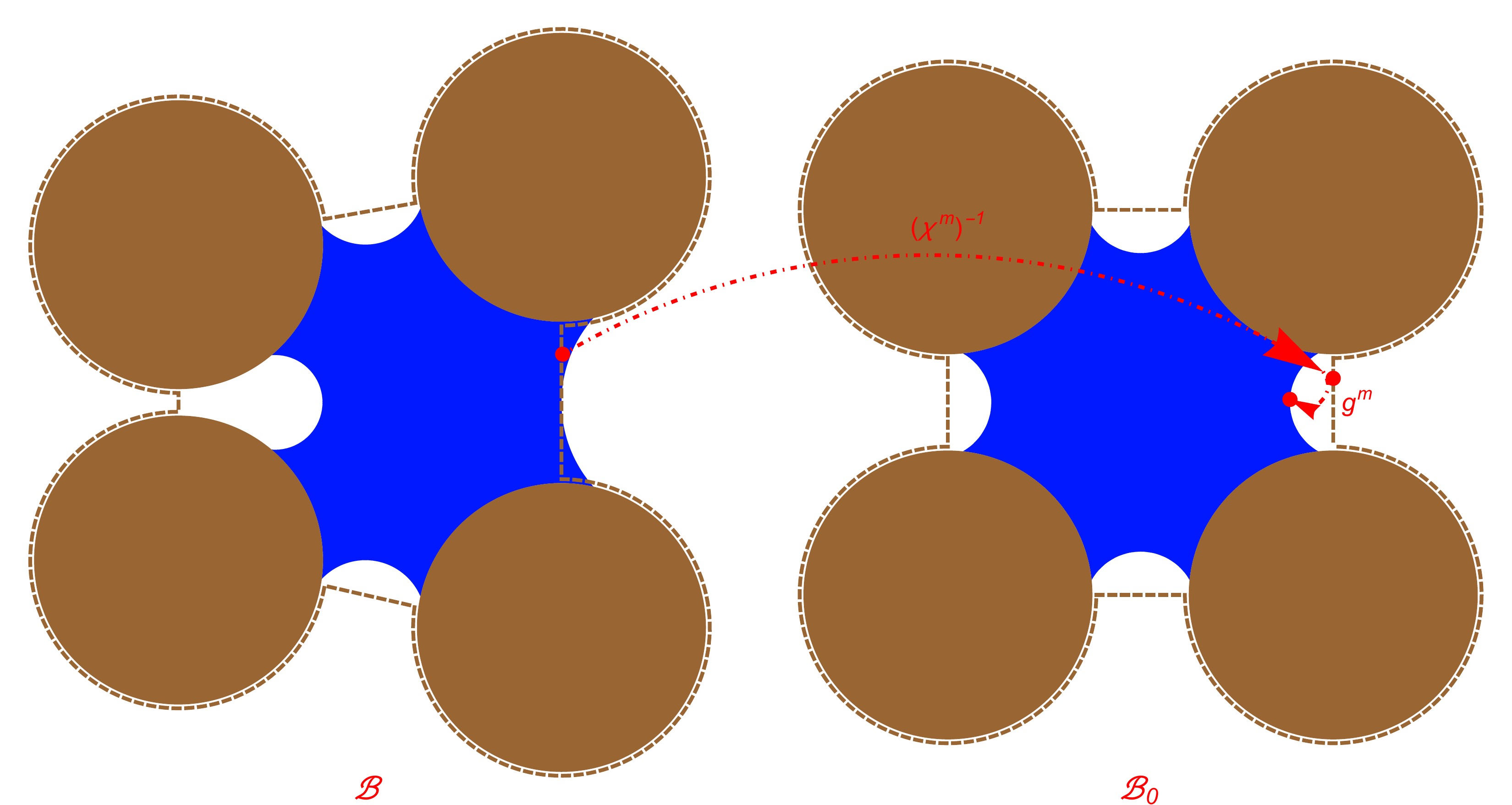}
\caption{On the left a possible RVE, say $\mathcal B$, on the right its referential counter image, say $\mathcal B^0$. The map $g^{\textrm{m}}$ associates to each point of the reference configuration, which is mapped by $\chi^{\textrm{m}}$ into the current position of a liquid particle, the corresponding position in the initial configuration of the liquid.}
\label{fig:chim&gm}
\end{figure}
Let $\chi^\textrm{m}$ and $g^\textrm{m}=(\chi_{liq}^\textrm{m})^{-1}\circ\chi^\textrm{m}$ be the micro-scale placement of the solid and the pull-back, in the solid reference configuration, of the inverse of the micro-scale placement of the liquid $\chi_{liq}^\textrm{m}$, respectively; they are diffeomorphisms over the above mentioned reference configuration, similarly to the corresponding macro-scale maps introduced in \S~\ref{S:Kin}. In particular $\chi^\textrm{m}$, which is naturally defined over the reference configuration of the solid $\mathcal D_s^0$, is extended over the pore space network. On the other hand the domain of $g^\textrm{m}$ is the subset $\mathcal D^0_{liq}$, of the reference configuration, mapped by $\chi^{\textrm{m}}$ into the inter-granular space occupied by the liquid in the current configuration. Notice that $\mathcal D^0_{liq}$ does not coincide with the initial configuration of the liquid, but it is the pull-back of the current shape of the liquid into the reference (initial) configuration of the solid grains, see Figure~\ref{fig:chim&gm}.

\begin{figure}[h]
\centering
\begin{subfigure}[b]{.22\textwidth}
\includegraphics[width=\textwidth]{fig2_3.pdf}
\caption{$J_{,1}=0$}
\label{fig:micro_Srmin_J0}
\end{subfigure}
\quad
\begin{subfigure}[b]{.22\textwidth}
\includegraphics[width=\textwidth]{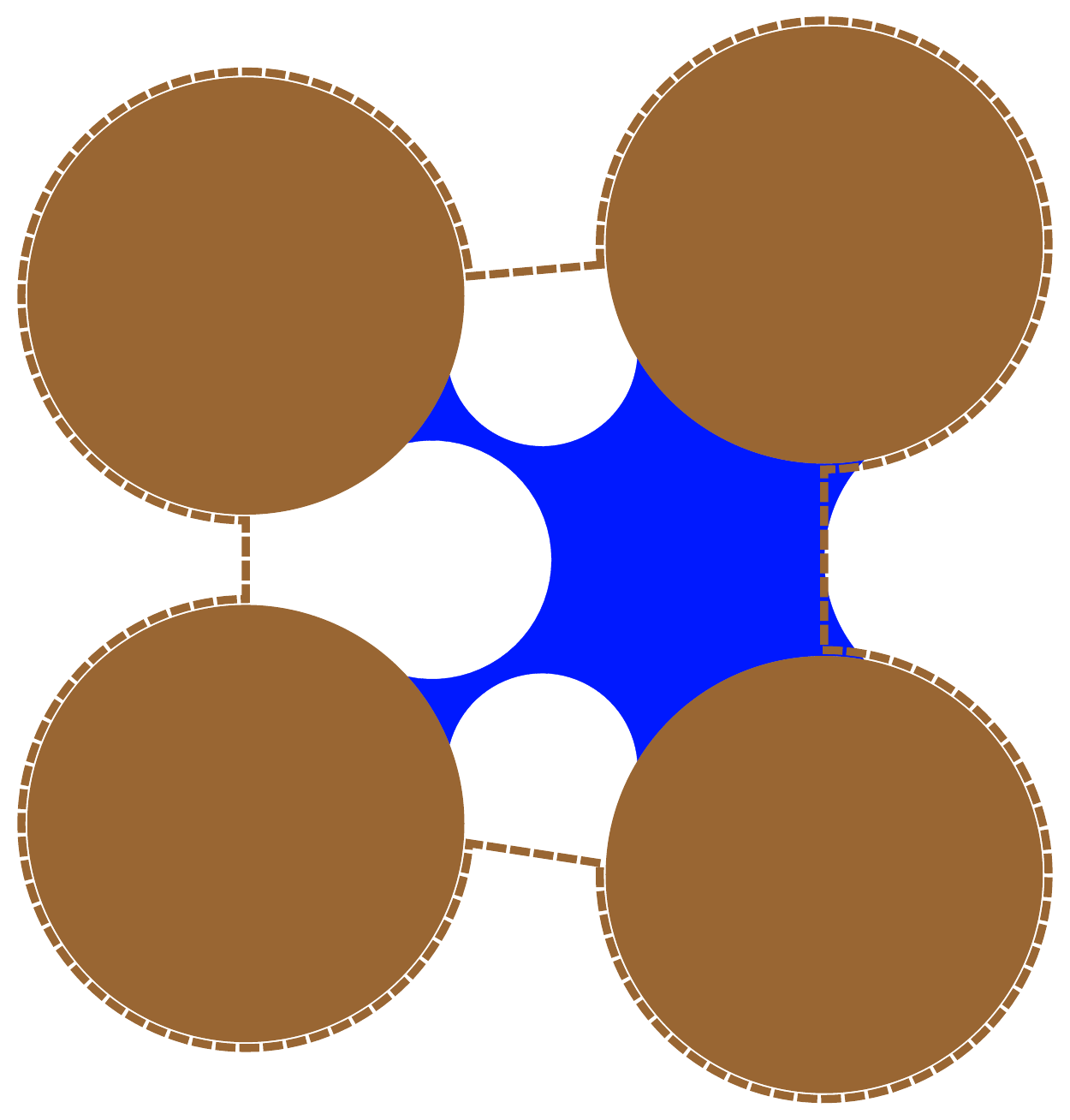}
\caption{$J_{,1}=0.173618$}
\label{fig:micro_Srmin_J2}
\end{subfigure}
\quad
\begin{subfigure}[b]{.22\textwidth}
\includegraphics[width=\textwidth]{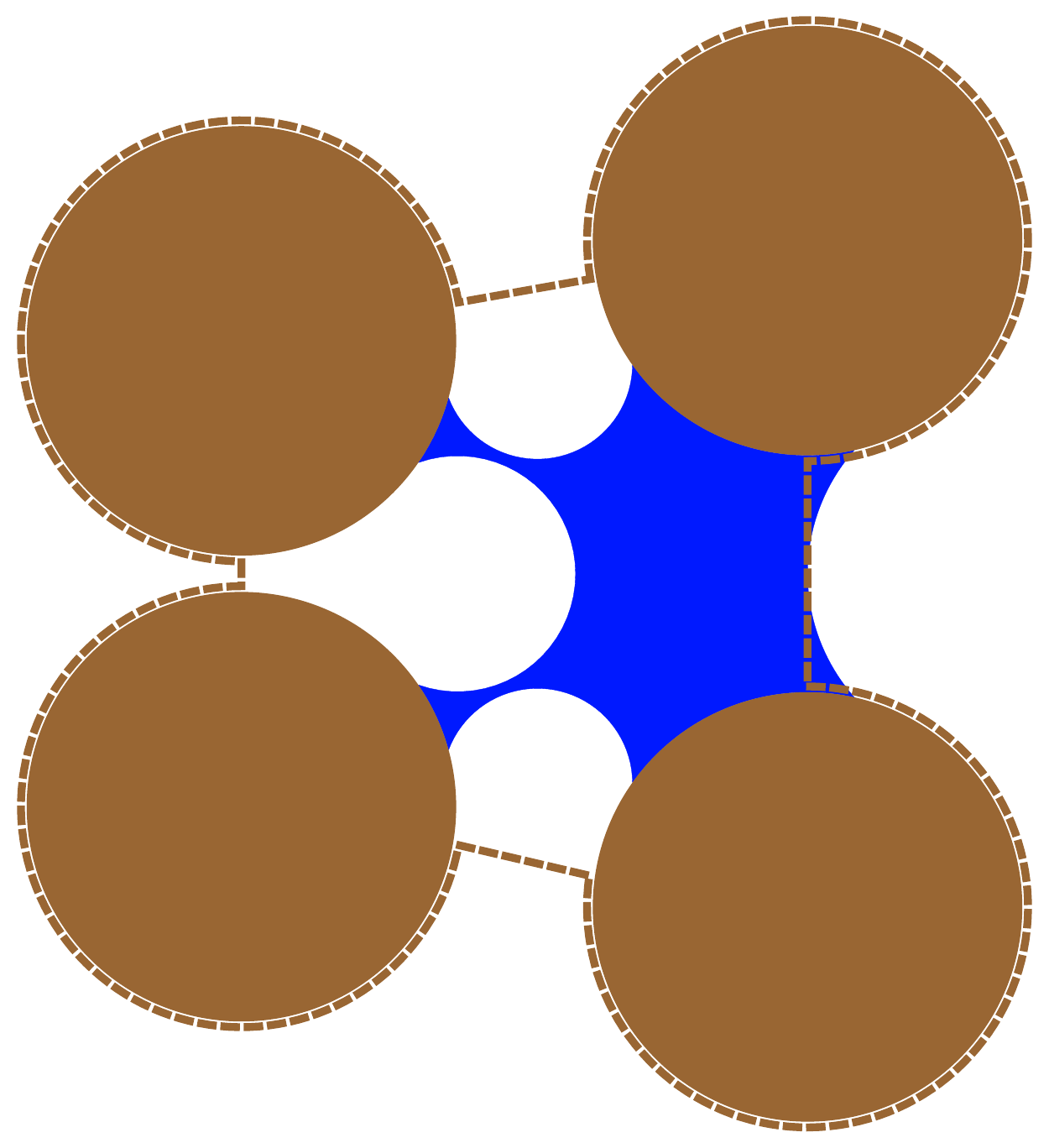}
\caption{$J_{,1}=0.347235$}
\label{fig:micro_Srmin_J3}
\end{subfigure}
\quad
\begin{subfigure}[b]{.22\textwidth}
\includegraphics[width=\textwidth]{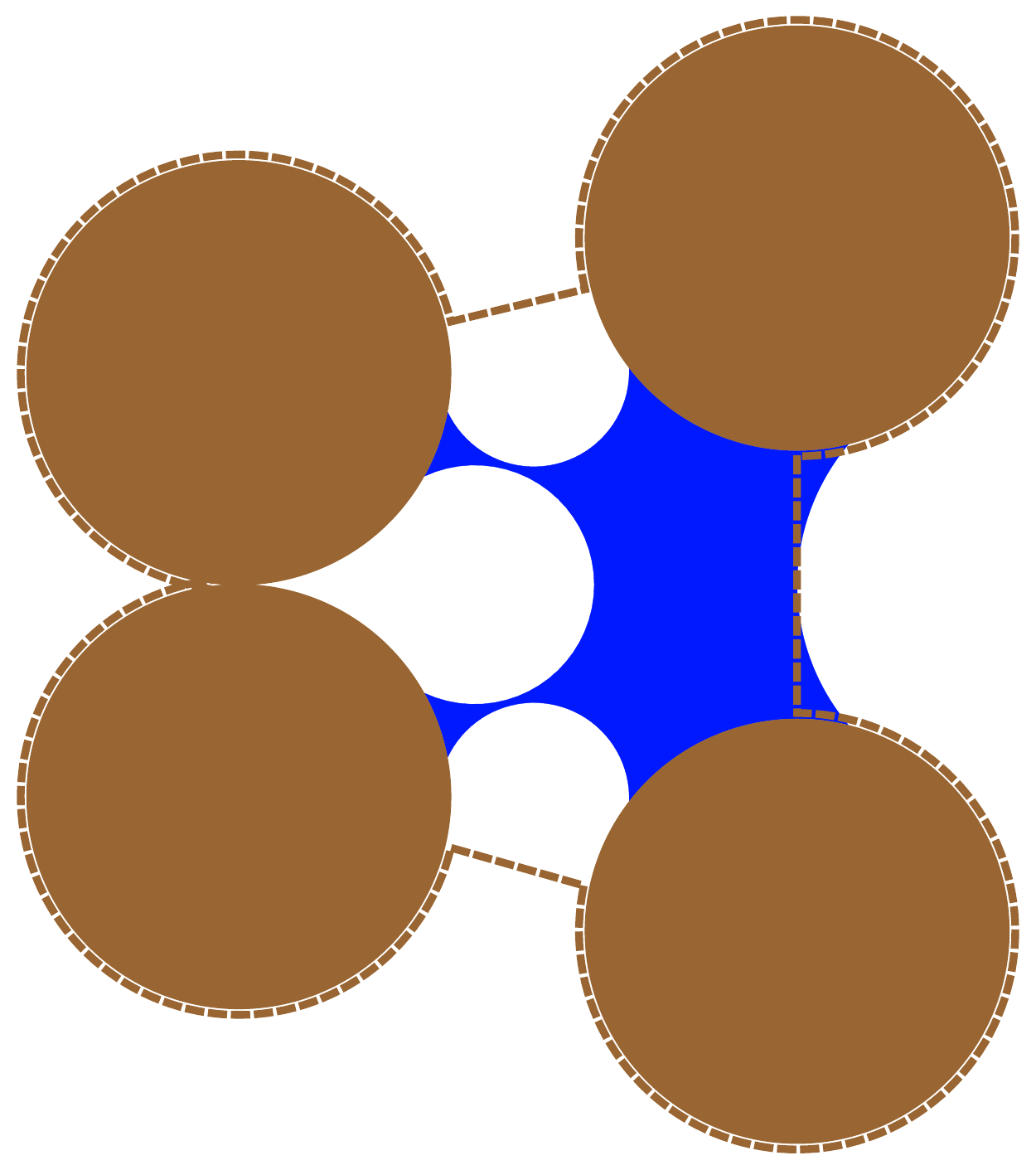}
\caption{$J_{,1}=0.471248$}
\label{fig:micro_Srmin_J4}
\end{subfigure}
\caption{Admissible distributions of the liquid phase within different RVEs, which are obtained as the current configurations of the reference shape reported in panel (a).
The micro-scale displacement of the solid beads is parametrized by the value of $J_{,1}$ via its dependence on the macro-scale second gradient of deformation. During the deformation the 
saturation ratio is kept constant: $S_r=0.441132$ (see Figure~\ref{fig:microSr1}).}
\label{fig:micro_Srmin}
\end{figure}

\begin{figure}[h]
\centering
\begin{subfigure}[b]{.22\textwidth}
\includegraphics[width=\textwidth]{fig2_2.pdf}
\caption{$J_{,1}=0$}
\label{fig:micro_Srmed_J0}
\end{subfigure}
\quad
\begin{subfigure}[b]{.22\textwidth}
\includegraphics[width=\textwidth]{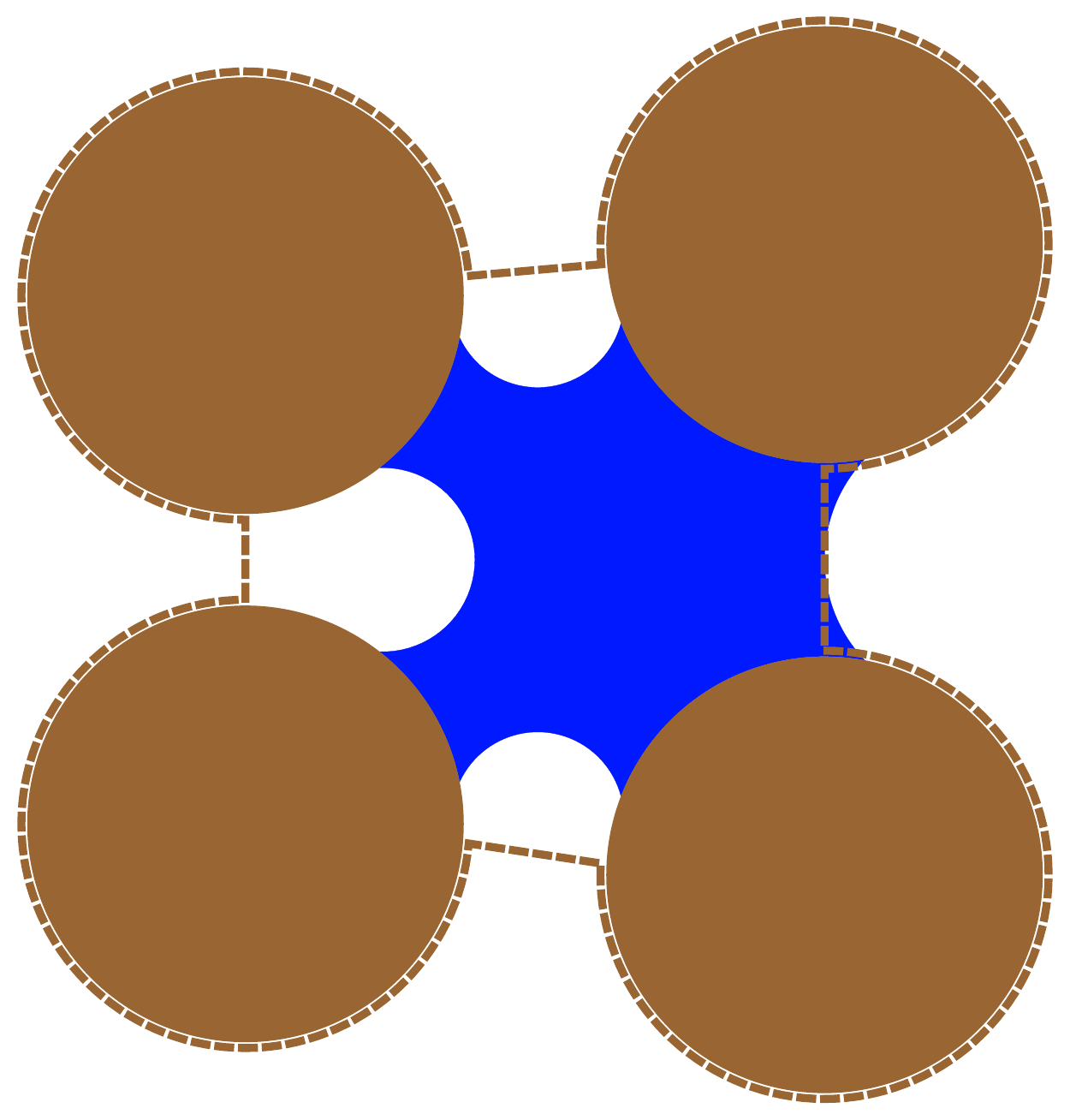}
\caption{$J_{,1}=0.173618$}
\label{fig:micro_Srmed_J2}
\end{subfigure}
\quad
\begin{subfigure}[b]{.22\textwidth}
\includegraphics[width=\textwidth]{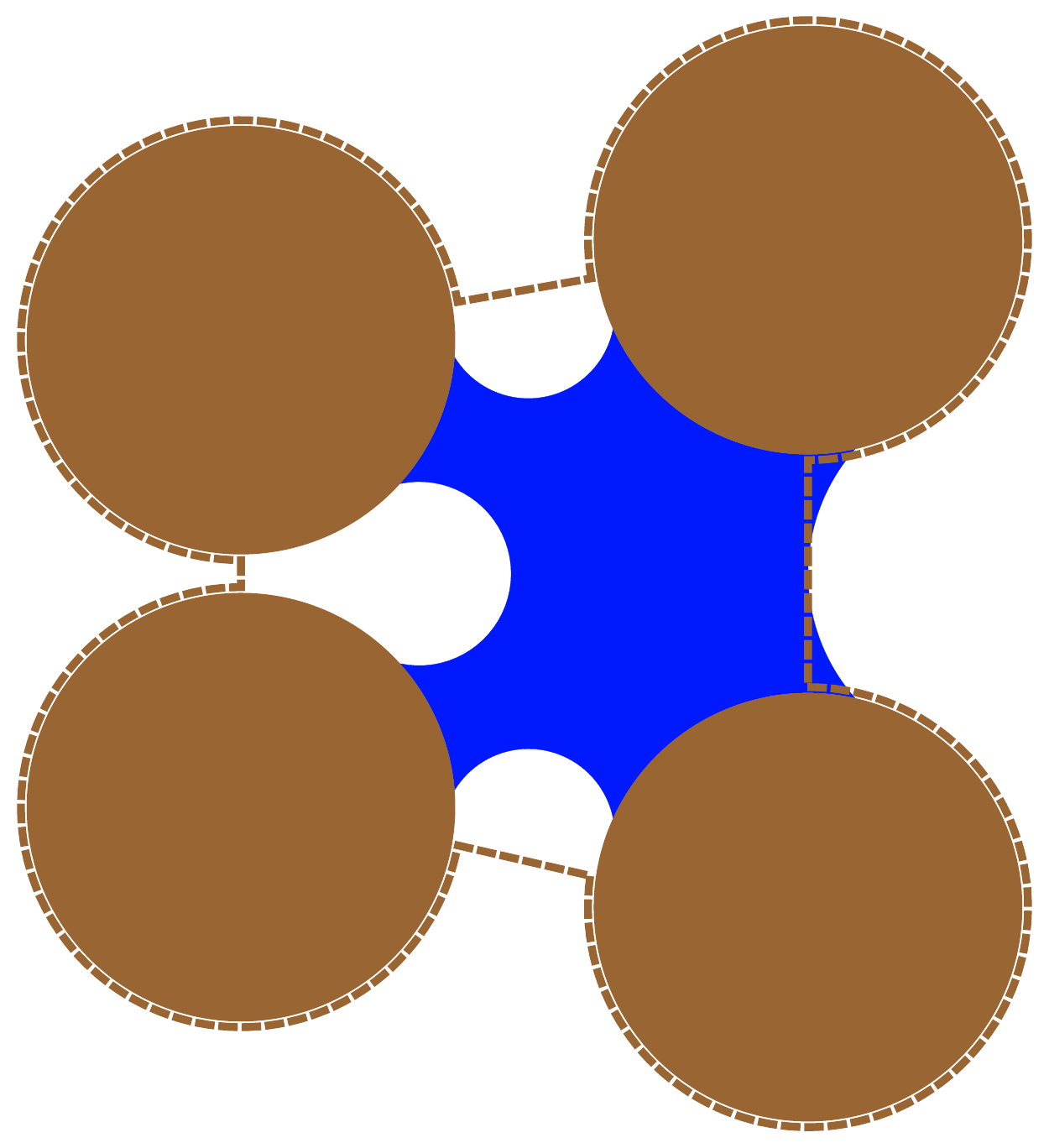}
\caption{$J_{,1}=0.347235$}
\label{fig:micro_Srmed_J3}
\end{subfigure}
\quad
\begin{subfigure}[b]{.22\textwidth}
\includegraphics[width=\textwidth]{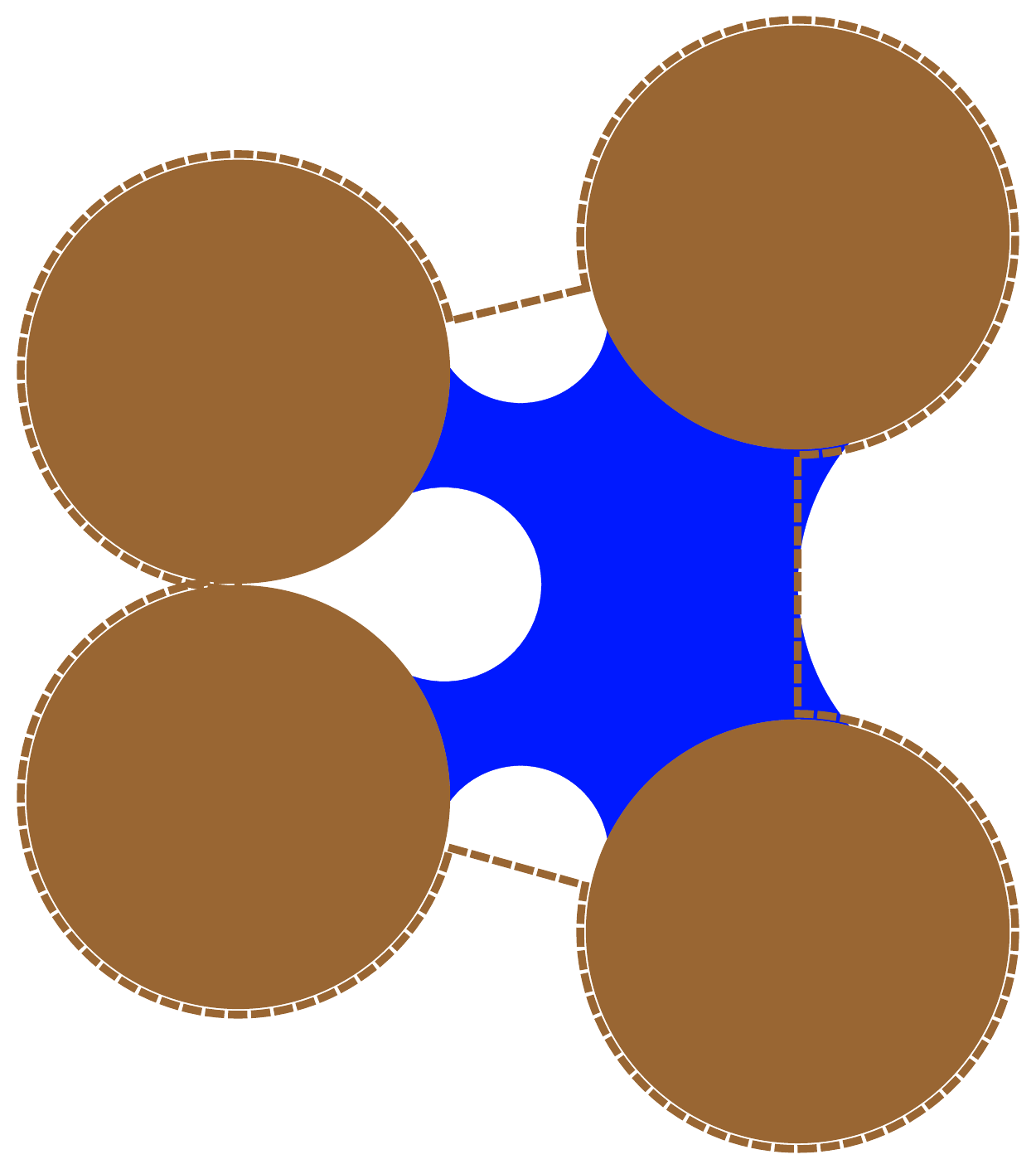}
\caption{$J_{,1}=0.471248$}
\label{fig:micro_Srmed_J4}
\end{subfigure}
\caption{Admissible distributions of the liquid phase within different RVEs, which are obtained as the current configurations of the reference shape reported in panels (a) and (e).
The micro-scale displacement of the solid beads is parametrized by the value of $J_{,1}$ via its dependence on the macro-scale second gradient of deformation. During the deformation the 
saturation ratio is kept constant: $S_r=0.667298$ (see Figure~\ref{fig:microSr2}).}
\label{fig:micro_Srmed}
\end{figure}

\begin{figure}[h]
\centering
\begin{subfigure}[b]{.22\textwidth}
\includegraphics[width=\textwidth]{fig2_1.pdf}
\caption{$J_{,1}=0$}
\label{fig:micro_Srmax_J0}
\end{subfigure}
\quad
\begin{subfigure}[b]{.22\textwidth}
\includegraphics[width=\textwidth]{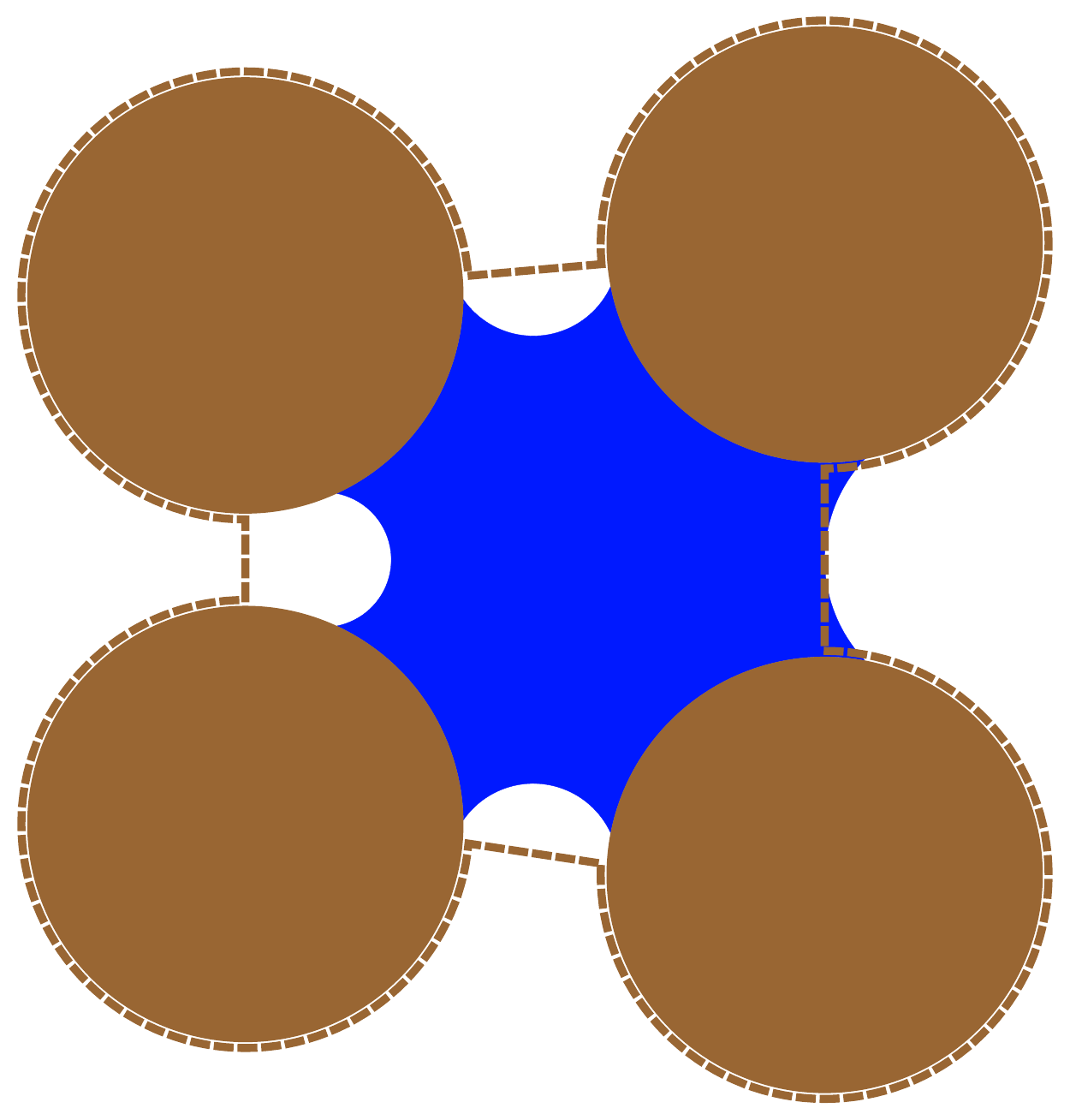}
\caption{$J_{,1}=0.173618$}
\label{fig:micro_Srmax_J2}
\end{subfigure}
\quad
\begin{subfigure}[b]{.22\textwidth}
\includegraphics[width=\textwidth]{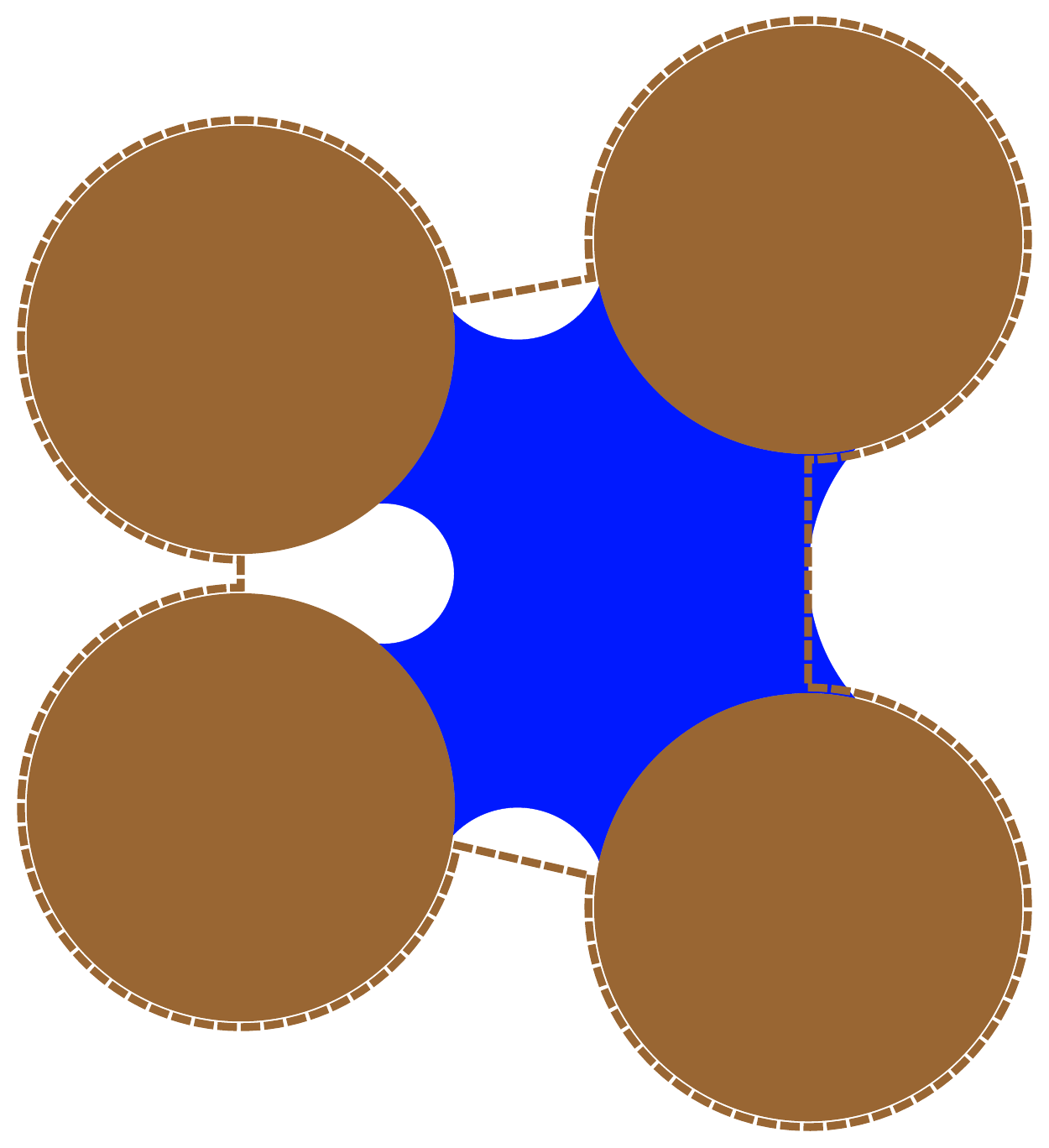}
\caption{$J_{,1}=0.347235$}
\label{fig:micro_Srmax_J3}
\end{subfigure}
\quad
\begin{subfigure}[b]{.22\textwidth}
\includegraphics[width=\textwidth]{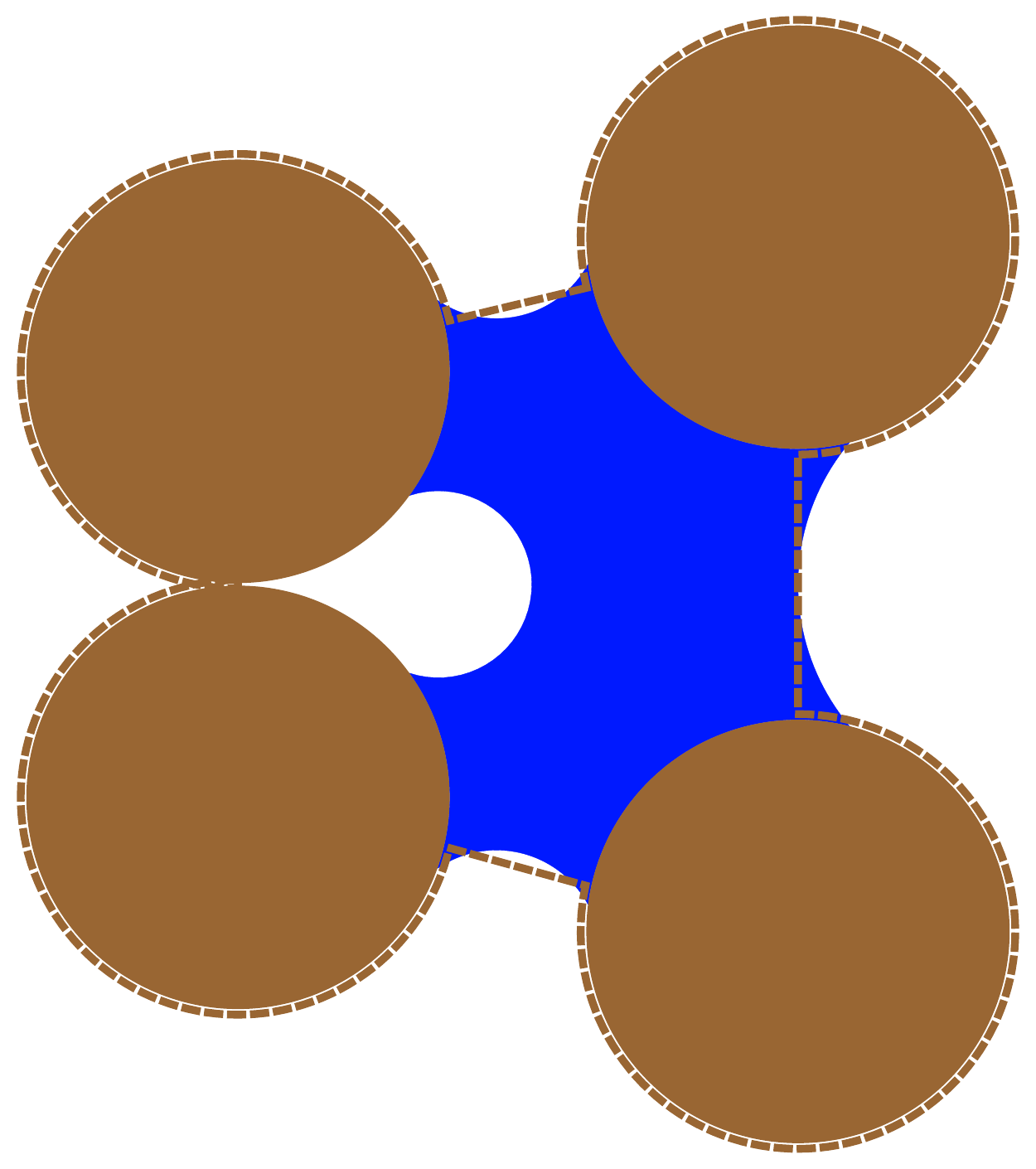}
\caption{$J_{,1}=0.471248$}
\label{fig:micro_Srmax_J4}
\end{subfigure}
\caption{Admissible distributions of the liquid phase within different RVEs, which are obtained as the current configurations of the reference shape reported in panels (a) and (e).
The micro-scale displacement of the solid beads is parametrized by the value of $J_{,1}$ via its dependence on the macro-scale second gradient of deformation. During the deformation the 
saturation ratio is kept constant: $S_r=0.882695$ (see Figure~\ref{fig:microSr3}).}
\label{fig:micro_Srmax}
\end{figure}

Let $x$, which indicates the macro-scale placement in the current configuration of the porous medium, be the centroid of the considered RVE and let $z$ be the micro-scale current position of a solid or a liquid particle within the RVE; the corresponding placements in the reference configuration of the solid are $X$ and $Z$, respectively. In Figures~\ref{fig:micro_Srmin}-\ref{fig:micro_Srmax} different RVEs are depicted, which are obtained by deforming the prototype initial configurations, reported in Figure~\ref{fig:microSr}, by the above mentioned loading path driven by the gradient of $J$. In particular the deformation is achieved by the relative motion of the beads. For the sake of simplicity a quadratic displacement field, defined over the referential domain enclosed by the centers of the beads, is assumed for describing this deformation mechanism, even if the beads suffer just mutual translations. 
Considering a local coordinate system, with the origin in the centroid of the reference shape of the RVE, the micro-scale displacement is therefore assigned as 
\begin{equation}
\label{displ_micro}
u_i^{\textrm{m}}(Z)=\chi_i^\textrm{m}(Z)-Z_i=a\, Z_i+\frac{1}{2}\mathbb{A}_{ijk}\,Z_j Z_k.
\end{equation}
A representation formula for the displacement of the liquid, with respect to its initial configuration, is not stated, but suitable assumptions on its admissible current configuration are required.
As already noticed the liquid is squeezed out of the pore due to the mutual displacement of the beads. In Appendix \ref{section_App_C} the parametrization of the domain occupied by the liquid at each loading step is illustrated in details.
The micro-scale parametrization \eqref{displ_micro} of the displacement of the grains and the hypotheses on the displacement of the liquid provide a micro-scale interpretation of the macro-scale gradients of the Jacobian determinant $J$ and the  volume liquid content $\phi S_r$, by recalling the notion of averaging operators relative to the solid and the liquid constituent, see for more details \cite{Marle82}, and \cite{Dormieuxbook2006}. As a matter of fact the macroscopic gradients of $J$ and $\phi S_r$, introduced in equation \eqref{Pc_corrected}, can be explicitly calculated, starting from the micro-scale data, as the integrals over the counter image under $\chi^\textrm{m}$ of those parts of the boundary of the RVE occupied by solid or liquid particles, if any:
\begin{equation}
\label{microIterpretJk}
J_{,k}=J F_{\alpha i,k}\, F^{-1}_{i \alpha};\quad  F_{\alpha i}(X):=\left(\dint_{\mathcal D_s^0\cap\mathcal B^0} {\hskip -1.115cm -} {\hskip .775cm} F^\textrm{m}_{\alpha i} (Z) \right),\quad  F_{\alpha i,k}(X):= \left(\dint_{\partial(\mathcal D_s^0\cap\mathcal B^0) \backslash \mathcal I^0_s} {\hskip -1.945cm -} {\hskip 1.7cm} F^\textrm{m}_{\alpha i} (Z) \,\nu_k^0 \right),
\end{equation}
\begin{equation}
\label{microIterpretSk}
(\phi S_r)_{,k} = (\phi S_r)\, G_{li,k}\, G^{-1}_{il};\quad  G_{l i}(X):=\left(\dint_{\mathcal D_{liq}^0\cap\mathcal B^0} {\hskip -1.275cm -} {\hskip .875cm} G^\textrm{m}_{l i} (Z) \right),\quad G_{l i,k}(X):= \left(\dint_{\partial(\mathcal D_{liq}^0\cap\mathcal B^0) \backslash \mathcal I^0_f} {\hskip -2.125cm -} {\hskip 1.8cm} G^\textrm{m}_{l i} (Z) \,\nu_k^0 \right).
\end{equation}
In equations \eqref{microIterpretJk}-\eqref{microIterpretSk} $\int\hskip -.33cm -$ indicates the average integral over the reference configuration of the RVE, \textit{i.e.} the integral divided by the volume of $\mathcal B^0$, moreover $F^\textrm{m}$ and $G^\textrm{m}$ are the gradients of the maps $\chi^\textrm{m}$ and $g^\textrm{m}$, respectively. At the micro-scale, $\mathcal D_s^0\cap\mathcal B^0$ ($\mathcal D_{liq}^0\cap\mathcal B^0$) is the intersection between the reference shape of the solid grains, say $\mathcal{D}^0_s$, (the pull-back of the current shape of the liquid in the reference configuration of the solid, say $\mathcal{D}^0_{liq}$) with the reference configuration of the RVE, which corresponds to that part of the reference shape occupied by the beads (liquid). On the other hand $\partial(\mathcal D_s^0\cap\mathcal B^0)\setminus\mathcal I^0_s$ ($\partial(\mathcal D_{liq}^0\cap\mathcal B^0)\setminus\mathcal I^0_f$) is the counter image of the boundary of the RVE occupied by the solid (liquid) particles, in the reference configuration. A detailed deduction of equations \eqref{microIterpretSk} is developed in Appendix \ref{section_App_D}, together with a remark on a possible way to estimate the macro-scale field $G_{li,k}$ in terms of the micro-scale gradient of $g^{\mathrm{m}}$.

The scalar quantity $a$ and the third order tensor $\mathbb{A}_{ijk}$, in equation \eqref{displ_micro}, coincide with the spherical part of $F_{\alpha i}$ and the macro-scale second gradient of deformation $F_{\alpha i,k}$, respectively. This last is supposed to be characterized by a unique non-vanishing contribution $\mathbb{A}_{221}=F_{22,1}$, which implies the gradient of $J$ to be along the abscissa $X_1$. The spherical part of the deformation gradient on the other hand is determined so as to keep constant the degree of saturation and to allow the liquid phase to form a meniscus concave towards the gaseous phase outside the RVE and tangent to its boundary. The liquid is displaced along the same direction of the gradient of the Jacobian determinant, see Figures~\ref{fig:micro_Srmin_J2}-\ref{fig:micro_Srmin_J4}, \ref{fig:micro_Srmed_J2}-\ref{fig:micro_Srmed_J4} and \ref{fig:micro_Srmax_J2}-\ref{fig:micro_Srmax_J4}.

As already noticed the contact angle between the solid and the liquid is kept constant, during the deformation process, and the interfaces between the wetting and the non-wetting phase maintain a circumferential shape. The smaller the saturation $S_r$ is the less realistic this assumption will be, see Figure~\ref{fig:microSr1}. Being required to simplify the calculations, it therefore remains valid only for values of $S_r$ which do not allow thin capillary bridges to form. 

Equations \eqref{microIterpretJk} and \eqref{microIterpretSk} motivate the conjecture assumed at the beginning of this section on the effects of the average interfacial area $a_{nw}$ on the macroscopic capillary pressure. As a matter of fact they imply that the corrective terms in equation \eqref{Pc_corrected} are not zero only if the integrals over specific parts of the boundary of the reference configuration of the RVE do not vanish. In other words the fluid trapped among the beads must attain the boundary in order for the gradient of the liquid content to not fade away. Thus in order to prove a functional correlation between gradient of $\phi S_r$ and interfacial area, all the micro-scale deformations which do not drive the liquid to the boundary of the RVE should not be taken into account.

\begin{figure}[h]
\centering
\includegraphics[width=.6\textwidth]{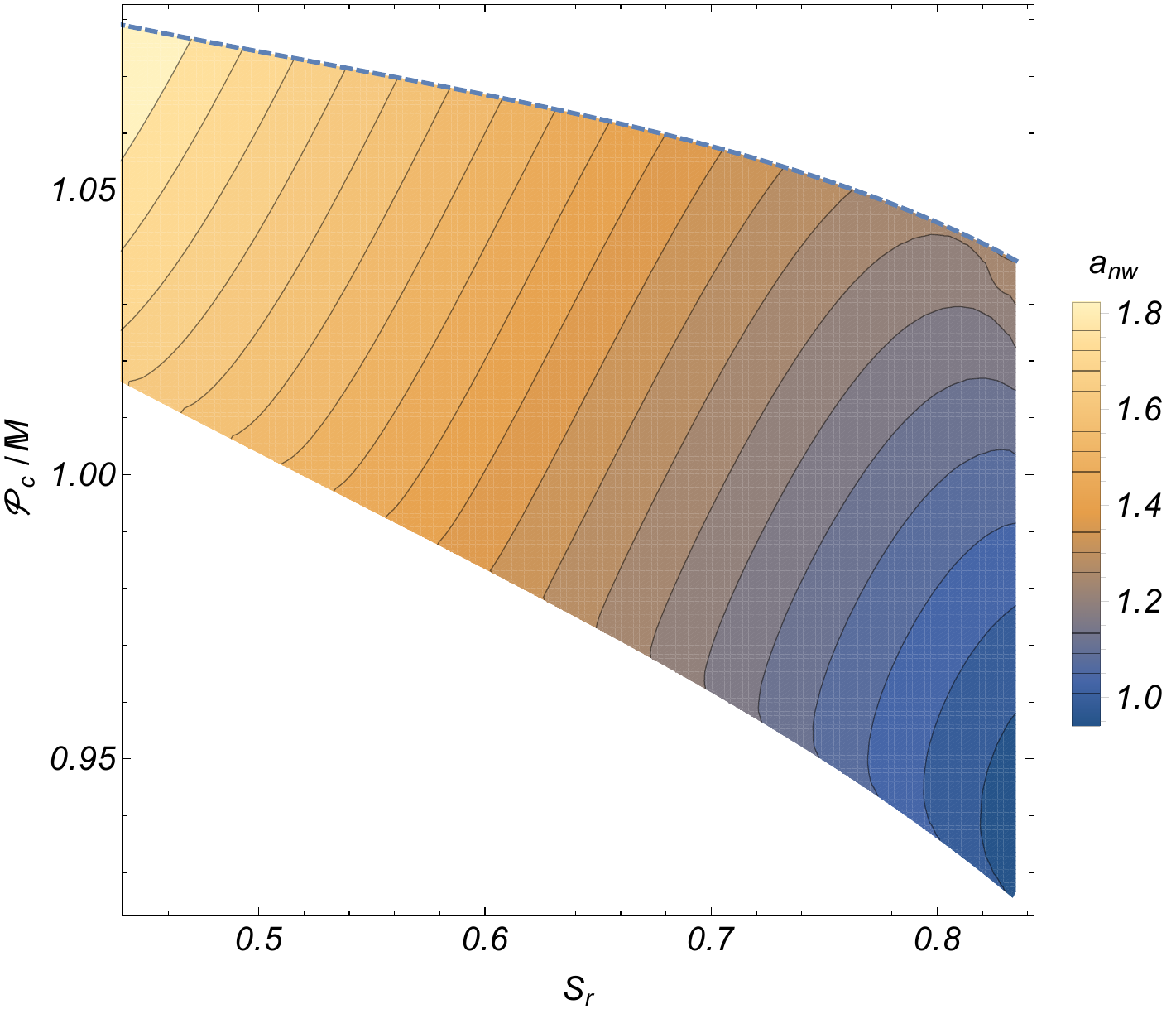}
\caption{The capillary pressure $\mathcal P_c$ is represented as a function of the saturation ratio $S_r$, within the interval delimited by the maximum and the minimum value of $S_r$ reported in Figure~\ref{fig:microSr} and parametrized by the interfacial area $a_{nw}$. The dashed line denotes to the locus of maximum values of $\mathcal P_c$ due to the gradient corrective terms.}
\label{fig:WRC}
\end{figure}

Considering the geometrical data reported in the caption of Figure~\ref{fig:microSr} and the above mentioned assumptions on the maps $\chi^{\textrm{m}}$ and $g^{\textrm{m}}$, the corrective terms of capillary pressure introduced in equation \eqref{Pc_corrected} can be explicitly calculated by means of equations \eqref{microIterpretJk}-\eqref{microIterpretSk}. At the same time the average interfacial area consistent with the envisaged liquid distribution, can be estimated for each value of $J_{,1}$. A contour plot of the constitutive law \eqref{Pc_corrected} parametrized by the obtained value of $a_{nw}$ could therefore be drawn, see Figure~\ref{fig:WRC}. 
Indeed it qualitatively resembles the one traced in \cite[Figure 11]{Joekar2010}, for the values of the saturation degree which have been considered in the present analysis. However it is worth to notice that the considered deformation and flow regimes are totally different with respect to those simulated by \cite{Joekar2010}, as they are driven by the gradient of the Jacobian $J$, at different values of $S_r$, whilst those ones of \cite{Joekar2010} are deduced simulating many scanning loops of drainage and imbibition. This is indeed an interesting point which corroborates the idea, introduced in the seminal paper of \cite{HassanizadehGray93}, of resolving the hysteresis of the capillary pressure, between drainage and imbibition, introducing a suitable micro-structural parameter, say $a_{nw}$, which tunes the value of $\mathcal P_c$ for each value of $S_r$. 

\begin{figure}[h]
\centering
\includegraphics[width=.6\textwidth]{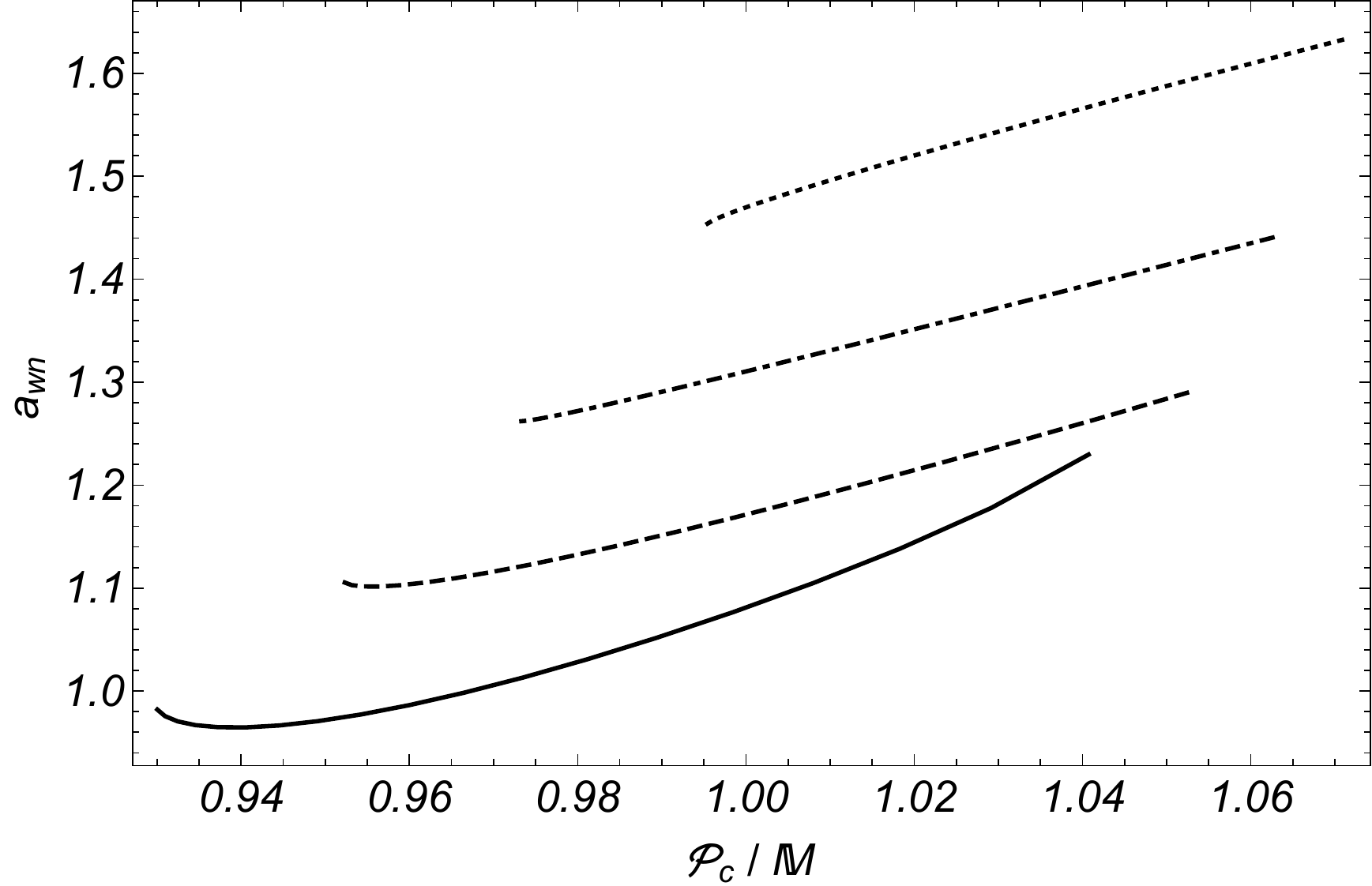}
\caption{Average interfacial area, calculated from the liquid spatial distribution within the RVE, vs capillary pressure, calculated on the basis of equation \eqref{Pc_corrected}, taking in  due account the definitions \eqref{microIterpretJk}-\eqref{microIterpretSk} of macro-scale gradients in terms of the micro-scale maps $\chi^{\textrm{m}}$ and $g^{\textrm{m}}$. The solid line corrsponds to $S_r=0.8197$, the dashed line to $S_r=0.7394$, the dot-dashed line to $S_r=0.6479$ and the dotted line to $S_r=0.5411$.}
\label{fig:aPc}
\end{figure}

\begin{figure}[h]
\centering
\includegraphics[width=.5\textwidth]{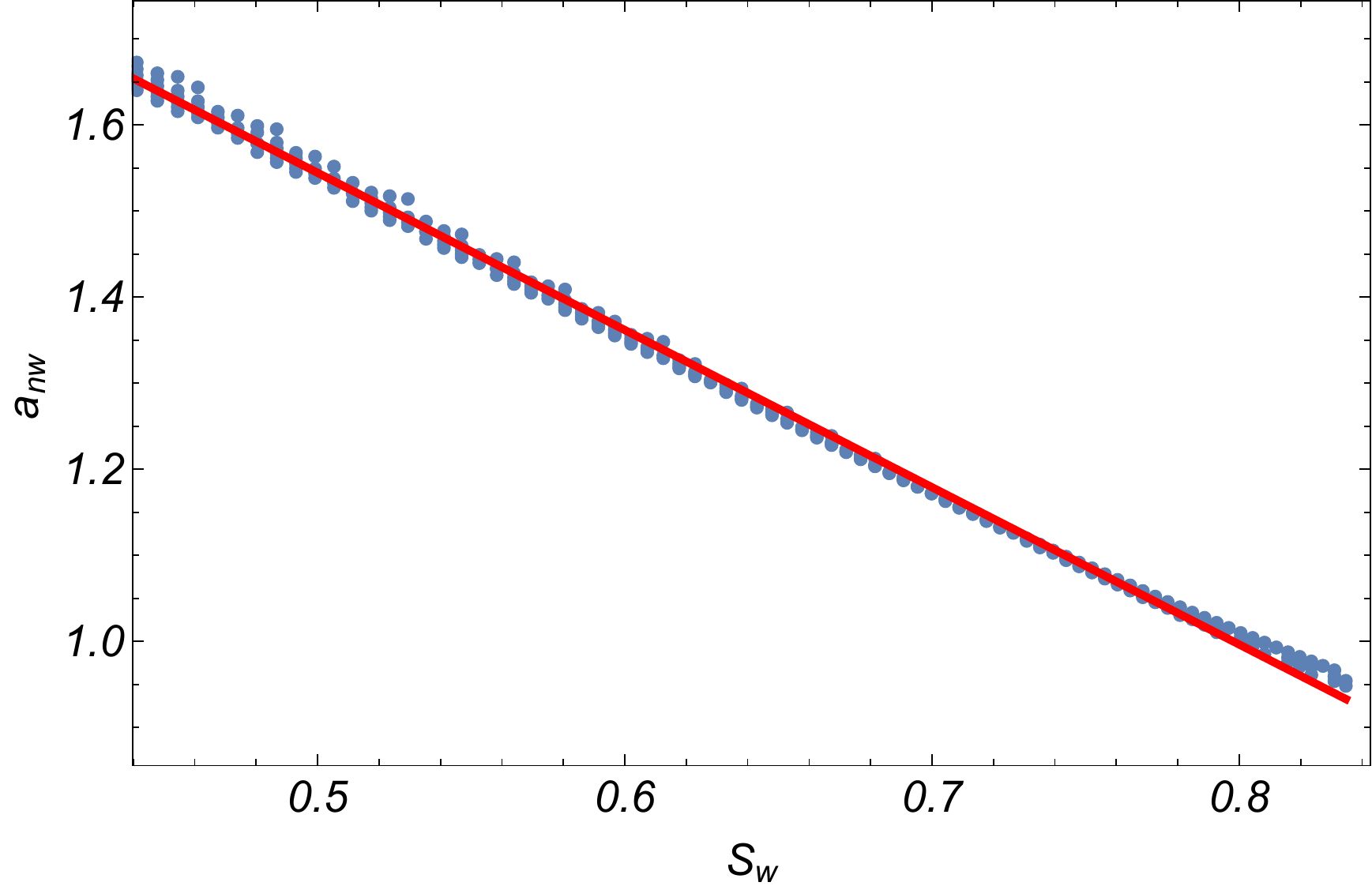}
\caption{Linear regression of data relative to the calculated admissible values of $S_r$ and $a_{nw}$, deduced from the micro-scale analysis.}
\label{fig:aSr}
\end{figure}

For the considered values of the saturation degree, the distribution of the interfacial area is plotted against the capillary pressure in Figure~\ref{fig:aPc}; again the result is consistent with the experimental evidence reported in \cite{Costanza2011}, \cite{Culligan2004} and \cite{Dalla2002}. Comparing the envisaged relation with that one deduced by \cite[Figure 10b]{Dalla2002} and \cite[Figure 7b]{Culligan2004} one can notice that consistency with the experimental data is achieved when the capillary pressure is not too large, in particular not larger than a suitable inversion point, see \cite[Figure 10]{Dalla2002}, 
in Figure~\ref{fig:aSr} the projection of the $\mathcal P_c$-$S_r$-$a_{nw}$ relation on the $S_r$-$a_{nw}$ plane is finally depicted for the considered distribution of the liquid phase within the RVE during the deformation process. Again a comparison with the experimental result of \cite{Dalla2002} and \cite{Culligan2004} confirms the validity of the obtained results for the considered range of values of $S_r$ and $\mathcal P_c$.

The considered micro-scale analysis showed an a-posteriori relation between the average interfacial area $a_{nw}$ and the gradient of the liquid content, in the presence of a non-trivial flow through the boundary of the RVE. As the average interfacial area is typically parametrized by the pore size distribution, the higher order constitutive parameter $\mathrm{C}_k$ should also depend on it so that the more the collection of pores, forming the RVE, is non-uniform, the more the micro-scale deformations of the skeleton will affect the retention properties of the porous material. The pore size polydispersity is therefore expected to highly affect the corrective term to $\mathcal P_c$ reported in equation \eqref{Pc_corrected} (the higher the polydispersity, the larger the gradient corrective term).

It is worth to underline that the considered micro-scale analysis stems from the assumption of non-negligible strain gradients, which in fact parametrize the deformation process, even if only small strain of the solid skeleton are considered. If, on the other hand, this assumption were not valid, equation \eqref{retention_&_porosity}$_1$ would abruptly simplify into the standard constitutive prescription of capillary pressure.


\section{Governing equations}
Restrictions due to thermodynamics have provided a consistent representation of the overall free energy of the porous continuum $\Psi$, as the sum of the free energy of the skeleton and the interfaces $\Psi_s$ and that of the bulk fluid $\Psi_f$, weighted by Lagrangian porosity, in terms of strain, Lagrangian porosity, saturation degree, strain gradient and gradient of liquid content. This characterization has been deduced postulating the balance of momentum for the overall porous medium, and exploiting the Clausius-Duhem inequality; the constitutive prescription of the local fluid flux, say the generalized Darcy law, has also been obtained. However no explicit deduction of the boundary conditions relative to the partial differential equations associated to the overall equilibrium \eqref{overall_eq} and the generalized Richards equation \eqref{Richards} or \eqref{Richards_rephrased} have been stated.

In this section the governing equations, and the corresponding boundary conditions, for an unsaturated porous medium, within the framework of the adopted regularized phase field model, are deduced by means of a variational approach. To do this the independent variations of $\chi_{\alpha}$, $g_k$ and $\phi$, $\forall X\in\mathcal{D}_{0}$, are taken into account. Let
\begin{equation}\label{varied_place}
\chi_{\alpha}^{\ast}\left(  X,t\right)  =\chi_{\alpha}\left(X,t\right)  +\delta\chi_{\alpha}\left(X,t\right), \quad g_j^{\ast}\left(X,t\right)  =g_j\left(X,t\right)  +\delta g_j\left(X,t\right),\quad \phi^{\ast}\left(X,t\right)  =\phi\left(X,t\right)  +\delta\phi\left(X,t\right),
\end{equation}
be the varied fields, and $\delta\chi_{\alpha}$, $\ \delta g_k$ and $\delta\phi$ the corresponding arbitrary variations. The physical meaning of the variation $\delta\chi_{\alpha}$ is well known in continuum mechanics and stands for the virtual displacement (deformation) of the solid skeleton. In the following this variation will also be indicated as $\delta u$, which corresponds to the varied displacement field $u^{\ast}$. The variation $\delta g_k$, instead, accounts for the virtual relative displacement of a fluid material particle with respect to a solid one. The variation $\delta\phi$ finally indicates the local change of Lagrangian porosity. All these variations are evaluated keeping fixed $X\in\mathcal{D}_{0}$, therefore $\delta$ commutes with the Lagrangian gradient operator and can enter the integral over $\mathcal{D}_{0}$.

Following the statements of classical mechanics, the principle of virtual working reads
\begin{equation}
\delta\mathcal{A=}-\delta\mathcal{L}^{\mathrm{ext}}-\delta\mathcal{L}^{\mathrm{diss}}
\label{VirtualWorks}
\end{equation}
where $\delta\mathcal{A}$\ represents the Lagrangian variation of the action functional defined, neglecting inertia effects, as
\begin{equation}
\mathcal{A}:=-\dint_{\mathcal{D}_{0}}\left(\Psi_s +m_f \psi_f\right), \label{Action}
\end{equation}
while $\delta\mathcal{L}^{\mathrm{diss}}$ and $\delta\mathcal{L}^{\mathrm{ext}}$ are the virtual working due to dissipative forces and the virtual working of external forces, respectively. The virtual working of dissipative forces is
\begin{equation}\label{virtual diss}
\delta\mathcal{L}^{\mathrm{diss}}= \dint_{\mathcal D_0} \left(\mathrm{det} G\right) (1/\rho_L) A_{k l} M_l  \left(G^{-1} \right)_{k j} \delta g_{j},
\end{equation}
see \cite{Sciarra08}, whilst the virtual working of external forces can be stated in a similar form as that in equation \eqref{Wext}. The virtual velocities $v_{\alpha}^s$ and $v_{\alpha}^f$ are replaced by the virtual displacements $\delta\chi_{\alpha}$ and $\delta\chi_{\alpha}^f=\delta\chi_{\alpha}-F_{\alpha k} (G^{-1})_{kj} \delta g_j$, respectively:
\begin{equation}\label{virtual ext}
\begin{array}{ll}
\delta\mathcal{L}^{\mathrm{ext}} &= \dint_{\partial\mathcal D_0}\left\lbrace t_{\alpha}^0\delta\chi_{\alpha} + \tau_{\alpha}^0 \delta\chi_{\alpha,k}\, n_k- \left[t_k^{0 f} \left(G^{-1} \right)_{k j} \delta g_j +\tau^{0 f}_k \left(G^{-1} \right)_{k j} \delta g_{j,l}\, n_l\right]\right\rbrace\medskip\\
 &+\dint_{\mathcal E_0}\left[f_{\alpha}^0\delta\chi_{\alpha} - f_k^{0 f} \left(G^{-1} \right)_{k j} \delta g_j \right] +\dint_{\mathcal D_0}\left[b_{\alpha}^0\delta\chi_{\alpha} - b_k^{0 f} \left(G^{-1} \right)_{k j} \delta g_j \right],
\end{array}
\end{equation}
where the referential bulk forces ($b^0_{\alpha}$ and $b^{0 f}_{k}$), the surface tractions ($t^0_{\alpha}$ and $t^{0 f}_{k}$), the surface double forces ($\tau^0_{\alpha}$ and $\tau^{0 f}_{k}$) and the line tractions ($f^0_{\alpha}$ and $f^{0 f}_{k}$) are defined in terms of the corresponding forces acting on the current shape of the porous continuum as follows, for the overall porous medium
\begin{equation}
\label{ref_solid_forces}
\begin{array}{ll}
t_{\alpha}^0   =&\!\!\! J^s t_{\alpha}-\left[J^s \tau_{\alpha} Q_{Bl} \left(F^{-1} \right)_{l\beta} \left(F^{-1} \right)_{k\beta} n_k\right]_{,B},\quad b_{\alpha}^0=J b_{\alpha}, \medskip\\
\tau_{\alpha}^0=&\!\!\! J^s \tau_{\alpha} \left(F^{-1} \right)_{l\beta} n_l \left(F^{-1} \right)_{k\beta} n_k,\medskip\\
f_{\alpha}^0   =&\!\!\! J^l f_{\alpha} + \left[\!\!\left[J^s\tau_{\alpha} Q_{Bl } \left(F^{-1} \right)_{l\beta} \left(F^{-1} \right)_{k\beta} n_k\, \nu_{B}\right]\!\!\right];
\end{array}
\end{equation}
and for the fluid
\begin{equation}
\label{ref_fluid_forces}
\begin{array}{ll}
t_k^{0 f}   =& \!\!\! J^s \left[F_{\alpha k} t_{\alpha}^f + \tau_{\alpha}^f \left(F_{\alpha r} \left(G^{-1}\right)_{rq} \right)_{,l}\left(F^{-1} \right)_{l\beta}\left(F^{-1} \right)_{p\beta} n_p G_{qk}\right]+\medskip\\
             &  - \left[J^s F_{\alpha r} \tau_{\alpha}^f \left(G^{-1}\right)_{rq} Q_{Bl} \left(F^{-1} \right)_{l\beta} \left(F^{-1} \right)_{p\beta} n_p \right]_{,B} G_{qk}, \quad b_k^{0 f}=J F_{\alpha k} b_{\alpha}^f,\medskip\\
\tau_k^{0 f}=& \!\!\! J^s F_{\alpha k} \tau_{\alpha}^f \left(F^{-1} \right)_{l\beta} n_l \left(F^{-1} \right)_{p\beta} n_p,\medskip\\
f_k^{0 f}   =& \!\!\! J^l F_{\alpha k} f_{\alpha}^f +\left[\!\!\left[ J^s F_{\alpha k} \tau_k^f  Q_{lB} \left(F^{-1} \right)_{l\beta} \left(F^{-1} \right)_{p\beta} n_p\, \nu_B\right]\!\!\right].
\end{array}
\end{equation}
In equations \eqref{ref_solid_forces} and \eqref{ref_fluid_forces} $J^s$ and $J^l$ indicate the determinant of the surface and line restrictions of $F$ over the boundary of $\mathcal D_0$ and its edges.
According to equations \eqref{Action} and \eqref{statefunct} the variation of the free energy is
\begin{equation}\label{varied Action}
\begin{array}{ll}
\delta\mathcal{A}=-\dint_{\mathcal D_0}\!\!\!\!\!&\left(\dfrac{\partial \Psi_s}{\partial E_{i j}}\delta E_{i j} + \dfrac{\partial \Psi_s}{\partial E_{i j,k}}\delta E_{i j,k}+\dfrac{\partial \Psi_s}{\partial \phi}\delta \phi + \dfrac{\partial \Psi_s}{\partial S_r}\delta S_r \right.\medskip\\
&\left.+ \dfrac{\partial \Psi_s}{\partial \left(\phi S_r \right)_{,l}}\delta \left(\phi S_r \right)_{,l}+\,\rho_L\psi_f \delta \left(\phi S_r \right)+\phi S_r\dfrac{\partial\psi_f}{\partial(1/\rho_f)}\delta\left(\dfrac{1}{S_r} \right)\right),
\end{array}
\end{equation}
where the following identities hold true:
\begin{equation}\label{varied strain}
\begin{array}{ll}
 \delta E_{i j}&=\dfrac{1}{2}\left(\delta F_{\alpha i} F_{\alpha j}+F_{\alpha i}\delta F_{\alpha j}\right),\medskip \\
 \delta E_{i j, k}&=\dfrac{1}{2}\left(\delta F_{\alpha i, k} F_{\alpha j}+\delta F_{\alpha j, k} F_{\alpha i}\right)+\dfrac{1}{2}\left(\delta F_{\alpha i} F_{\alpha j, k}+\delta F_{\alpha j} F_{\alpha i, k}\right),\medskip\\
 \delta \left(\phi S_r \right) & = \mathrm{det} G \left(G^{-1}\right)_{k j} \delta g_{j,k},\quad \delta S_r = \dfrac{1}{\phi} \delta\left(\phi S_r\right) - \dfrac{S_r}{\phi}\delta\phi.
\end{array}
\end{equation}

The local governing equations are deduced by requiring the validity of equation \eqref{VirtualWorks} for every kinematically admissible variation $\delta\chi_{\alpha}$, $\delta g_{j}$ and $\delta\phi$. To this aim, the distinction between essential ($e$) and natural ($n$) boundary conditions needs to be generalized, following the scheme proposed in \cite{fdigssv}. 
On every part $\left(\mathcal L_{x}^{y}\right)_c \subset\partial\mathcal D_0$, with $x=e,n$, $y=e,n$, and for each constituent $c=\{s,f\}$, one may impose four different kind of boundary conditions, see Table~\ref{tab:bcs_solid} and Table~\ref{tab:bcs_fluid}.
\begin{table}
\begin{center}
\begin{tabular}{lccll}\hline
$\left(\mathcal L_e^e\right)_s$ & $u_{\alpha}=\overline{u}_{\alpha},$ & $u_{\alpha,k} n_{k}=\overline{w}_{\alpha}$ & $u$-essential &$D_n u$-essential \\ \hline
$\left(\mathcal L_e^n\right)_s$ &  $u_{\alpha}=\overline{u}_{\alpha},$ & $\tau_{\alpha}=\overline{\tau}_{\alpha}$ & $u$-essential & $\tau$-natural  \\ \hline
$\left(\mathcal L_n^e\right)_s$ & $t_{\alpha}=\overline{t}_{\alpha},$ & $u_{\alpha,k} n_{k}=\overline{w}_{\alpha}$ & $t$-natural &$D_n u$-essential \\ \hline
$\left(\mathcal L_n^n\right)_s$ & $t_{\alpha}=\overline{t}_{\alpha},$ & $\tau_{\alpha}=\overline{\tau}_{\alpha}$ & $t$-natural &$\tau$-natural \\ \hline
\end{tabular}
\captionof{table}{Generalized essential and natural boundary conditions relative to the porous skeleton.}
\label{tab:bcs_solid}
\end{center}
\end{table}
\begin{table}
\begin{center}
\begin{tabular}{lccll}\hline
$\left(\mathcal L_e^e\right)_f$ & $ g_j=\overline{g}_j,$ & $g_{j,k} n_k=\overline{\omega}_j$ & $g$-essential &$D_n g$-essential \\ \hline
$\left(\mathcal L_e^n\right)_f$ & $ g_j=\overline{g}_{j},$ & $\tau_{\alpha}^f=\overline{\tau}^f_{\alpha}$ & $g$-essential & $\tau$-natural  \\ \hline
$\left(\mathcal L_n^e\right)_f$ & $t^f_{\alpha}=\overline{t}^f_{\alpha},$ & $g_{j,k} n_k=\overline{\omega}_{j} $ & $t$-natural &$D_n g$-essential \\ \hline
$\left(\mathcal L_n^n\right)_f$ & $t^f_{\alpha}=\overline{t}^f_{\alpha},$ & $\tau_{\alpha}^f=\overline{\tau}^f_{\alpha}$ & $t$-natural &$\tau$-natural \\ \hline
\end{tabular}
\captionof{table}{Generalized essential and natural boundary conditions relative to the non-uniform fluid.}
\label{tab:bcs_fluid}
\end{center}
\end{table}
Here the displacement $u$ rather than the placement $\chi$ of the solid particles is used and $(D_n u)_{\alpha}:=u_{\alpha,k} \, n_{k}$ and $(D_n g)_j:=g_{j,k}\, n_k$ indicate the normal derivatives of $u$ and $g$ along the outward normal to $\partial\mathcal D_0$. Subscripts distinguish $(u,g)$-essential from $t$-natural boundary conditions, whilst superscripts distinguish $D_n (u,g)$-essential from $\tau$-natural boundary conditions; moreover on every part of the $h$-th edge $\mathcal E_h$ the standard distinction between essential and natural conditions holds true. The varied maps $u^*$ and $g^*$ are said to be kinematically admissible if they satisfy the same essential boundary conditions as $u$ and $g$; as a consequence the variations $\delta u$ and $\delta g$ together with their normal derivative will vanish on $\left(\partial\mathcal D_0^c\right)^{\star}_{e}:=\left(\mathcal L^e_e\right)_c\cup\left(\mathcal L^n_e\right)_c$ and $\left(\partial\mathcal D_0^c\right)^e_{\star}:=\left(\mathcal L^e_e\right)_c\cup\left(\mathcal L^e_n\right)_c$, $c=\{s,f \}$, respectively.
Replacing equations \eqref{varied strain} into \eqref{varied Action}, and using standard localization arguments provides the following bulk equations in $\mathcal D_0$
\begin{equation}\label{balance}
\begin{array}{l}
\left[ F_{{\alpha}i }\,\left(S_{i j }-P_{i j k ,k }\right)\right]_{,j }+ b^0_{\alpha} = 0,\medskip\\
\dfrac{\partial\Psi_s}{\partial\phi} - S_r \dfrac{\partial U}{\partial S_r} - \mathcal P=0,\medskip\\
\left[\dfrac{\partial (\Psi_f+U)}{\partial S_r} - \left(\dfrac{\partial \Psi_{s}}{\partial (\phi S_r)_{,l}} \right)_{,l} \right]_{,k}  = -\dfrac{1}{\rho_L} A_{kl} M_l + \dfrac{b_k^{0f}}{\phi S_r},
\end{array}
\end{equation}
where the notion of specific free energy of the fluid $\Psi_f$ (per unit volume) has been used together with that of capillary energy $U$. Equation \eqref{balance}$_1$ states the equilibrium of the porous skeleton, equation \eqref{balance}$_2$ the constitutive prescription of the thermodynamic pressure, which is the same as equation \eqref{energy_s_const}$_1$, and equation \eqref{balance}$_3$ the generalized Darcy law \eqref{Darcy_gen_energy}. At the same time the corresponding boundary conditions are provided, in particular traction boundary conditions are prescribed on $\left(\partial\mathcal D^c_0\right)^{\star}_n$, $c=\{s,f \}$
\begin{equation}
\begin{array}{l}
F_{{\alpha}i }\,\left(S_{i j }-P_{i j k ,k }\right)n_{j }-\left(Q_{B j} F_{\alpha i} P_{i j k } n_{k } \right)_{,B} = t^0_{\alpha},\medskip\\ 
-\left[\dfrac{\partial (\Psi_f+U)}{\partial S_r} - \left(\dfrac{\partial \Psi_{s}}{\partial (\phi S_r)_{,l}} \right)_{,l} \right] n_k + \left[ \phi S_r\,  Q_{B j} \left(G^{-1} \right)_{j\ell} \left(\dfrac{\partial \Psi_s}{\partial (\phi S_r)_{,l}} n_l\right)\right]_{,B}  \dfrac{G_{\ell k}}{\phi S_r}= \dfrac{t^{0 f}_k}{\phi S_r}
\end{array}
\end{equation}
double force boundary conditions are prescribed on $\left(\partial\mathcal D^c_0\right)^n_{\star}$, $c=\{s,f \}$
\begin{equation}
F_{\alpha i} P_{i j k} n_{k } n_{j} = J_S \tau_\alpha,\quad  -\phi S_r\left(\dfrac{\partial \Psi_s}{\partial (\phi S_r)_{,l}} n_l \right) n_k = \tau^{0 f}_k,
\end{equation}
and edge forces are prescribed on the $h$-th edge $\left(\mathcal E^c_{0h}\right)^n$ of $\mathcal D_0$,
\begin{equation}
\left[\!\left[ Q_{B j} F_{\alpha i} P_{i j k } n_{k } \nu_B\right]\!\right] =J_L f_{\alpha},\quad -\left[\!\!\left[\phi S_r Q_{B k} \left(\dfrac{\partial \Psi_s}{\partial (\phi S_r)_{,l}} n_l\right) \nu_B \right]\!\!\right] = f^{0 f}_k.
\end{equation}
As already mentioned these natural boundary conditions must be used together with the essential boundary conditions
\begin{equation}
 \left\lbrace\!\!\! 
 \begin{array}{ll}
 u_{\alpha}=\overline{u}_{\alpha}\quad & \textrm{on}\,\, \left(\partial\mathcal D^s_0\right)^{\star}_{e},\medskip\\
 g_{\ell}\,\,=\overline{g}_{\ell}\quad & \textrm{on}\,\, (\partial\mathcal D^f_0)^{\star}_{e},
 \end{array}
\right.\,
\left\lbrace\!\!\!
\begin{array}{ll}
 u_{\alpha,k}\, n_{k}=\overline{w}_{\alpha},\quad & \textrm{on}\,\, \left(\partial\mathcal D^s_0\right)^e_{\star},\medskip\\
 g_{\ell,k}\, n_k\,\,=\overline{g}_{\ell},\quad & \textrm{on}\,\, (\partial\mathcal D^f_0)^e_{\star},
\end{array}
\right.\,
\left\lbrace
\begin{array}{ll}\!\!\!
u_{\alpha}=\overline{u}_{\alpha},\quad & \textrm{on}\,\, (\mathcal E^s_{0h})^e,\medskip\\ 
g_{\ell}=\overline{g}_{\ell},\quad & \textrm{on}\,\, (\mathcal E^f_{0h})^e.
\end{array}
\right.
\end{equation}
We underline that by definition $Q_{Bj}:=\partial \hat X_B/\partial X_j$, $\hat X_B$ is a system of coordinates which locally parametrizes $\partial\mathcal D_0$, moreover $\nu_B$ indicates the component of the Darboux tangent-normal vector to each edge of $\partial\mathcal D_0$.

\section{Conclusions}
In this paper gradient theory of poromechanics endowed with phase field modeling have been used to describe partial saturation. The spatial distribution of saturation and strain is spatially regularized, in the presence of phase coexistence, within the non-uniform fluid, and in the neighbors of possible heterogeneities in the porous skeleton. To do this the free energy of the overall porous medium has been regarded as a function not only of strain and saturation but also on their (Lagrangian) gradients. 

As usual in gradient theories, the governing partial differential equations relative, in this case, to the solid skeleton and the non-uniform fluid are deduced using integration by parts twice, which implies the equations to be, in the general case, of the fourth order. In particular an enhanced version of classical Richards' equation has been deduced from generalized Darcy's law, which in the considered regime of partial saturation does not depend on the capillary pressure, but on the so-called generalized effective chemical potential. 

Moreover a novel constitutive characterization of the capillary pressure is established and a micro-scale interpretation is provided, which is consistent with that originally formulated by \cite{Hassanizadeh1990,HassanizadehGray93}. A comparison with experimental results also confirms that the contribution to capillary pressure due to saturation gradient allows for recovering similar effects as those captured by the average interfacial area.

In particular the model proposed in this paper aims at bridging the regularizing effects provided by the gradient of strain and the gradient of saturation in order to capture on one hand the response of the porous skeleton, due to micro-scale multi-phase flow, and on the other hand the heterogeneous and possibly anisotropic flow of the saturating fluids due to localized strains, induced by micro-structural remodeling processes. Further developments will be carried out in order to account for damaging and plastic strain.

\section*{Acknoledgements} 
\noindent The research has been partially supported by INdAM (Italian National Institute for Advanced Mathematics), within the framework of the \lq\lq Young Researchers GNFM (National Group of Mathematical Physics)\rq\rq\, program. The author gratefully acknowledges this institution.

\noindent The author is indebted to the reviewers for their constructive comments on the manuscript.

\begin{appendices}
\numberwithin{equation}{section}

\section{}\label{section_App_A}
Equation \eqref{CD_complete} is rephrased identifying the coefficients of the time derivative of the Lagrangian volumetric liquid content, $J \theta =\phi S_r$, as well as of the Lagrangian filtration vector and its spatial derivatives, $M_k$ and $M_{l,k}$ respectively. To do this the following chain of equalities are used:
\begin{equation}
\label{App_A_identities_1}
\begin{array}{rl}
\left[\left(J p^f+\gamma_{l,l}\right) \dfrac{M_k}{m_f} \right]_{,k} = & \dfrac{1}{\rho_L} \left(\dfrac{p^f}{n S_r} +\dfrac{\gamma_{l,l}}{\phi S_r}\right)_{,k} M_k - \left(\dfrac{p^f}{n S_r} +\dfrac{\gamma_{l,l}}{\phi S_r}\right) \dfrac{d(\phi S_r)}{dt},\medskip\\
\left[\dfrac{\gamma_k}{J} \left(J \dfrac{M_l}{m_f} \right)_{,l} \right]_{,k} = & \left[\dfrac{1}{\rho_L}\dfrac{\gamma_k}{J} \left(\dfrac{1}{n S_r} \right)_{,l} M_l-\left(\dfrac{\gamma_k}{\phi S_r}\right)  \dfrac{d(\phi S_r)}{dt}\right]_{,k}=\medskip\\
=&\!\!\!\!\!-\left[\left(\dfrac{\gamma_k}{\phi S_r} \right)_{,k}+\dfrac{1}{\textrm{tr}I}\dfrac{\gamma_k}{J}\left(\dfrac{1}{n S_r} \right)_{,k} \right]\dfrac{d(\phi S_r)}{dt}-\dfrac{\gamma_k}{\phi S_r} \dfrac{d(\phi S_r)_{,k}}{dt}\,+\medskip\\
& \!\!\!\!\!\!\!\!\!\!\!+\dfrac{1}{\rho_L} \left[\dfrac{\gamma_k}{J} \left(\dfrac{1}{n S_r}\right)_{,l} \overline M_{l,k} +\left(\dfrac{\gamma_k}{J} \left(\dfrac{1}{n S_r}\right)_{,l}\right)_{,k} M_l \right],
\end{array}
\end{equation}
and\begin{equation}
\label{App_A_identities_2}
\left[J \left(F^{-1} \right)_{k \beta} \Sigma'^f_{\alpha \beta} F_{\alpha q} \dfrac{M_q}{m_f}\right]_{,k}=\dfrac{1}{\rho_L}\left\lbrace \left[\dfrac{1}{n S_w}\left(F^{-1} \right)_{k \beta} \Sigma'^f_{\alpha \beta} F_{\alpha q}\right]_{,k} M_q+\dfrac{1}{n S_w}\left(F^{-1} \right)_{k \beta} \Sigma'^f_{\alpha \beta} F_{\alpha q} \overline{M}_{q,k}\right\rbrace,
\end{equation}
together with the fluid mass conservation \eqref{fluid mass Coussy}. In equations \eqref{App_A_identities_1}-\eqref{App_A_identities_2} $\overline{M}_{l,k}:=M_{l,k}-1/(\textrm{tr} I) M_{j,j} \delta_{lk}$ is the deviatoric component of the spatial gradient of the filtration vector $M_l$, whilst $\Sigma'^f_{\alpha \beta}$ is the deviatoric component of the fluid stress tensor, see equation \eqref{fluid_const}.

Replacing these identities into equation \eqref{CD_complete} and keeping into account equation \eqref{Helm_skel}, yields
\begin{equation}
\label{CD_App}
\begin{array}{l}
S_{ij} \dot E_{ij} +P_{ijk} \dot E_{ij,k}+\left[  \left(\dfrac{p^f}{n S_r} +\dfrac{\gamma_{l,l}}{\phi S_r}\right) - \left(\dfrac{\gamma_k}{\phi S_r} \right)_{,k}-\dfrac{1}{\textrm{tr}I}\dfrac{\gamma_k}{J}\left(\dfrac{1}{n S_r} \right)_{,k}+\dfrac{1}{n S_r}\left(\kappa_f+\dfrac{1}{2}\mathcal E_{sf} \right)\right]\dfrac{d(\phi S_r)}{dt}+\medskip\\
-\dfrac{\phi}{S_r} \mathcal P\, \dfrac{d S_r}{dt}-\dfrac{\gamma_k}{\phi S_r}\dfrac{d(\phi S_r)_{,k}}{dt}+\dfrac{1}{\rho_L}\left[\dfrac{\gamma_k}{J} \left(\dfrac{1}{n S_r} \right)_{,l}+\dfrac{1}{n S_r}\left(F^{-1} \right)_{k \beta}\Sigma'^f_{\alpha \beta} F_{\alpha l}\right]\overline{M}_{l,k}-\left(e_{f,k}-T s_{f,k}\right)M_k+\medskip\\
-\dfrac{1}{\rho_L}\left[\dfrac{p^f}{n S_r}+ \dfrac{\gamma_{l,l}}{\phi S_r}+\dfrac{1}{n S_r}\left(\kappa_f+\dfrac{1}{2}\mathcal E_{sf} \right)\right]_{,k} M_k+\dfrac{1}{\rho_L}\left[\dfrac{\gamma_k}{J} \left(\dfrac{1}{n S_r} \right)_{,l} + \dfrac{1}{n S_r}\left(F^{-1} \right)_{k \beta}\Sigma'^f_{\alpha \beta} F_{\alpha l}\right]_{,k} M_l\medskip\\
+\dfrac{1}{\rho_L}\dfrac{b_k^{0f}}{\phi S_r}\, M_k-\dfrac{\mathfrak q_k}{T} T_{,k}-\mathbb S_s \dfrac{dT}{dt}-\dfrac{d\Psi_s}{dt}\geq 0.
\end{array}
\end{equation}
Considering the relation between the fluid stress and the pull-back of the fluid hyper-stress, deduced from equation \eqref{fluid_const}, say:
\begin{equation}
\label{fluid_Lagr_const}
p^f = n\, \mathcal P-\kappa_f -\dfrac{\gamma_k}{J}\left(1+\dfrac{1}{\mathrm{tr} I} \right)\dfrac{1}{n S_r}(n S_r)_{,k}, \quad \Sigma'^f_{\alpha \beta}=\dfrac{1}{n S_r} F_{\beta q}\left[\dfrac{\gamma_q}{J} (n S_r)_{,p}-\dfrac{1}{\textrm{tr}I}\dfrac{\gamma_j}{J} (n S_r)_{,j} \delta_{qp}\right]\left(F^{-1}\right)_{p \alpha}
\end{equation}
the coefficients of $d\phi/dt$, $d S_r/dt$ appearing in the Clausius-Duhem inequality \eqref{CD_App} can be reduced to the following form
\begin{equation}
\label{C_phi/C_Sr}
\begin{array}{rl}
\mathtt{C}^{\phi}:= & \dfrac{p_f}{n}+\dfrac{\gamma_{k,k}}{\phi}-\left(\dfrac{\gamma_k}{\phi S_r} \right)_{,k} S_r -\dfrac{1}{\textrm{tr}I}\dfrac{\gamma_k}{J}\left(\dfrac{1}{n S_r} \right)_{,k} S_r+\dfrac{1}{n}\left(\kappa_f +\dfrac{1}{2} \mathcal E_{sf}\right)\medskip\\
= & \dfrac{1}{n} \left(p^f+\kappa_f +\dfrac{1}{2} \mathcal E_{sf}\right) - \gamma_k \left(1+ \dfrac{1}{\textrm{tr}I}\right) \left(\dfrac{1}{\phi S_r} \right)_{,k} S_r-\dfrac{1}{\textrm{tr}I}\, \dfrac{\gamma_k}{\phi}\, \dfrac{J_{,k}}{J}\medskip\\
= & \mathcal P +\dfrac{1}{2 n}\, \mathcal E_{sf}+\dfrac{\gamma_k}{\phi}\, \dfrac{J_{,k}}{J},\medskip\\
\mathtt{C}^{S_r}:= & \dfrac{\phi}{S_r}\left[\dfrac{p_f}{n}-\mathcal P+\dfrac{\gamma_{k,k}}{\phi}-\left(\dfrac{\gamma_k}{\phi S_r} \right)_{,k} S_r -\dfrac{1}{\textrm{tr}I}\dfrac{\gamma_k}{J}\left(\dfrac{1}{n S_r} \right)_{,k} S_r+\dfrac{1}{n}\left(\kappa_f +\dfrac{1}{2} \mathcal E_{sf}\right) \right]\medskip\\
= & \dfrac{\phi}{S_r}\left(\mathtt{C}^{\phi} -\mathcal P\right)= \dfrac{\phi}{S_r} \left(\dfrac{1}{2 n}\, \mathcal E_{sf}+\dfrac{\gamma_k}{\phi}\, \dfrac{J_{,k}}{J} \right).
\end{array}
\end{equation}
Analogously equation \eqref{fluid_Lagr_const} implies the deviatoric component of the coefficient of $\overline{M}_{l,k}$ to vanish and the coefficient of $M_k$ to reduce:
\begin{equation}
\label{C_dissip}
\begin{array}{rl}
\mathtt{C}_{l k}^{\nabla M}:= &\dfrac{1}{\rho_L}\left[\dfrac{1}{n S_r}\left(F^{-1} \right)_{k \beta}\Sigma'^f_{\alpha \beta} F_{\alpha l}-\left(\dfrac{1}{n S_r}\right)^2\left(\dfrac{\gamma_k}{J} (n S_r)_{,l}-\dfrac{1}{\textrm{tr}I}\dfrac{\gamma_j}{J}(n S_r)_{,j}\right)\delta_{kl}\right]=0,\medskip\\
\mathtt{C}_k^M := &\!\!\!\! -\dfrac{1}{\rho_L}\left[ \dfrac{\mathcal P}{S_r} -\dfrac{\gamma_l}{J}\left(1+\dfrac{1}{\mathrm{tr} I} \right)\left(\dfrac{1}{n S_r}\right)^2(n S_r)_{,l} \right]_{,k}-\dfrac{1}{\rho_L}\left(\dfrac{\gamma_{l,l}}{\phi S_r} \right)_{,k}+\dfrac{1}{\rho_L}\left[\dfrac{\gamma_l}{J} \left(\dfrac{1}{n S_r} \right)_{,k}\right]_{,l}+\dfrac{\mathcal P}{\rho_L}\left(\dfrac{1}{S_r} \right)_{,k}\medskip\\
 & \!\!\!\! + \dfrac{1}{\rho_L}\left[\left(\dfrac{1}{n S_r}\right)^2 \left(\dfrac{\gamma_l}{J} (n S_r)_{,k} -\dfrac{1}{\textrm{tr}I}\dfrac{\gamma_j}{J} (n S_r)_{,j} \, \delta_{l k}\right) \right]_{,l}-\dfrac{1}{\rho_L} \left(\dfrac{1}{2\, n S_r} \mathcal E_{sf}\right)_{,k}+\dfrac{1}{\rho_L} \dfrac{b_k^{0f}}{\phi S_r}\medskip\\
= &\!\!\!\! -\dfrac{1}{\rho_L}\left\lbrace\dfrac{1}{S_r}\mathcal P_{,k} +\left[\left(\dfrac{\gamma_l}{\phi S_r}\right)_{,l} +\dfrac{1}{S_r}\left(\dfrac{1}{2 n} \mathcal E_{sf} +\dfrac{\gamma_{l}}{\phi} \dfrac{J_{,l}}{J}\right)\right]_{,k}\right\rbrace+\dfrac{1}{\rho_L} \dfrac{b_k^{0f}}{\phi S_r}.
\end{array}
\end{equation}
As a consequence one gets equation \eqref{CD_upgraded}:
\begin{equation}
\begin{array}{l}
S_{ij} \dot E_{ij}+P_{ijk} \dot E_{ij,k}+\mathtt{C}^{\phi} \dfrac{d\phi}{dt}+\mathtt{C}^{S_r}\dfrac{d S_r}{dt}-\left(\dfrac{\gamma_k}{\phi S_r}\right) \dfrac{d(\phi S_r)_{,k}}{dt}-\mathbb S_s \dfrac{dT}{dt}-\dfrac{d\Psi_s}{dt} \medskip\\
+\,\mathtt{C}^M_k M_k+\mathtt{C}_{l k}^{\nabla M}\, \overline{M}_{l,k}-\dfrac{\mathfrak q_k}{T} T_{,k}\geq 0.
\end{array}
\end{equation}

\section{ }\label{section_App_B}
Incompressibility of the solid grains implies: $J=1+\phi-\phi_0$; consequently considering the following identities
\begin{equation}\label{AppB_identities}
\begin{array}{rl}
\dot J &= J \left(F^{-1} \right)_{i \beta} \left(F^{-1} \right)_{j \beta}\, \dot E_{ij},\medskip\\
J_{,k} &= J \left(F^{-1} \right)_{i \beta} \left(F^{-1} \right)_{j \beta}\, E_{ij,k},\medskip\\
\dot J_{,k} &= \dot J \left(F^{-1} \right)_{i \beta} \left(F^{-1} \right)_{j \beta} E_{ij,k} + J \dfrac{d}{dt} \left[\left(F^{-1} \right)_{i \beta} \left(F^{-1} \right)_{j \beta} \right] E_{ij,k}+J \left(F^{-1} \right)_{i \beta} \left(F^{-1} \right)_{j \beta} \dot E_{ij,k}\medskip\\
            &=J \left(F^{-1} \right)_{i \beta} \left(F^{-1} \right)_{j \beta} \dot E_{ij,k}+ J \left(F^{-1} \right)_{i \beta} \left(F^{-1} \right)_{j \beta} E_{lm,k}\, \left(F^{-1} \right)_{l \alpha} \left(F^{-1} \right)_{m \alpha} \dot E_{ij}+\medskip\\
            &- 2 J \left(F^{-1} \right)_{i \beta} \left(F^{-1} \right)_{l \beta} E_{lm,k} \left(F^{-1} \right)_{m \alpha} \left(F^{-1} \right)_{j \alpha} \dot E_{ij}\medskip\\
            &= J  \left(F^{-1} \right)_{i \beta}  \left[\left(F^{-1} \right)_{j \eta} \left(F^{-1} \right)_{l \gamma}  \left(F^{-1} \right)_{m \alpha}   E_{lm,k} \left(\delta_{\alpha \gamma} \delta_{\eta \beta}-2 \delta_{\beta \gamma} \delta_{\alpha \eta} \right) \dot E_{ij}+ \left(F^{-1} \right)_{j \beta} \dot E_{ij,k}\right],
\end{array}
\end{equation}
the time derivatives of the Lagrangian porosity and the Lagrangian gradient of the liquid volume content can be explicitated in terms of the corresponding derivative of strain, strain gradient, degree of saturation and gradient of degree of saturation:
\begin{equation}
\label{AppB_deriv}
\dfrac{d\phi}{dt}=\dfrac{dJ}{dt}, \qquad  \dfrac{d\left(\phi S_r \right)_{,k}}{dt}= (n S_r) \dfrac{d J_{,k}}{dt}+ J_{,k} \dfrac{d (n S_r)}{dt}+(n S_r)_{,k} \dfrac{d J}{dt}+ J\, \dfrac{d (n S_r)_{,k}}{dt}
\end{equation}
Replacing the identities \eqref{AppB_identities} into equations \eqref{AppB_deriv} and these last into the dissipation relative to the solid skeleton, given by equation \eqref{Phi_s}, provides the form of $\Phi_s$ consistent with the hypothesis of incompressibility of the solid grains:
\begin{equation}
\label{Phi_s_Bishop}
\begin{array}{ll}
\Phi_s &= \left\lbrace S_{ij}+ \left[\left(\mathcal P -S_r \mathcal P_c \right) +\dfrac{\gamma_k}{J} \left(\dfrac{J_{,k}}{J} -\dfrac{(n S_r)_{,k}}{n S_r}\right)\right] J  \left(F^{-1} \right)_{i \beta}  \left(F^{-1} \right)_{j \beta}-\gamma_k\, \left(F^{-1} \right)_{i \beta}  \left(F^{-1} \right)_{j \eta} \times\right.\medskip\\
& \qquad \left. \times \left(F^{-1} \right)_{l \gamma}  \left(F^{-1} \right)_{m \alpha}   E_{lm,k} \left(\delta_{\alpha \gamma} \delta_{\eta \beta}-2 \delta_{\beta \gamma} \delta_{\alpha \eta} \right) \right\rbrace \dot E_{ij} -\phi \mathcal P_c  \dfrac{d S_r}{dt} -\dfrac{\gamma_k}{n S_r}\, \dfrac{d (n S_r)_{,k}}{dt}+\medskip\\
&\qquad + \left(P_{ijk}- \gamma_k \left(F^{-1} \right)_{i \beta} \left(F^{-1} \right)_{j \beta}  \right) \dot E_{ij,k}-\dfrac{d\Psi_s}{dt}.
\end{array}
\end{equation}
The extended form of Bishop's stress can therefore be defined as follows:
\begin{equation}
\label{Bishop_stress_complete}
\begin{array}{ll}
S'_{ij} & \!\!\!\! := S_{ij}+ \left[\left(\mathcal P -S_r \mathcal P_c \right) +\dfrac{\gamma_k}{J} \left(\dfrac{J_{,k}}{J} -\dfrac{(n S_r)_{,k}}{n S_r}\right)\right] J  \left(F^{-1} \right)_{i \beta}  \left(F^{-1} \right)_{j \beta}-\gamma_k  \left(F^{-1} \right)_{i \beta}  \left(F^{-1} \right)_{j \eta} \times\medskip\\
& \quad\, \times \left(F^{-1} \right)_{l \gamma}  \left(F^{-1} \right)_{m \alpha}   E_{lm,k} \left(\delta_{\alpha \gamma} \delta_{\eta \beta}-2 \delta_{\beta \gamma} \delta_{\alpha \eta} \right),\medskip\\
P'_{ijk} &\!\!\!\! :=P_{ijk}- \gamma_k \left(F^{-1} \right)_{i \beta} \left(F^{-1} \right)_{j \beta},
\end{array}
\end{equation}
which indeed coincides with that given in equation \eqref{Bishop_diss}, once the hypothesis of small deformations has been stated.

\section{ }\label{section_App_C}
The liquid, initially trapped within the beads, is assumed to be squeezed out because of the mutual displacement of the beads, so that the macro-scale gradient of $\phi S_r$ has the same direction as that of $J$ and the liquid reaches the boundary of the RVE. In particular the displaced configuration of the liquid, associated to a non-vanishing gradient of $J$ (along the $X_1$ direction) is obtained assuming the interfaces between the liquid and the wet air to keep a constant contact angle with the solid grains and to maintain a circumferential shape, see Figure~\ref{fig:chim&gm}. 

\begin{figure}[h]
\centering
\includegraphics[width=.7\textwidth]{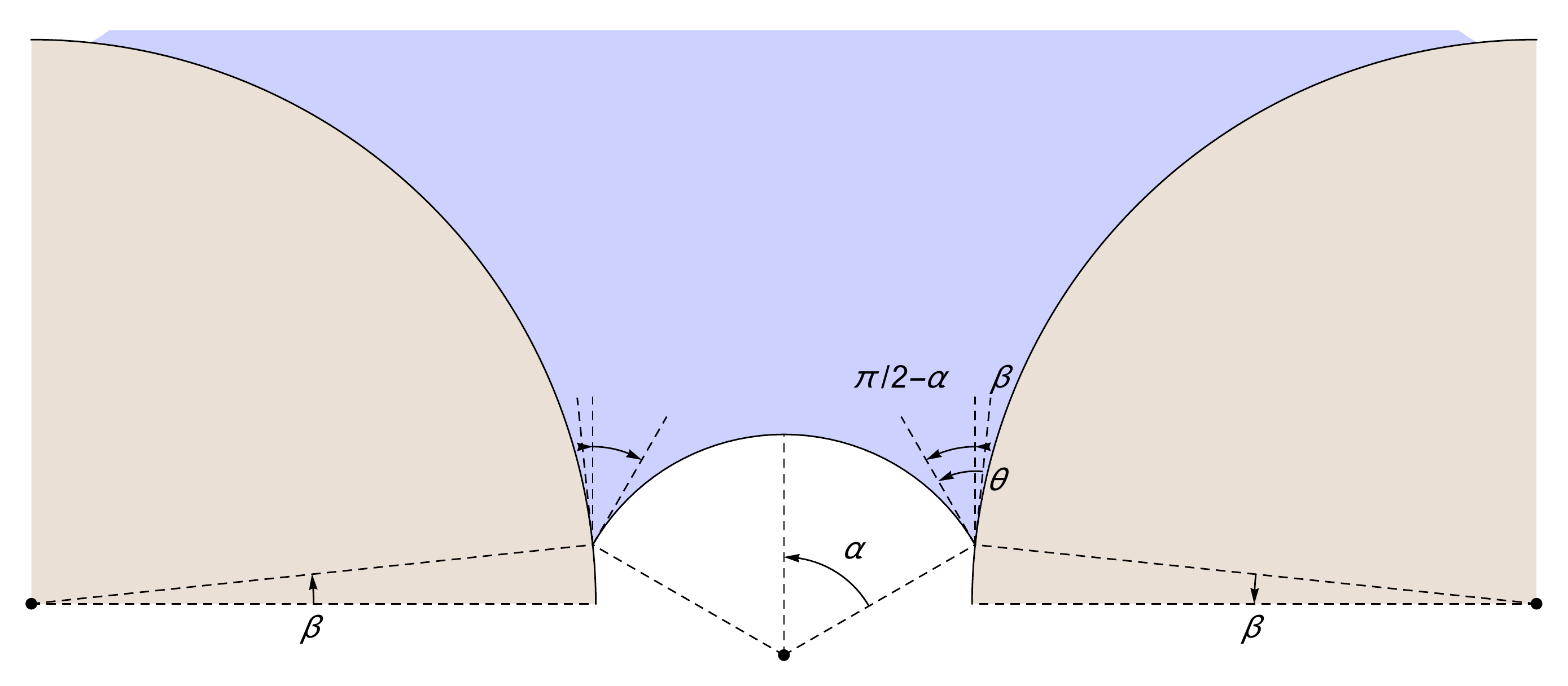}
\caption{Detail of the refence configuration $\mathcal B^0$ of the RVE; the domain occupied by the liquid is parametrized by the three non-independent angles $\alpha$, $\beta$ and $\theta$.}
\label{fig:AppendixMicro}
\end{figure}

Let $2\alpha$ be the central angle relative to any of the circumferential arches which characterize the reference shape of the liquid, moreover let $\beta$ (or $\pi-\beta$) be the azimuthal angle of the junction line among the liquid, the wet air and the solid grain, with respect to the center of each bead; the contact angle $\theta$ is therefore $\theta=\frac{\pi}{2}-\alpha+\beta$, see Figure~\ref{fig:AppendixMicro}. Apparently, once fixed the amount of the trapped liquid, the central angles $\alpha$ and the the angles $\beta$, relative to the four circumferential interfaces in the reference configuration of the RVE are the same.
For the sake of convenience the circumferential interfaces between the liquid and the wet air are labeled as North (N), East (E), South (S) and West (W)-Arch, respectively, turning clockwise from the North. 

Due to the displacement of the beads centers, given by equation \eqref{displ_micro}
the centers of the E-Arch and the W-Arch are displaced in the direction of the gradient of  $J$. Their curvatures change so as to keep constant the contact angle $\theta$ at the junction line, and in particular the curvature of the E-Arch decreases, as it is  pulled out of the intergranular space, whilst that of the W-Arch increases, as it is pushed inwards it. Notice that some liquid must be ejected from the pore chamber so as to make consistent the placement of the W-Arch with the requirement of displacing the liquid to the boundary of the RVE, keeping fixed the contact angle, see Figure~\ref{fig:chim&gm}. Finally in order to guarantee a constant value of the ratio between the amount of liquid (in volume) and the volume of the pore space, say of the degree of saturation, the N-Arch and the S-Arch are displaced and rotated keeping their curvatures almost constant during the deformation process.

Thus the problem of finding out the current placement of the liquid particles can be formulated as follows: given the third order tensor $\mathbb A$ determine the $\beta$-angles relative to the N-Arch, the E-Arch and the W-Arch as well as the value of the spherical part of the macro-scale deformation gradient, $a$, so that the saturation degree remains constant during deformation and the amount (in volume) of liquid initially trapped within the beads equals the amount (in volume) of the displaced liquid. Apparently in order to force the liquid to squeeze out of the pore chamber, a negative value of $a$ should be obtained.

\section{ }\label{section_App_D}
Following the microporomechanics approach summarized by \cite{Dormieuxbook2006} the macro-scale gradient of $G_{li}$ is calculated starting from the definition of $G_{li}$ as the apparent average of $G^{\textrm{m}}_{li}$ relative to the liquid phase, see equation \eqref{microIterpretSk}$_2$; consequently on has: 
\begin{equation}\label{AppD_macroderiv}
G_{li,k}=\dfrac{\partial}{\partial X_k}\dint\! f(Z-X)\, H^0_{liq} (Z)\, G^{\textrm{m}}_{li} (Z),
\end{equation}
where $f(Z)$ is a $C^{\infty}$ function defined, at the micro-scale, on the entire Euclidean space of positions, which is equal to zero outside the reference configuration of the RVE, and is normalized so that its integral over the entire space is equal to one. The integral in equation \eqref{AppD_macroderiv} is calculated with respect to $Z$ variables. In equations \eqref{microIterpretJk}-\eqref{microIterpretSk} this function has been assumed to tend to the characteristic function of $\mathcal B^0$ divided by its volume.
Finally $H^0_{liq} (Z)$ is the characteristic function of the liquid, in the reference configuration of the solid, which is also defined on the entire space of positions.

According with the definition of derivatives in the sense of distributions, the following chain of equalities hold true:
\begin{equation}\label{AppD_chain}
\begin{array}{rl}
G_{li,k}&=-\dint\! \dfrac{\partial f(Z-X)}{\partial Z_k}\, H^0_{liq} (Z)\, G^{\textrm{m}}_{li} (Z)=\dint\! f(Z-X)\dfrac{\partial}{\partial Z_k} \left[ H^0_{liq} (Z)\, G^{\textrm{m}}_{li} (Z)\right]=\medskip\\
        &=\,\,\,\, \dint\! f(Z-X)\, H^0_{liq} (Z)\,\dfrac{\partial}{\partial Z_k} G^{\textrm{m}}_{li} (Z) + \dint\! f(Z-X)\, G^{\textrm{m}}_{li} (Z)\,\dfrac{\partial}{\partial Z_k}  H^0_{liq} (Z)=\medskip\\
        &=\,\,\,\, \left\vert \mathcal B^0 \right\vert^{-1} \left(\dint_{\mathcal D^0_{liq}\cap \mathcal B^0} \dfrac{\partial}{\partial Z_k} G^{\textrm{m}}_{li} (Z) \right)- \dint_{\mathcal D^0_{liq}} \dfrac{\partial}{\partial Z_k}\left[ f(Z-X) G^{\textrm{m}}_{li} (Z)\right]=\medskip\\
        &=\,\,\,\, \left\vert \mathcal B^0 \right\vert^{-1} \left(\dint_{\partial (\mathcal D^0_{liq}\cap \mathcal B^0)} G^{\textrm{m}}_{li} (Z) n_k \right)- \dint_{\partial\mathcal D^0_{liq}} f(Z-X) G^{\textrm{m}}_{li}(Z) n_k=\medskip\\
        &=\,\,\,\, \left\vert \mathcal B^0 \right\vert^{-1} \left(\dint_{\partial (\mathcal D^0_{liq}\cap \mathcal B^0)} G^{\textrm{m}}_{li} (Z) n_k - \dint_{\partial\mathcal D^0_{liq}\cap \mathcal B^0} G^{\textrm{m}}_{li}(Z) n_k\right)=
        \left\vert \mathcal B^0 \right\vert^{-1} \left(\dint_{\partial (\mathcal D^0_{liq}\cap \mathcal B^0) \backslash \mathcal I^0_f} G^{\textrm{m}}_{li}(Z) n_k\right),
\end{array}
\end{equation}
where $\mathcal I^0_f$ is the  interface between the liquid and the grain and the liquid and the gas within the reference configuration of the RVE, so that $\partial (\mathcal D^0_{liq}\cap \mathcal B^0) \backslash \mathcal I^0_f$ represents that part of the liquid, pulled-back into the reference configuration of the solid, which overlaps the boundary of the reference configuration itself. $\left\vert \mathcal B^0 \right\vert$ is the volume of the reference configuration of the RVE.

According with equation \eqref{AppD_chain}, even if a complete statement of the displacement of the liquid is not explicitly formulated, the macroscopic gradient of the liquid content $\phi S_r$ can be deduced from the micro-scale, assuming a suitable form of the gradient of $g^{\mathrm{m}}$ only over $\partial (\mathcal D^0_{liq}\cap \mathcal B^0) \backslash \mathcal I^0_f$. Let  the RVE be that of the left panel of Figure~\ref{fig:chim&gm}, the only part of the boundary of $\mathcal B_0$ which provides a non-vanishing contribution to $G_{li,k}$, and therefore to $(\phi S_r)_{,k}$ is the right intergranular channel, which means that one only needs an instance on how the map $g^{\mathrm{m}}$ transforms a thin neighborhood of this part of the boundary into the corresponding domain of the initial configuration of the liquid.
\end{appendices}

\bigbreak\noindent
\section*{References}

\end{document}